\documentclass[preprint2]{aastex62}

\usepackage{amsmath}
\usepackage[capitalise]{cleveref} 
\usepackage{textcomp} % \textmu
\usepackage{array}
\usepackage[english]{babel}

\received{}
\revised{}
\accepted{}

\submitjournal{ApJ}

\shorttitle{Wind--driven chemistry in the atmosphere of HD~189733\MakeLowercase{b}}
\shortauthors{Drummond et al.}

\begin{document}

\title{The 3D thermal, dynamical and chemical structure of the atmosphere of HD 189733b: implications of wind-driven chemistry for the emission phase curve}

\correspondingauthor{Benjamin Drummond}
\email{b.drummond@exeter.ac.uk}

\author{Benjamin Drummond}
\affil{Astrophysics Group, University of Exeter, EX4 2QL, Exeter, UK}

\author{Nathan J. Mayne}
\affil{Astrophysics Group, University of Exeter, EX4 2QL, Exeter, UK}

\author{James Manners}
\affil{Met Office, Exeter, EX1 3PB, UK}
\affil{Astrophysics Group, University of Exeter, EX4 2QL, Exeter, UK}

\author{Isabelle Baraffe}
\affil{Astrophysics Group, University of Exeter, EX4 2QL, Exeter, UK}
\affil{Univ Lyon, Ens de Lyon, Univ Lyon1, CNRS, CRAL, UMR5574, F-69007, Lyon, France}

\author{Jayesh Goyal}
\affil{Astrophysics Group, University of Exeter, EX4 2QL, Exeter, UK}

\author{Pascal Tremblin}
\affil{Maison de la simulation, CEA, CNRS, Univ. Paris-Sud, UVSQ, Université Paris-Saclay, 91191 Gif-Sur-Yvette, France}

\author{David K. Sing}
\affil{Department of Earth and Planetary Sciences, Johns Hopkins University, Baltimore, MD, USA}

\author{Krisztian Kohary}
\affil{Astrophysics Group, University of Exeter, EX4 2QL, Exeter, UK}

% ABSTRACT
\begin{abstract}In this paper we present three-dimensional atmospheric simulations of the hot Jupiter HD~189733b under two different scenarios: local chemical equilibrium and including advection of the chemistry by the resolved wind. Our model consistently couples the treatment of dynamics, radiative transfer and chemistry, completing the feedback cycle between these three important processes. The effect of wind--driven advection on the chemical composition is qualitatively similar to our previous results for the warmer atmosphere of HD~209458b, found using the same model. However, we find more significant alterations to both the thermal and dynamical structure for the cooler atmosphere of HD~189733b, with changes in both the temperature and wind velocities reaching $\sim10\%$. We also present the contribution function, diagnosed from our simulations, and show that wind--driven chemistry has a significant impact on its three--dimensional structure, particularly for regions where methane is an important absorber. Finally, we present emission phase curves from our simulations and show the significant effect of wind--driven chemistry on the thermal emission, particularly within the 3.6 \textmu m Spitzer/IRAC channel. 
\end{abstract}

%%%%%%%%%%%%%%%%%%%%%%%%
% INTRODUCTION
%%%%%%%%%%%%%%%%%%%%%%%%
\section{Introduction}

Interpreting observations of exoplanet atmospheres requires comparison with theoretical models in order to infer the properties of the atmosphere, such as the thermal structure and the chemical composition. Whether this is done by a retrieval method \citep[e.g.][]{IrwTd08,MadS09,WalTR15} or by comparison with forward models \citep[e.g.][]{Barman2005,ForML05,MadBC11,Moses2011,GoyMS18} these theoretical tools are most often in the form of a one--dimensional (1D) model. However, the results of three--dimensional (3D) radiative-hydrodynamic codes \citep{ShoG02,DobL08,Showman2009,MenR09,HenMP11,MayBA14,AmuMB16} clearly show that the atmospheres of highly--irradiated tidally--locked planets are far from being horizontally symmetric.

For tidally---locked hot Jupiters, intense stellar irradiation heats the dayside to thousands of Kelvin while the nightside is significantly cooler, driving fast zonal winds ($\sim10^3$ ${\rm m}~{\rm s}^{-1}$) that redistribute heat around the planet \citep[e.g.][]{Showman2009,DobA13,AmuMB16}. High resolution transmission spectra revealing Doppler shifting of atomic absorption lines \citep{Snedd10,Louden2015} as well as measurements of the emission as a function of orbital phase \citep[e.g.][]{KnuLF12,ZelLK14} have since confirmed these theoretical predictions. However, \citet{DanCS18} recently reported a westward shift in the location of the hot spot, that contradicts the eastward shifts unanimously predicted by current 3D models.

Recently, \citet{DDC17} demonstrated the importance of considering the 3D structure of the atmosphere when interpreting the emission phase curve. They showed that the contribution function, the peak of which indicates the pressure level of the photosphere, varies with longitude. In general, they found that the photosphere typically lies at lower pressures on the dayside compared with the nightside, with the conclusion that the observed emission originates from different pressure levels of the atmosphere throughout the orbit of the planet.

Most 3D models of hot exoplanet atmospheres to date share a common limitation: the assumption of a cloud--free atmosphere that is in local chemical equilibrium. However, many investigations with 1D models have shown the importance of vertical transport and photochemistry that drive the chemistry away from equilibrium \citep[e.g.][]{LinLY2010,Moses2011,Venot2012,Zahnle2014,DruTB16,TsaLG2017}. Large dayside-nightside temperature contrasts naturally lead to large dayside-nightside contrasts in the chemical equilibrium composition. In addition, the properties of clouds have recently been investigated using 3D models of various complexity \citep[][]{LeeDH16,ParFS16,RomR17,LinMB18}.

\citet{CooS06} used simplified chemistry and radiative transfer schemes with a 3D hydrodynamics code and found that vertical transport from high pressure regions, which are horizontally uniform, leads to homogenisation of the chemistry at lower pressures. \citet{AguPV14} used a ``psuedo--2D'' code (in practice a 1D column with a time--varying temperature profile) and found that horizontal (zonal) transport has a more important effect than vertical transport and contaminates the nightside chemistry with  that of the hot dayside.

Recently, we coupled the simplified chemical scheme of \citet{CooS06} to our own 3D model, including a state-of-the-art radiative transfer scheme \citep{AmuMB16}, completing the feedback cycle between the dynamics, radiation and gas--phase chemistry \citep{DruMM18}. We found that a combination of horizontal (zonal and meridional) and vertical transport ultimately determine the abundance of methane in the atmosphere of HD~209458b. The overall effect of 3D transport is to increase the mole fraction of methane leading to more prominent methane absorption features in both the simulated transmission and emission spectra, compared with the chemical equilibrium case.

In this paper we present results from the same 3D radiative-hydrodynamics code, coupled with the same gas--phase chemical relaxation scheme \citep{CooS06,DruMM18}, applied to the specific case of HD~189733b. We compare our results with our previous simulations of the warmer atmosphere of HD~209458b and present the response of the atmospheric temperature and circulation to wind--driven chemistry, in Section \ref{section:chemistry} and Section \ref{section:temp}. We then present contribution functions diagnosed from our 3D simulations and compare with the results of \citet{DDC17}, where we find significant qualitative differences, in Section \ref{section:cf}. Finally we consider the effect of departures of the chemistry from chemical equilibrium on the emission phase curve in Section \ref{section:phase}.

%%%%%%%%%%%%%%%%%%%%%%%%
% MODEL DESCRIPTION
%%%%%%%%%%%%%%%%%%%%%%%%
\section{Model description}

\subsection{The Unified Model}
\label{section:um}

We use the Met Office Unified Model (UM) to simulate the atmosphere of HD~189733b. The UM has been used in previous works to simulate the exoplanet atmospheres of HD~209458b \citep{MayBA14,MayDB17,AmuMB16,LinMB18,DruMM18}, GJ~1214b \citep{DruMB18} and Proxima Centauri b \citep{BouMD17,LewLB18}.

The dynamical core of the UM (ENDGame) solves the deep-atmosphere, non-hydrostatic Navier-Stokes equations \citep{WooSW14,MayBA14b,MayDB17}. The heating rates are computed using the open-source SOCRATES\footnote{\url{https://code.metoffice.gov.uk/trac/socrates}} radiative transfer scheme \citep{Edw96,EdwS96} which has been updated and tested for the high-temperature, hydrogen-dominated conditions of hot Jupiter atmospheres \citep{AmuBT14,AmuMB16,AmuTM17}. The chemical composition is derived using the same methods as described in \citet{DruMM18}: an analytical formula to derive the chemical equilibrium abundances and a chemical relaxation scheme based on \citet{CooS06}.

The chemical relaxation method describes the time--dependence of a chemical species by relaxing the mole fraction toward a prescribed equilibrium profile on some chemical timescale \citep{CooS06,TsaKL17,DruMM18}. The equilibrium profile is taken to be the mole fraction corresponding to chemical equilibrium while the timescale is estimated or parameterised based on the elementary reactions involved in the interconversion of the relevant chemical species. The accuracy of the chemical relaxation method is primarily determined by the accuracy of the chosen chemical timescale \citep{TsaKL17}.

In  Appendix \ref{section:app3} we compare the results of the chemical relaxation method against a full chemical kinetics method within the framework of a 1D model and find a good agreement. We also demonstrate the sensitivity of our results to the chosen chemical timescale in Appendix \ref{section:test} by artificially increasing/decreasing the value of the chemical timescale by a factor of 10. We find that varying the timescale does not significantly effect the final mole fraction of methane over a large pressure range and therefore uncertainty in the precise value of the timescale is unlikely to effect our conclusions.

Tracer advection is handled using the extensively tested semi--implicit semi--Lagrangian scheme of the UM \citep{WooSW14}. In Appendix \ref{section:cons} we test the conservation of the global mass of elemental carbon and oxygen for our simulations. The model conserves the global mass of these elements to within better than 99.9\%.

\subsection{Model parameters and setup}
\label{sect:mod_param}
We use the planetary and stellar parameters of HD~189733b from \citet{Southworth2010} which we summarise in \cref{table:params}. The intrinsic temperature $T_{\rm int}$ is the blackbody temperature of the net intrinsic flux at the lower boundary, accounting for heat escaping from the interior \citep{AmuMB16}. A full description of the vertical damping terms ($R_w$ and $\eta_s$) can be found in \citet{MayBA14}.

For the stellar irradiation spectrum we use the Kurucz spectrum for HD~189733\footnote{\url{http://kurucz.harvard.edu/stars.html}}. We assume solar elemental abundances from \citet{Asplund2009}. For a complete description of the included opacities see \citet{AmuMB16}. For the main model integration the radiation spectrum is divided into 32 bands for the heating rate calculations \citep{AmuMB16}. However, for the simulated observation calculations we restart the model, starting from the state at 1000 days, and run for a small number of timesteps using 500 spectral bands, to obtain a higher spectral resolution \citep[as done in][]{BouMD17,DruMB18,DruMM18,LinMB18}.

The model setup is broadly the same as in our previous coupled radiative transfer simulations of hot Jupiters \citep{AmuMB16,DruMM18}. As in \citet{DruMM18} we perform two simulations that are identical except that one assumes local chemical equilibrium while the other includes the effect of advection, due to the large-scale resolved wind, and chemical evolution via the chemical relaxation method \citep{CooS06}. We refer to these two simulations as the ``equilibrium'' and ``relaxation'' simulations, respectively.

The model is initialised with zero winds and with a horizontally uniform thermal profile. For the latter we use a radiative-convective equilibrium temperature profile from the 1D ATMO model using the same stellar and planetary parameters \citep[][]{DruTB16}. The simulations are integrated for 1000 Earth days (hereafter days refers to Earth days) by which point the maximum wind velocities have ceased to evolve, though we note that the deep atmosphere ($P\gtrsim10^6$ Pa) has not reached a steady-state \citep{AmuMB16,MayDB17}, though this does not affect lower pressures. Unless otherwise stated, all figures show the simulations at 1000 days. The total axial angular momentum is conserved to within 99\% for the simulations presented here.

We note that throughout the results sections of this paper we regularly calculate the difference of several quantities (e.g. temperature) between the relaxation and equilibrium simulations, to aid interpretation. Unless otherwise stated, the difference we refer to is the absolute difference, $A^{\rm relaxation}-A^{\rm equilibrium}$. In some cases, where more useful, we consider the relative difference, $(A^{\rm relaxation}-A^{\rm equilibrium})/A^{\rm relaxation}$, instead.

\begin{table}
\centering
\setlength\extrarowheight{2pt}
\begin{tabular}{l r}
\hline\hline
Parameter & Value \\
\hline
Mass, $M_{\rm P}$ & 2.18$\times10^{27}$ kg \\
Radius, $R_{\rm P}$ & $8.05\times10^{7}$ m \\
Semi major axis, $a$ & 0.031 AU\\
Surface gravity, $g_{\rm surf}$ & 22.49 ms$^{-2}$ \\
Intrinsic temperature, $T_{\rm int}$ & 100 K \\
Lower boundary pressure, $P_{\rm lower}$ & $2\times10^7$ Pa \\
Rotation rate, $\Omega$ & $3.28\times10^{-5}$ s$^{-1}$ \\
Specific heat capacity, $c_P$ & 13 ${\rm kJ}~{\rm kg}^{-1}~{\rm K}^{-1}$ \\
Specific gas constant, $R$ & 3556.8 ${\rm J}~{\rm kg}^{-1}~{\rm K}^{-1}$ \\
Vertical damping coefficient, $R_w$ & 0.15 \\
Vertical damping extent, $\eta_s$ & 0.75 \\
Horizontal resolution & 144$\times$90 \\
Vertical resolution & 66 \\
Dynamical time step & 30 s\\
Radiative time step & 150 s \\
\hline
\end{tabular}
\caption{Key model parameters. Stellar and planetary parameters for HD~189733b adapted from \citet{Southworth2010}.}
\label{table:params}
\end{table}

%%%%%%%%%%%%%%%%%%%%%%%%
% Results
%%%%%%%%%%%%%%%%%%%%%%%%
%%%%%%%%%%%%%%%%%%%%%%%%%%%%%%%%%%%%
\section{Wind--driven chemistry in the atmosphere of HD~189733\MakeLowercase{b}}
\label{section:chemistry}
%%%%%%%%%%%%%%%%%%%%%%%%%%%%%%%%%%%%

In this section we present the thermal, dynamical and chemical structure of the atmosphere. We show the thermal and dynamical structure of the equilibrium simulation, and compare the chemical structure of the equilibrium and relaxation simulations.

% Winds and temperatures
\begin{figure*}
  \begin{center}
    \includegraphics[width=0.45\textwidth]{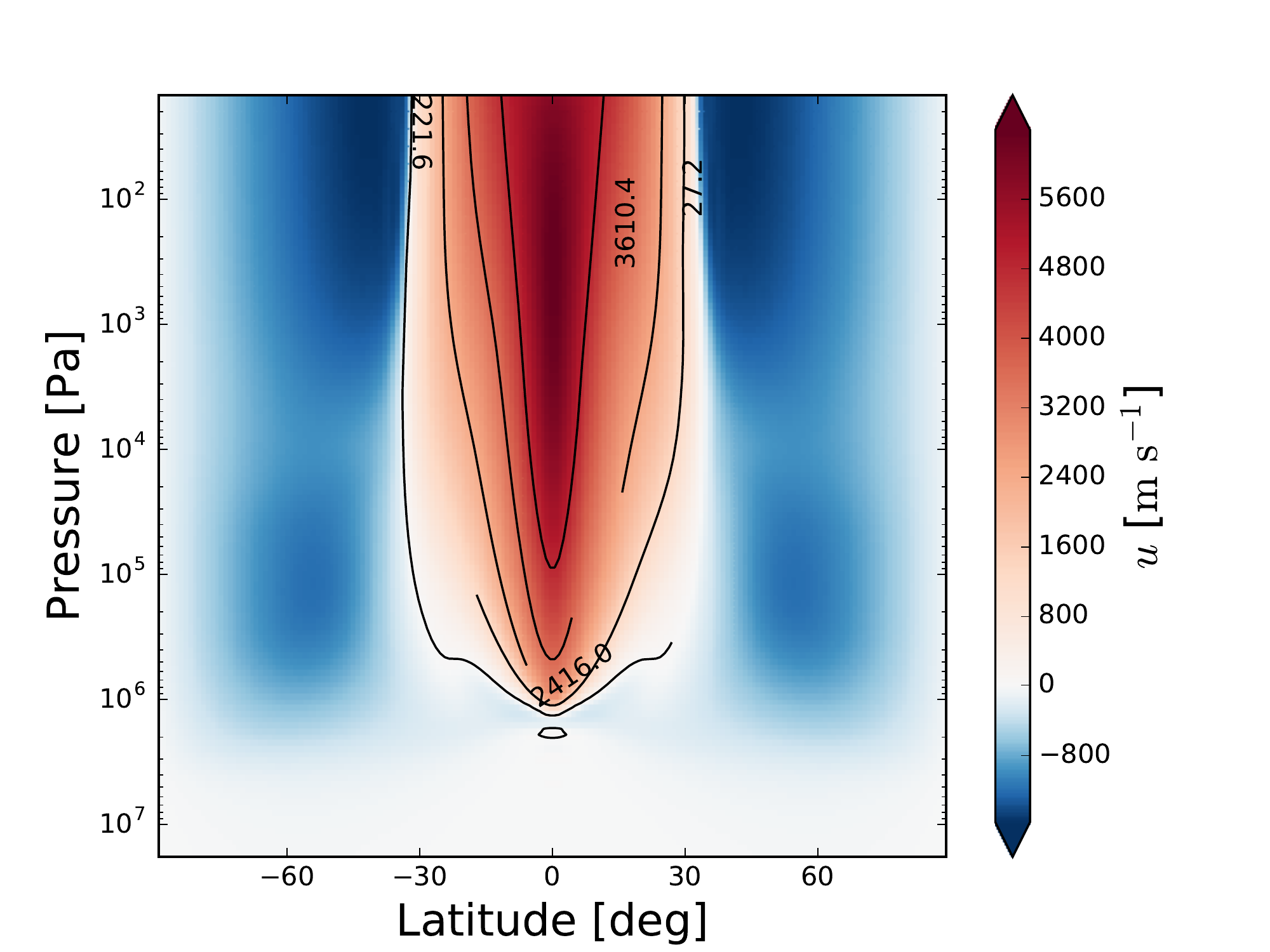}
    \includegraphics[width=0.45\textwidth]{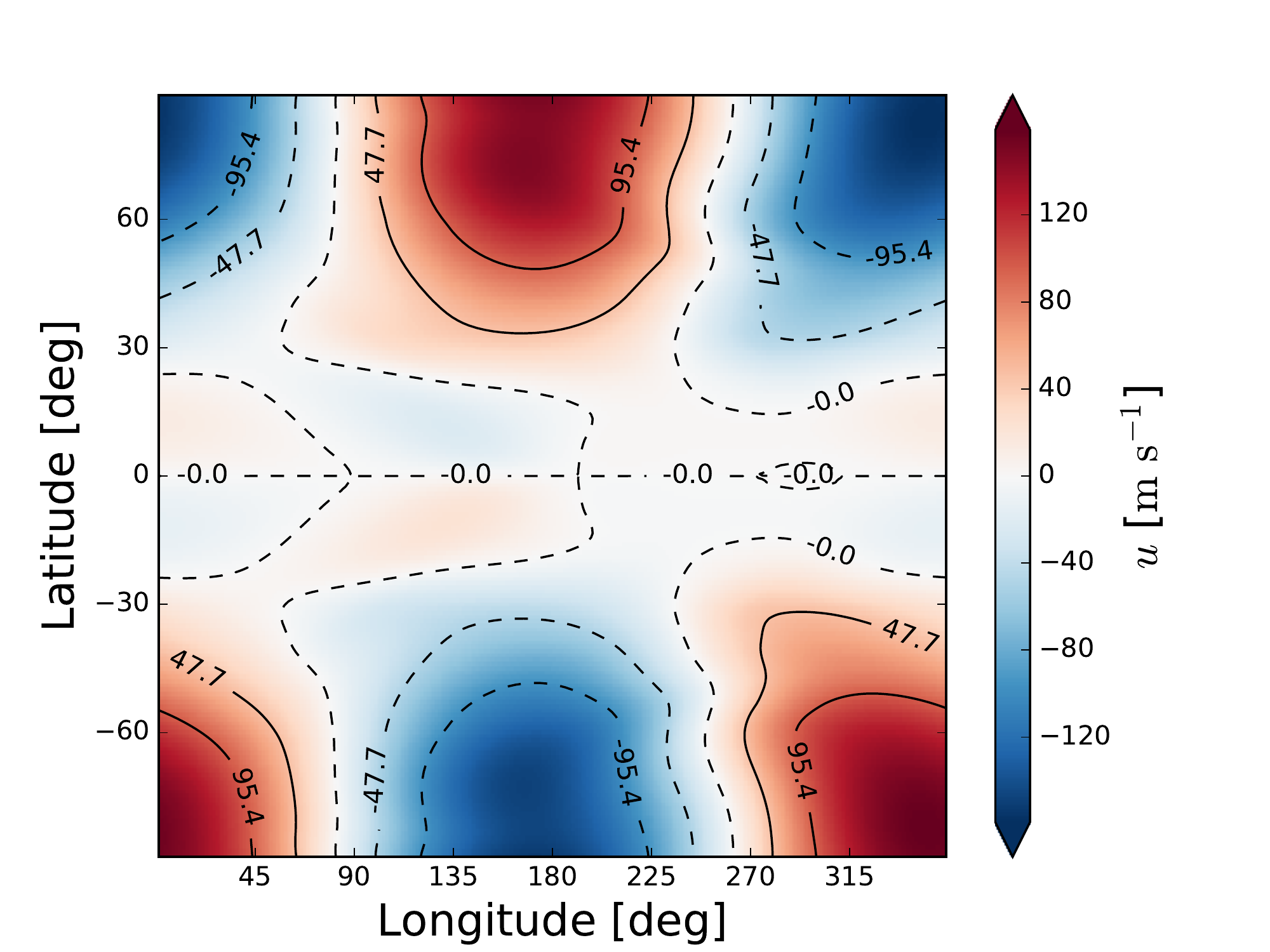} \\
    \includegraphics[width=0.45\textwidth]{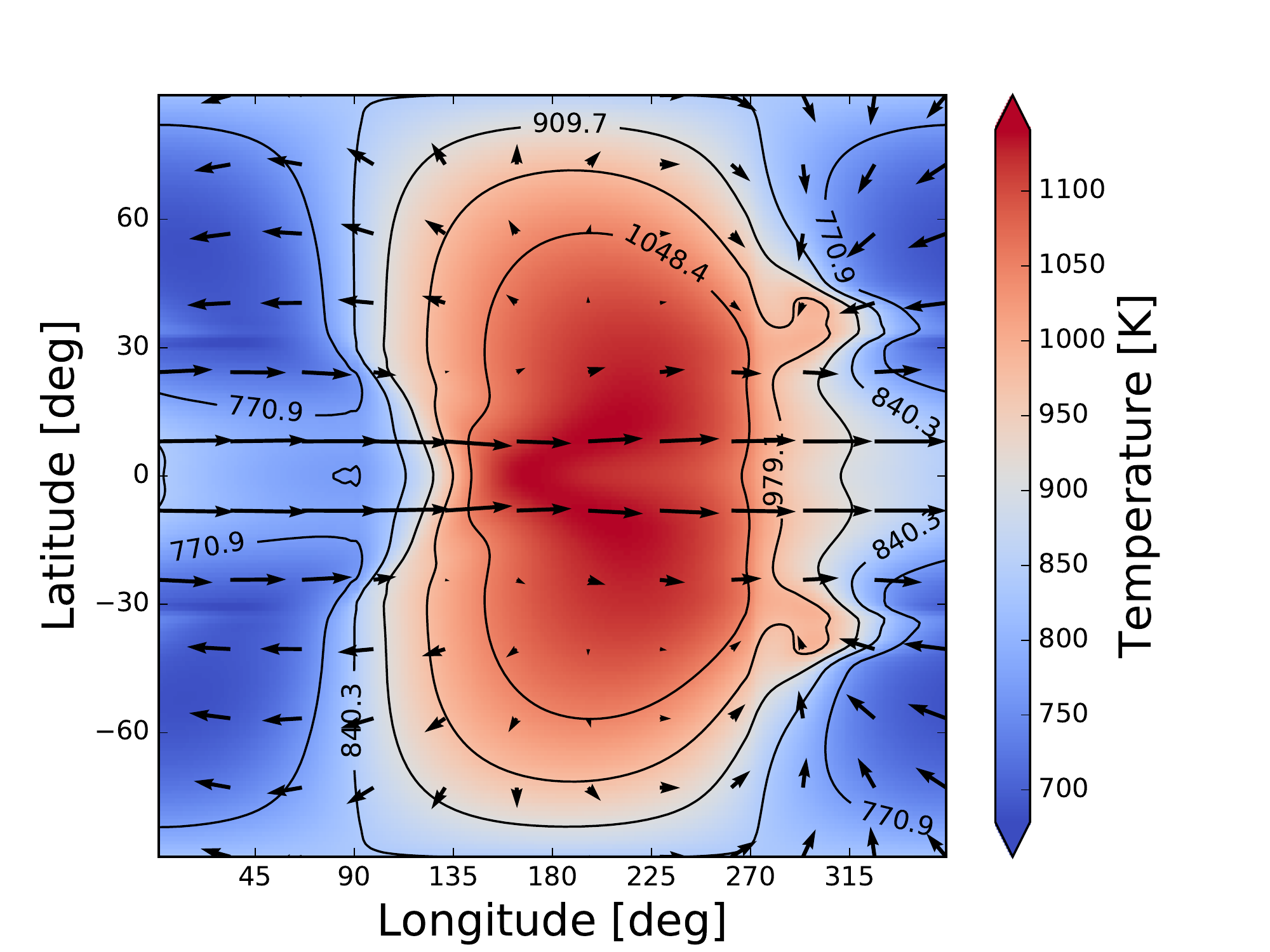}
    \includegraphics[width=0.45\textwidth]{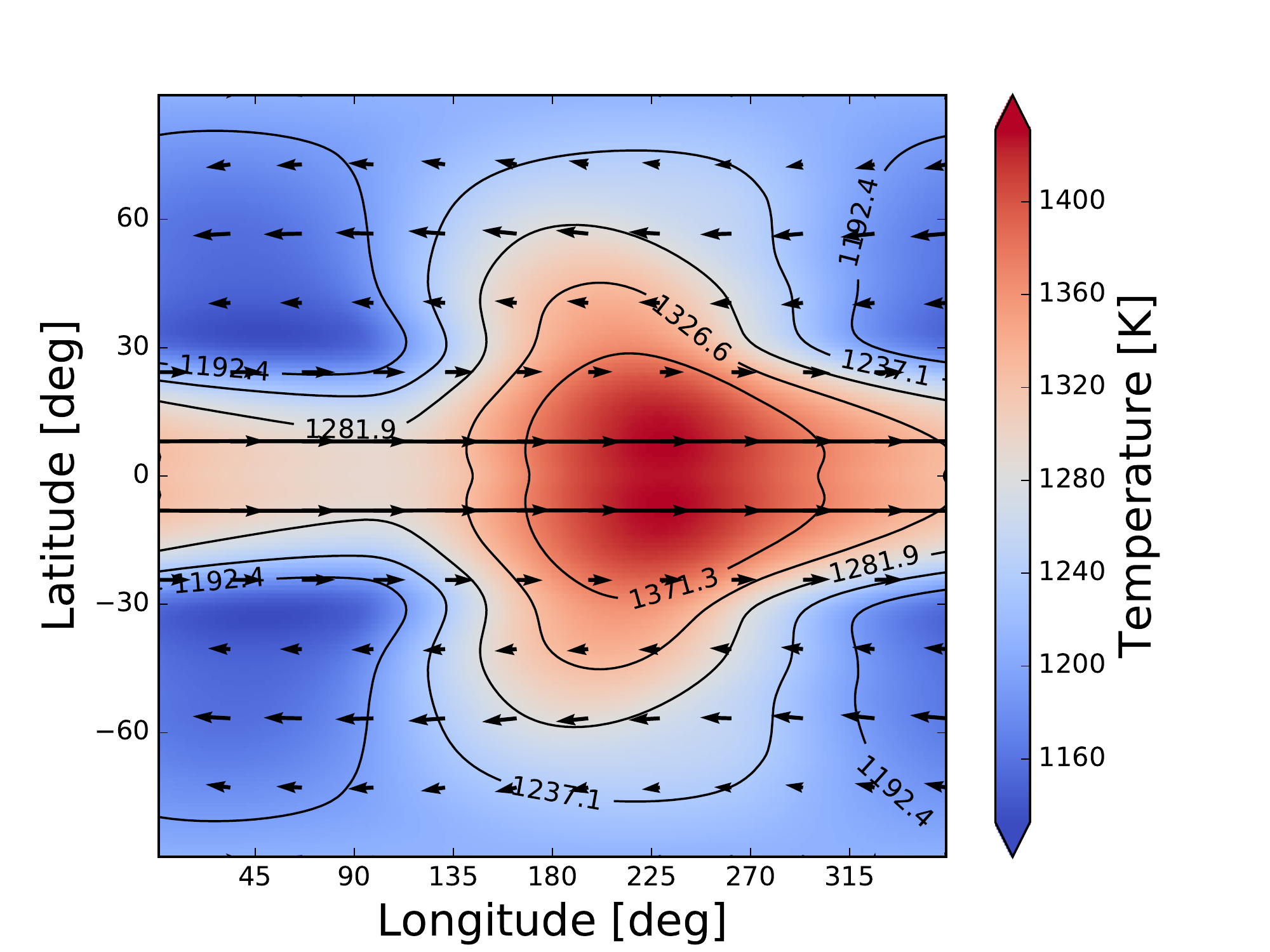} \\
    \includegraphics[width=0.45\textwidth]{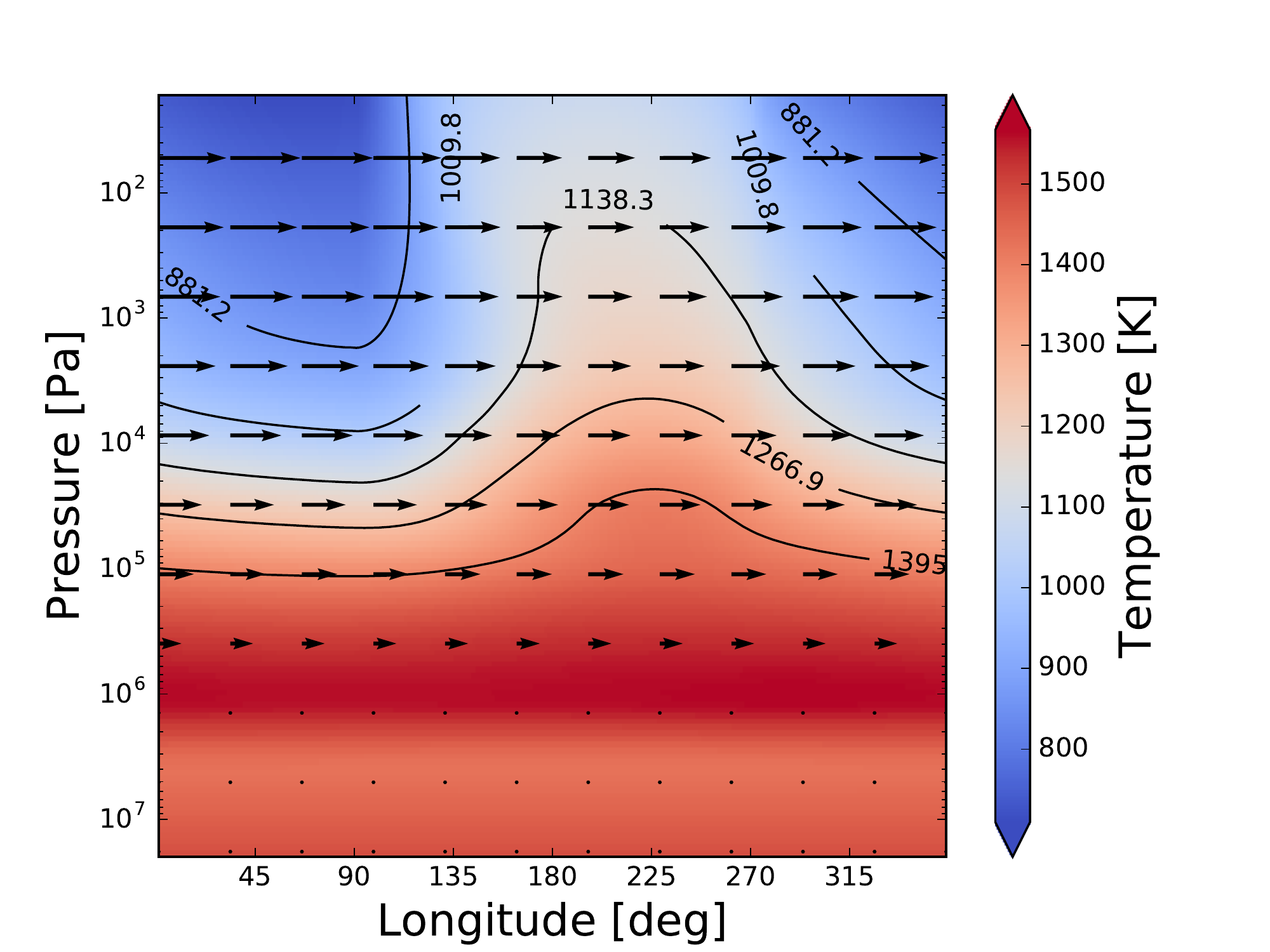}
   \includegraphics[width=0.45\textwidth]{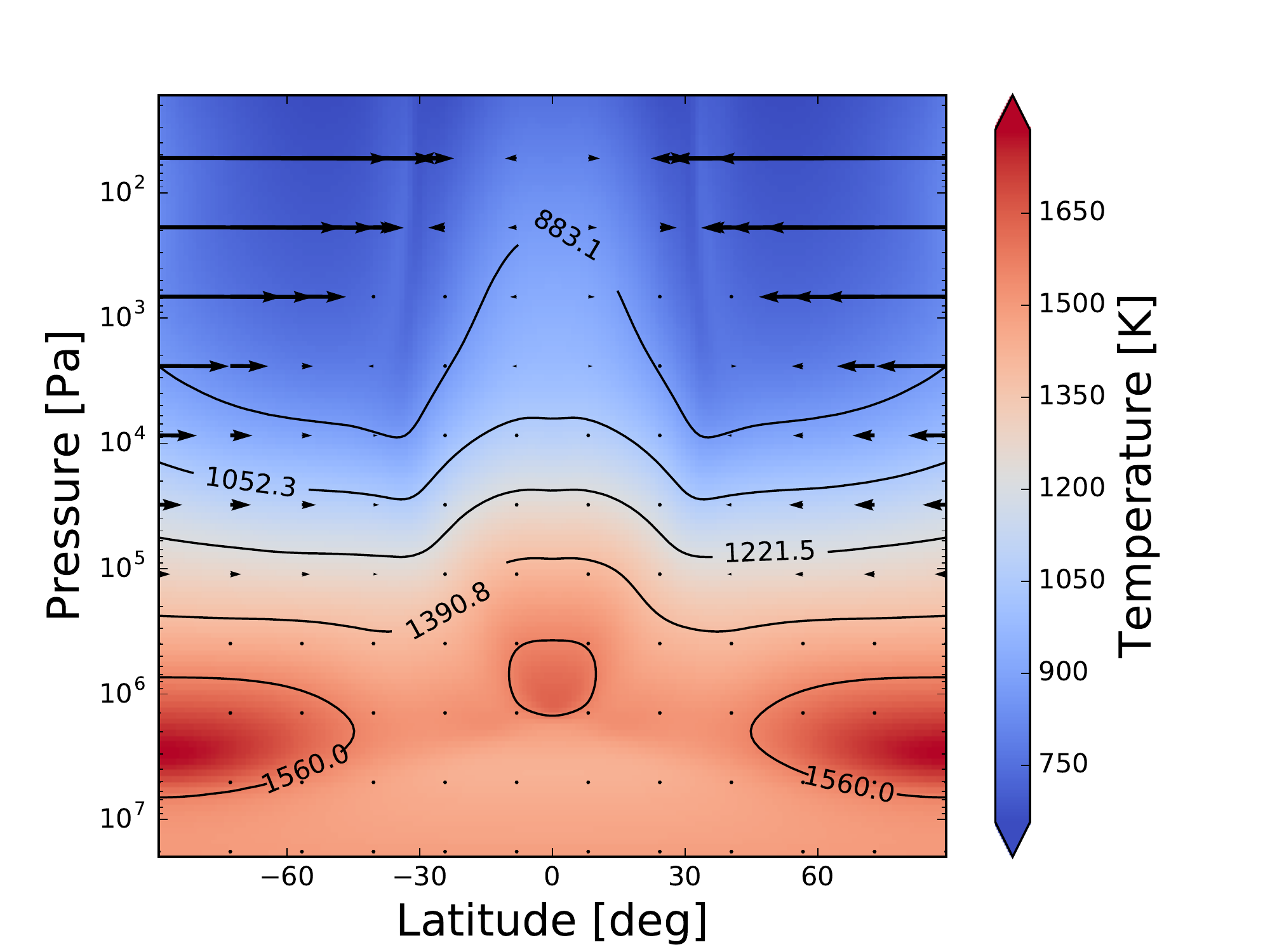}
  \end{center}
\caption{Top left panel: zonal-mean zonal wind velocity (colour scale and black contours) temporally averaged between 800 and 1000 days. Top right panel: meridional wind velocity (colour scale and black contours) on the 5$\times$10$^4$ Pa isobaric surface. Middle row: the atmospheric temperature (colour scale and black contours) on the 1$\times$10$^2$ Pa  isobaric surface (left) and on the 5$\times$10$^4$ Pa isobaric surface (right) with horizontal wind velocity vectors (black arrows). Bottom left: area--weighted meridional--mean (between $\pm20^{\circ}$ latitude) of the atmospheric temperature (colour scale and black contours) with vertical-zonal wind velocity vectors (black arrows). Bottom right panel: atmospheric temperature (colour scale and black contours) at a longitude of 0$^{\circ}$ with meridional--vertical wind vectors (black arrows).}
\label{figure:wind_temp}
\end{figure*}

\cref{figure:wind_temp} shows the zonal-mean zonal wind velocity, meridional wind velocity at $P=5\times10^4$ Pa and temperature structure for the chemical equilibrium simulation of HD~189733b. The wind velocities and temperatures are qualitatively similar to previous 3D simulations of HD~189733b \citep{Showman2009,DobA13,RauM13,KatSL16}. The circulation is characterised by an equatorial jet with a maximum wind velocity of $\sim6$ ${\rm km}$ ${\rm s}$$^{-1}$ with slower retrograde circulation at higher latitudes. The dayside-nightside temperature contrast (typically hundreds of Kelvin) increases with decreasing pressure while the hot spot moves closer to the substellar point, due to the decreasing radiative timescale with decreasing pressure \citep[e.g.][]{Iro2005}.

% Equilibrium fractions
\begin{figure*}
  \begin{center}
    \includegraphics[width=0.32\textwidth]{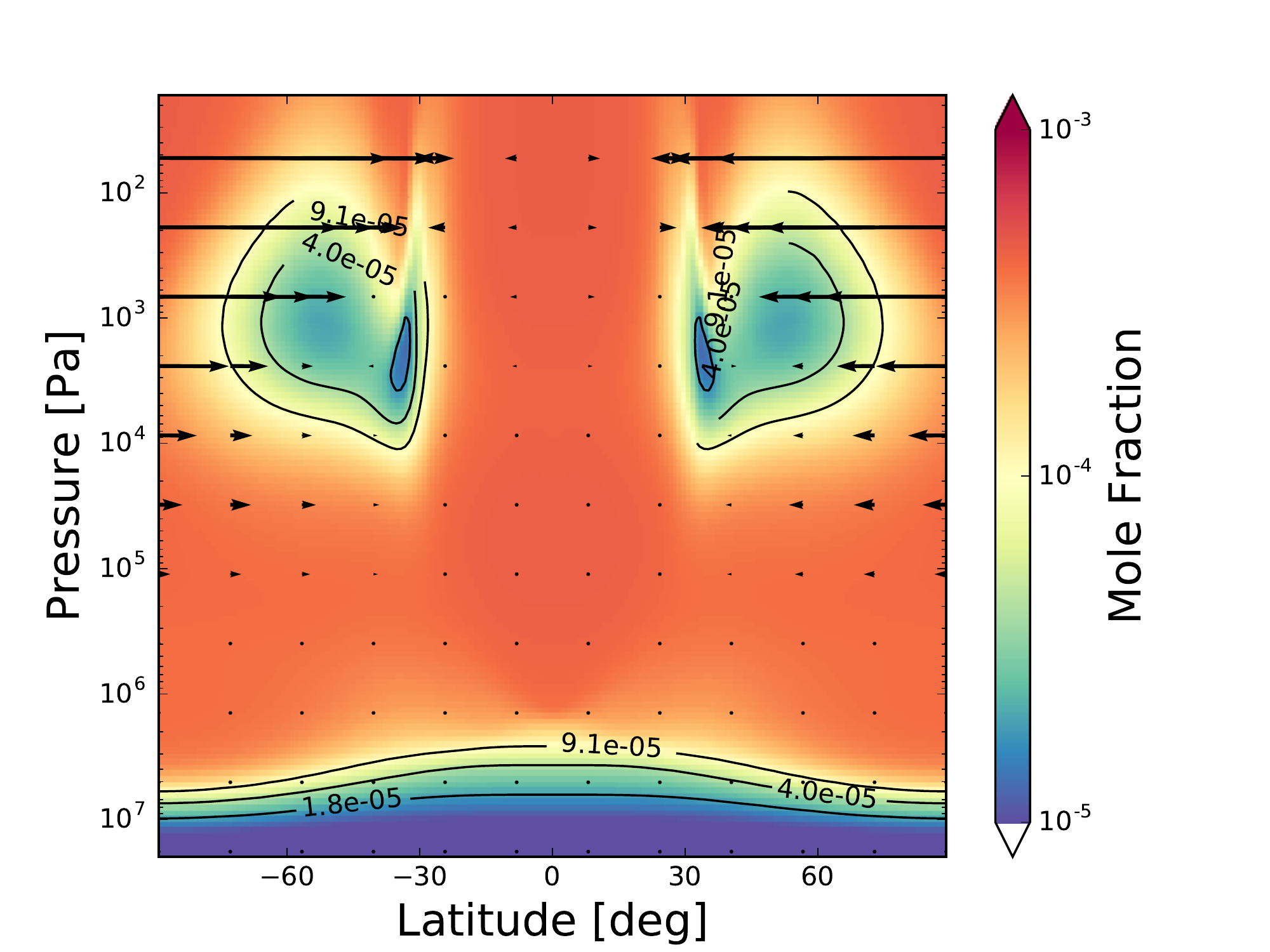} 
    \includegraphics[width=0.32\textwidth]{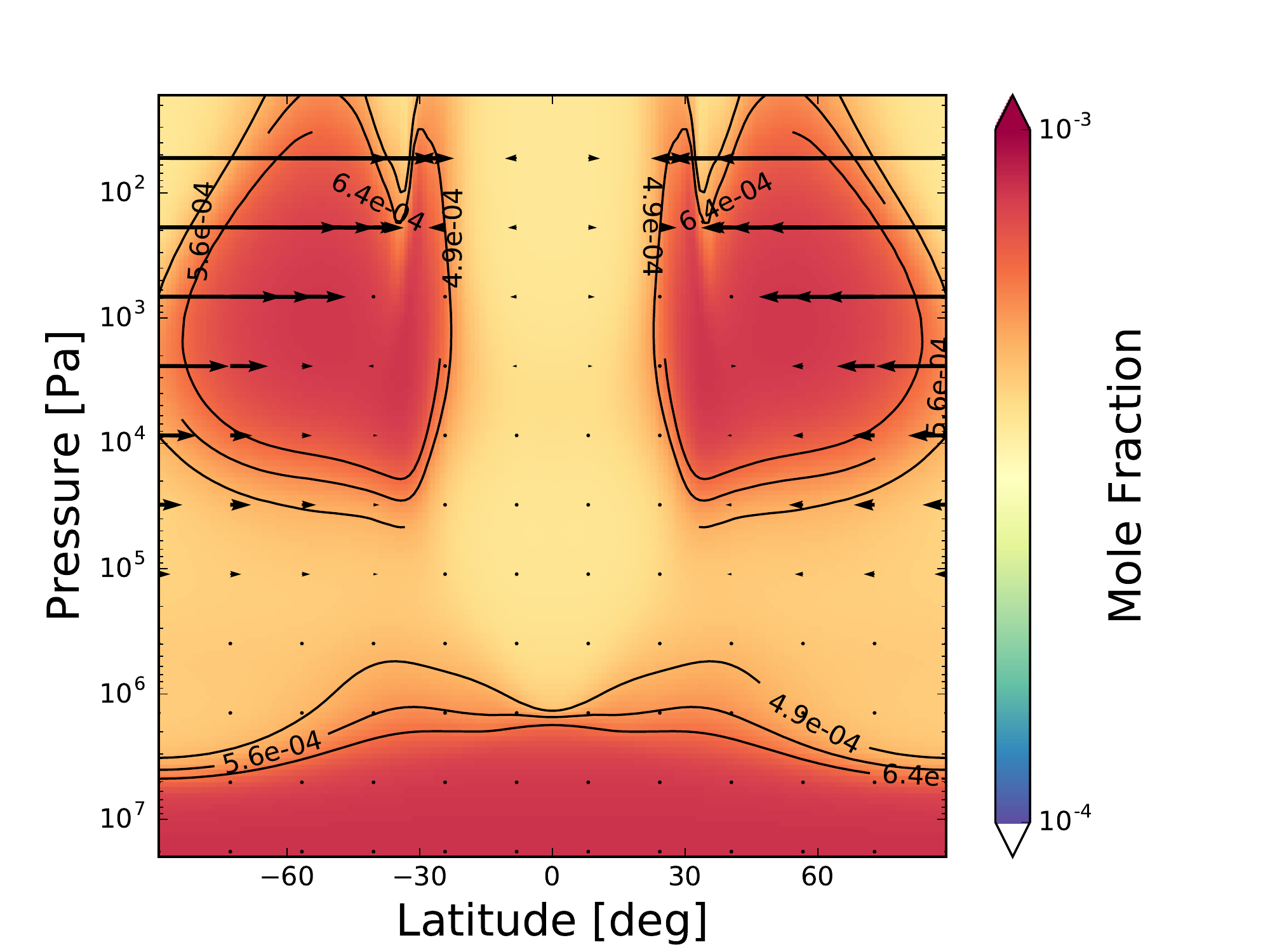} 
    \includegraphics[width=0.32\textwidth]{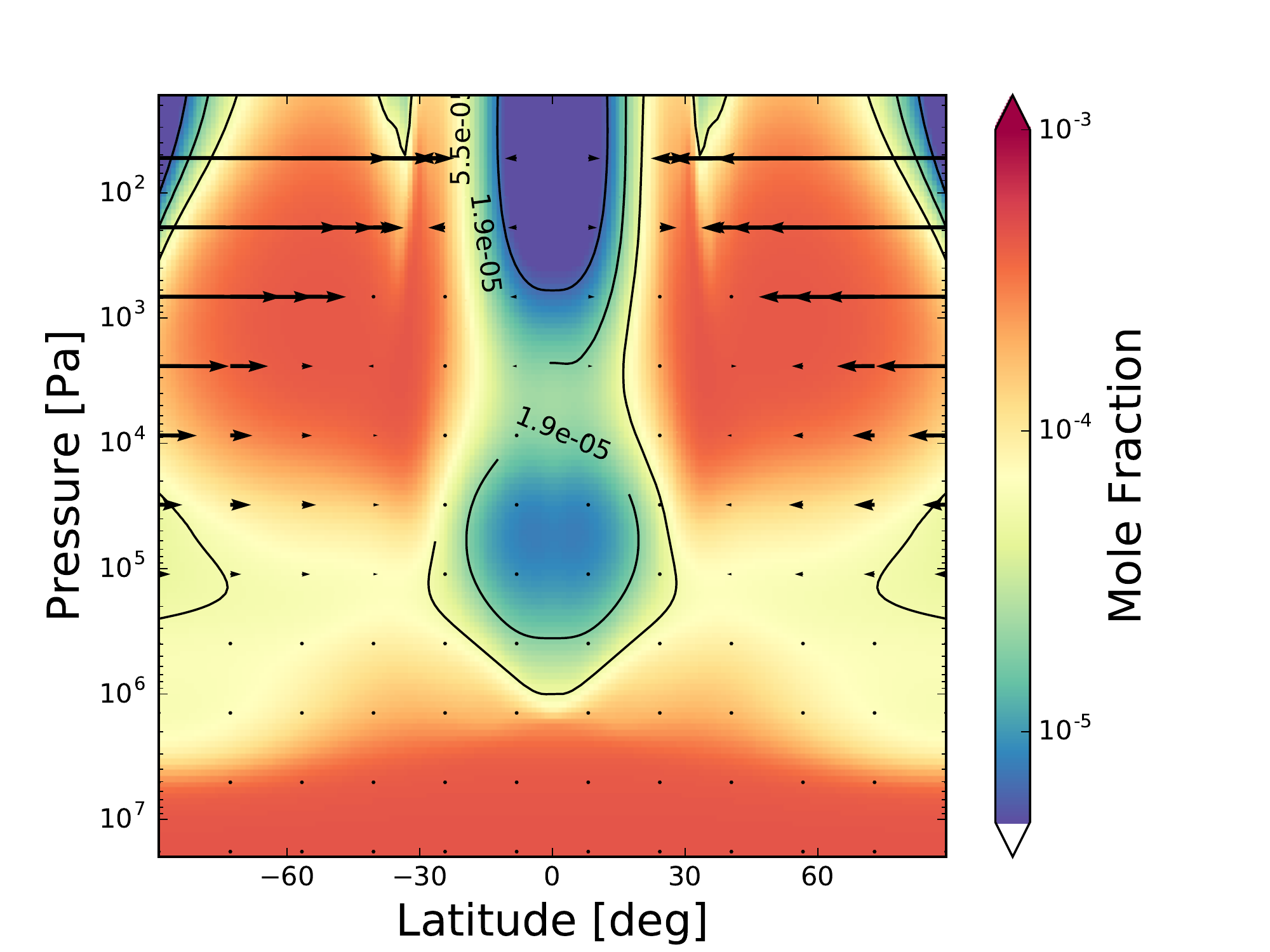} \\
    \includegraphics[width=0.32\textwidth]{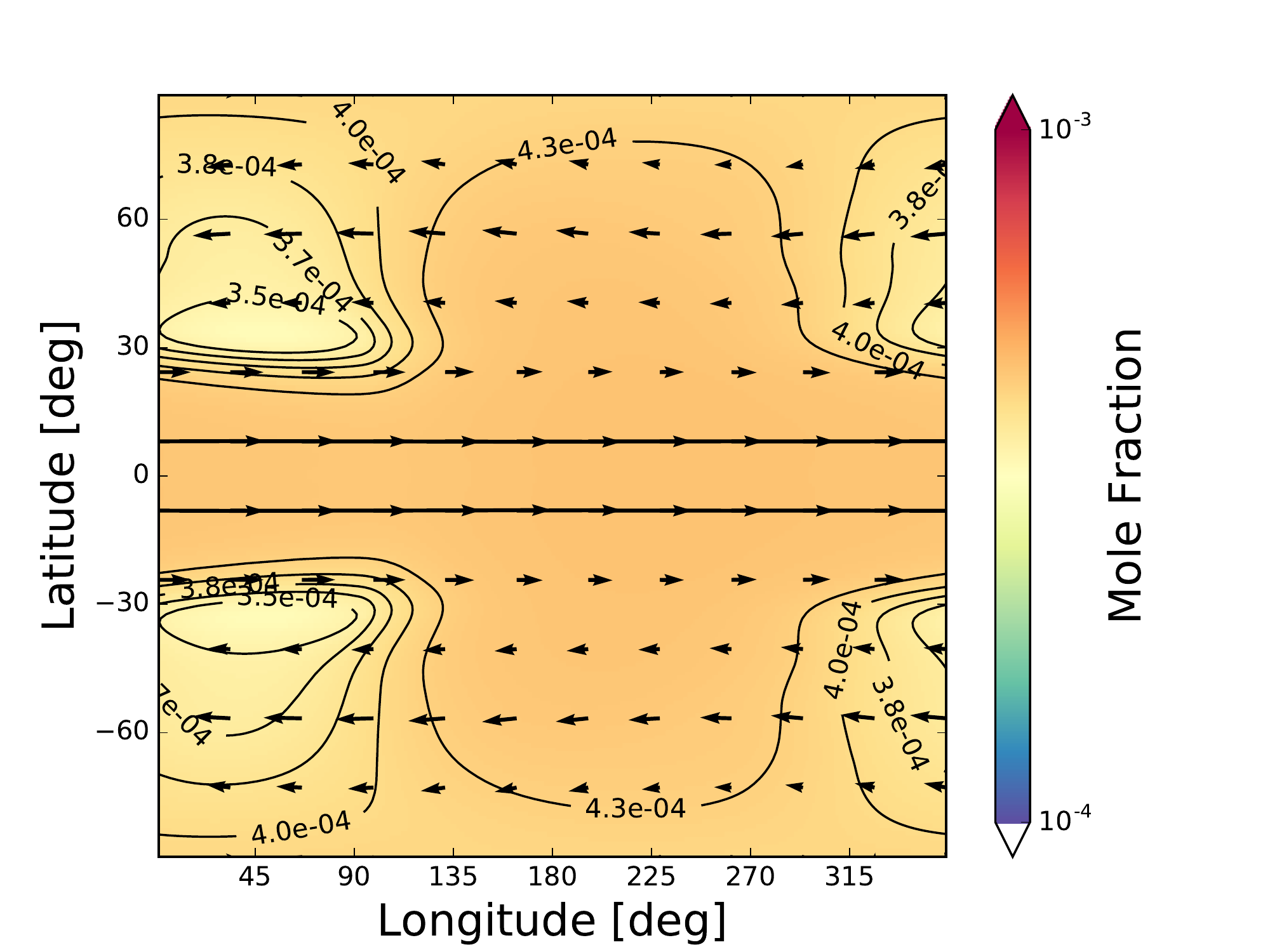} 
    \includegraphics[width=0.32\textwidth]{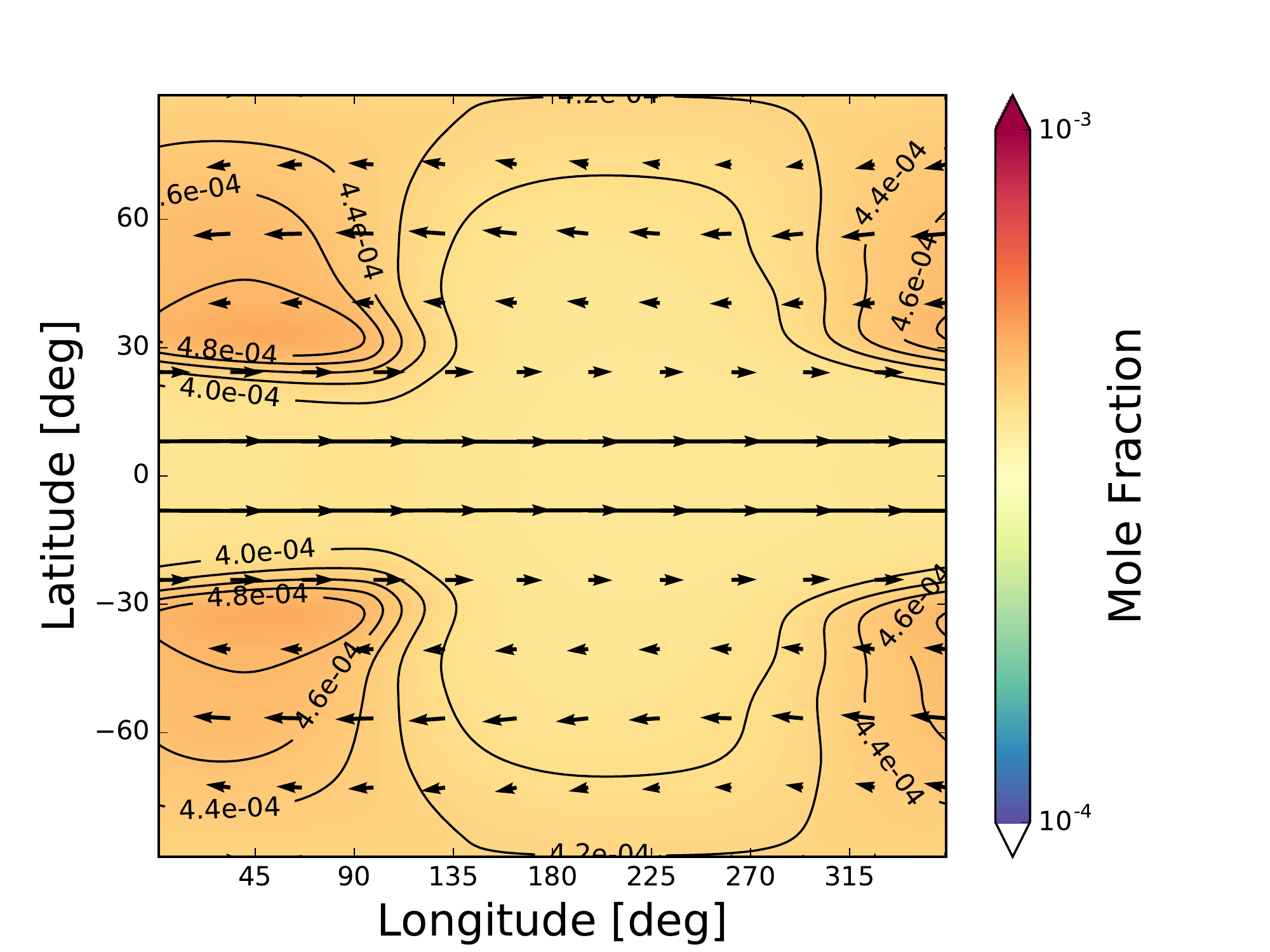} 
    \includegraphics[width=0.32\textwidth]{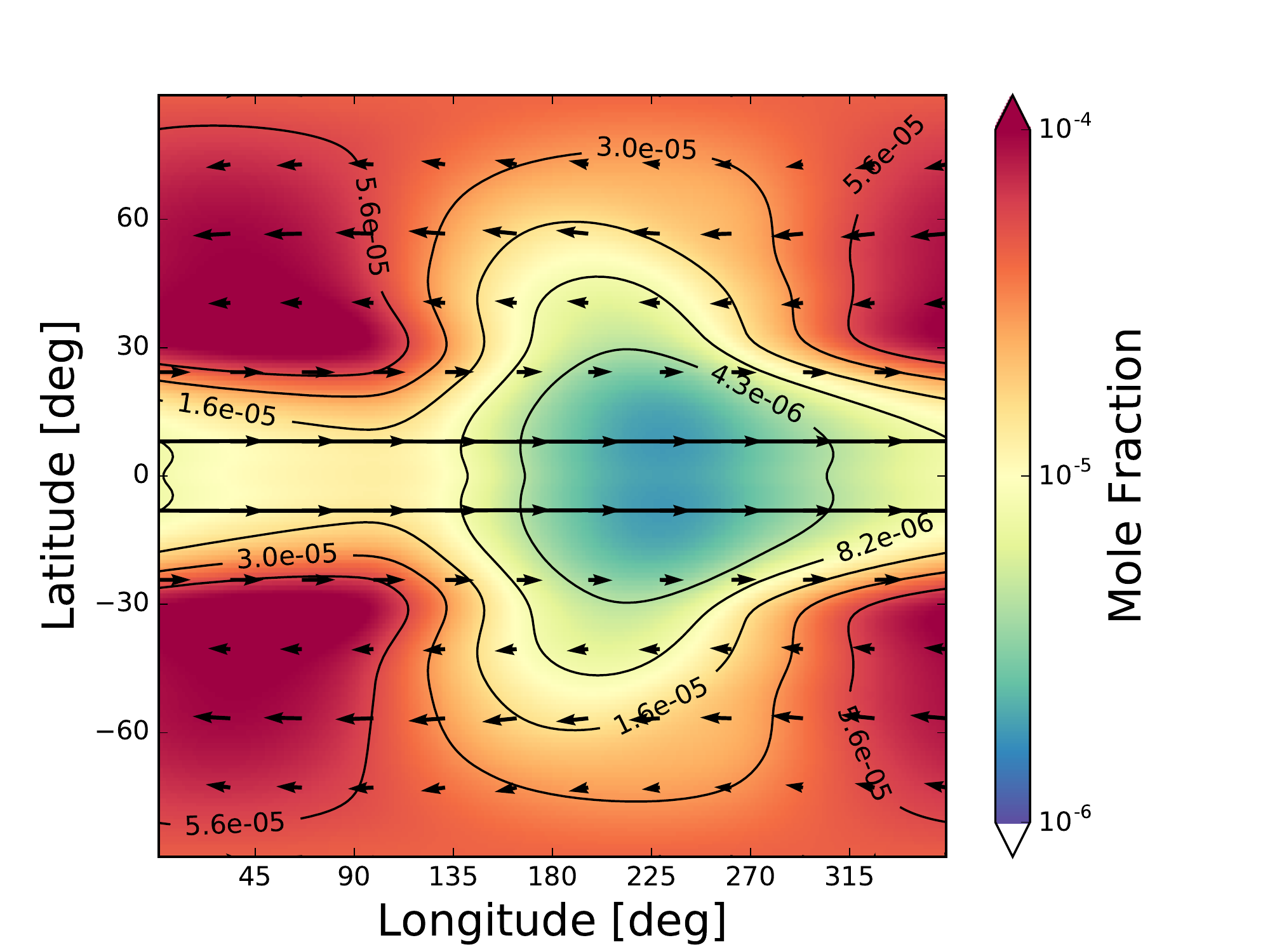} \\
    \includegraphics[width=0.32\textwidth]{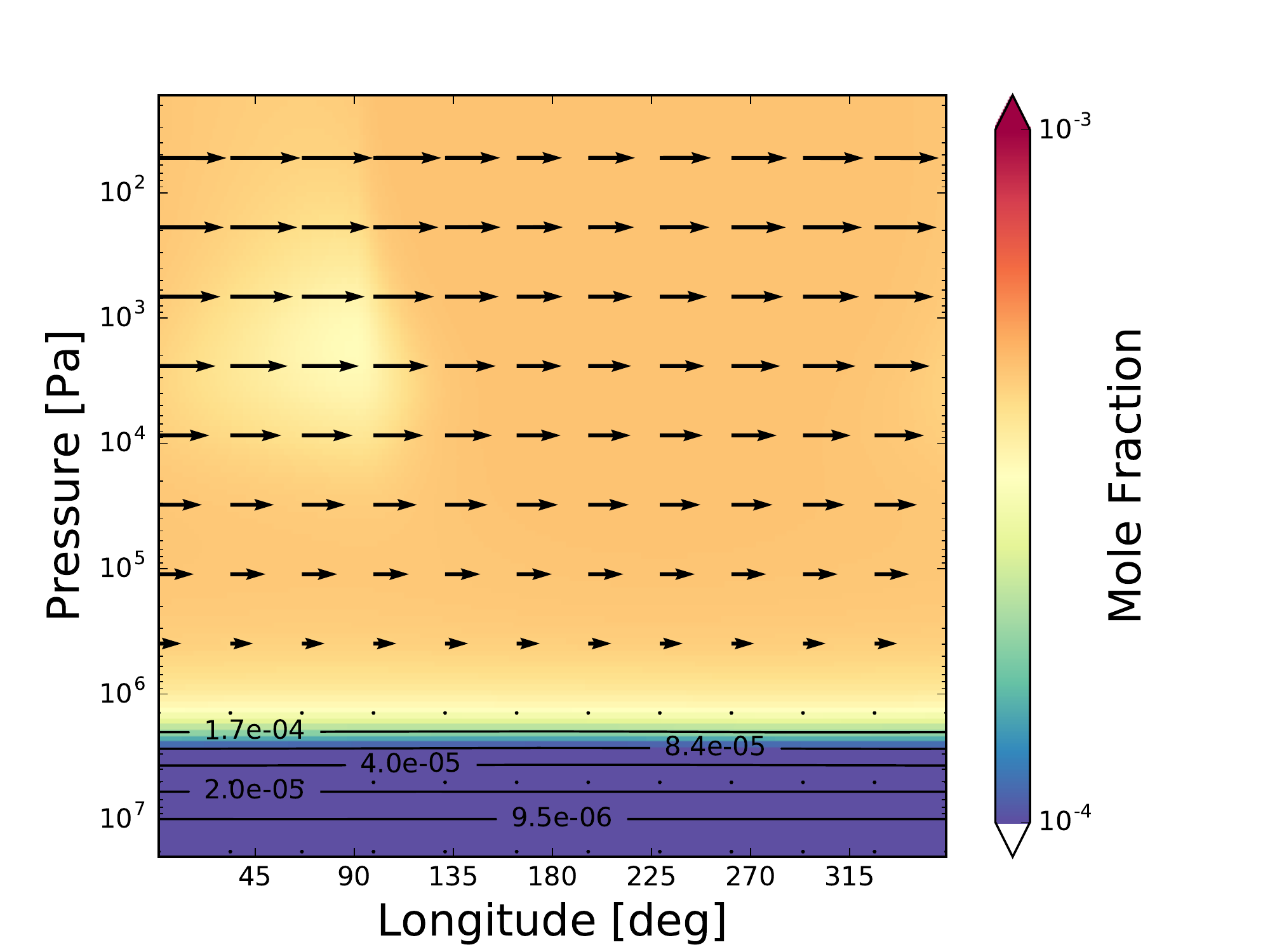} 
    \includegraphics[width=0.32\textwidth]{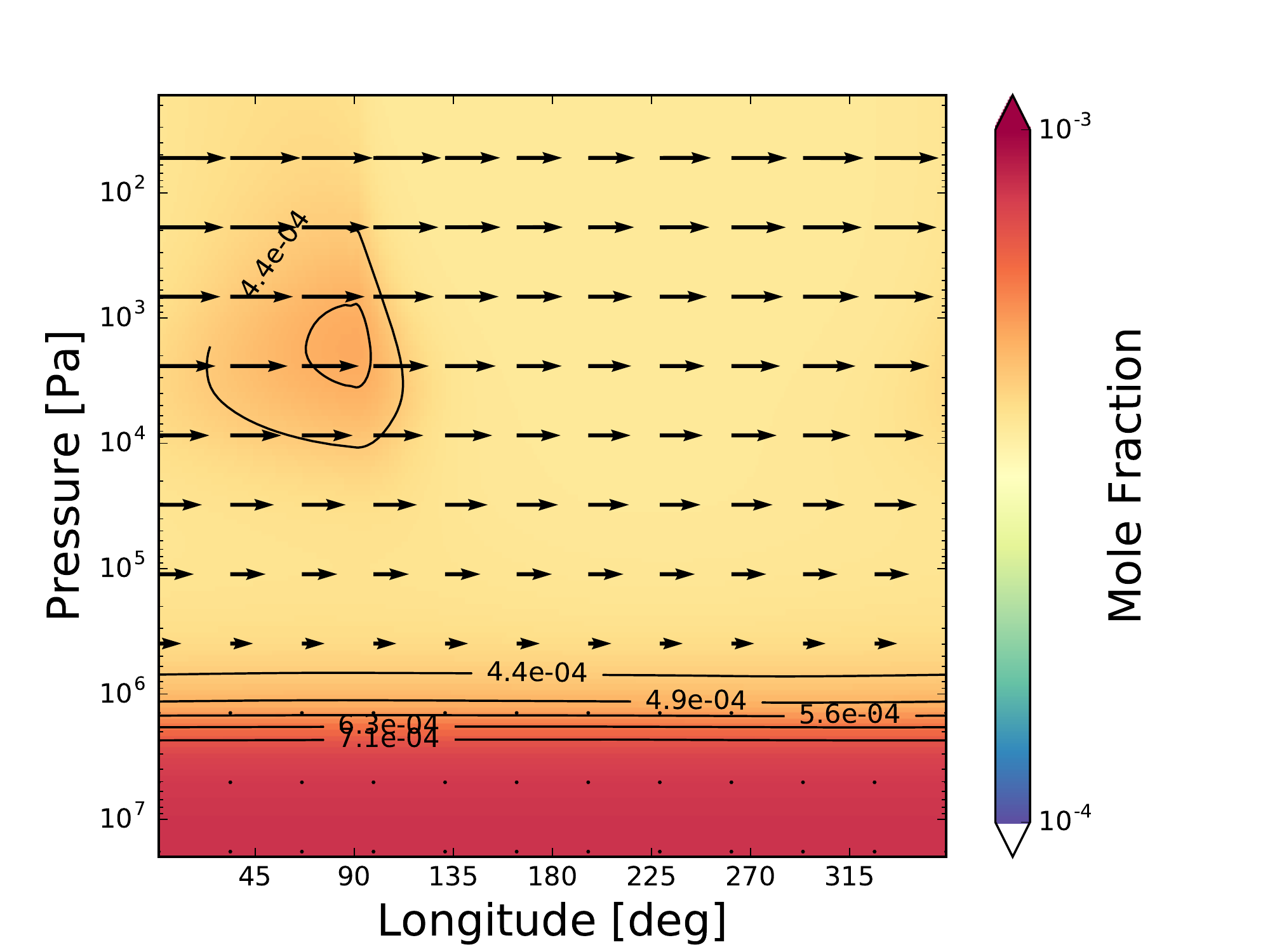} 
    \includegraphics[width=0.32\textwidth]{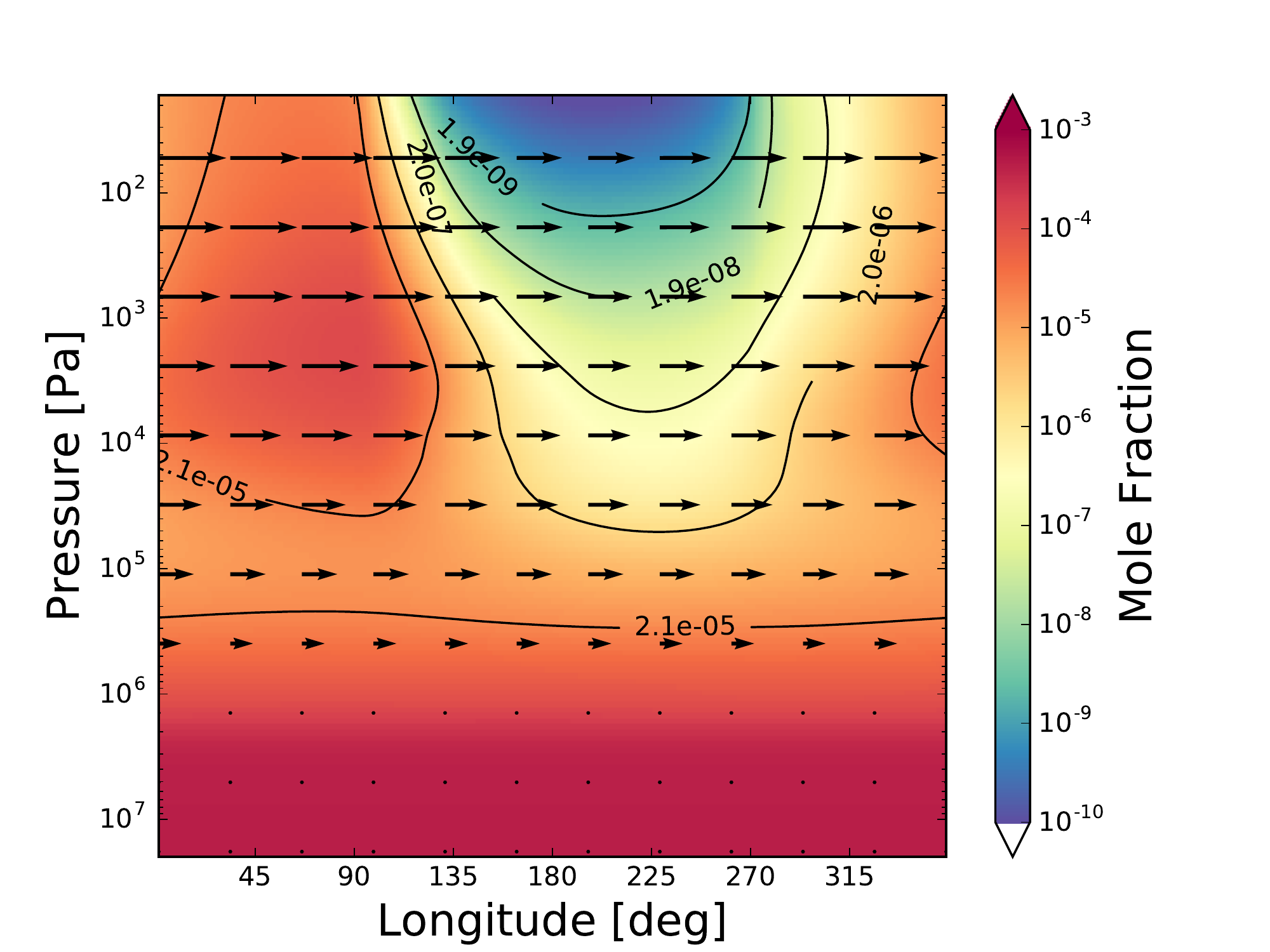} \\\
  \end{center}
\caption{Chemical equilibrium mole fractions (colour scale and black contours) and wind velocity vectors (black arrows) of carbon monoxide (left column), water (middle column) and methane (right column) at a longitude of 0$^{\circ}$ (top row), on the $5\times10^4$ Pa isobaric surface (middle row) and an area--weighted meridional mean ($\pm20^{\circ}$ latitude) around the equator (bottom row).}
\label{figure:eq}
\end{figure*}

% Relaxation mole fractions
\begin{figure*}
  \begin{center}
    \includegraphics[width=0.32\textwidth]{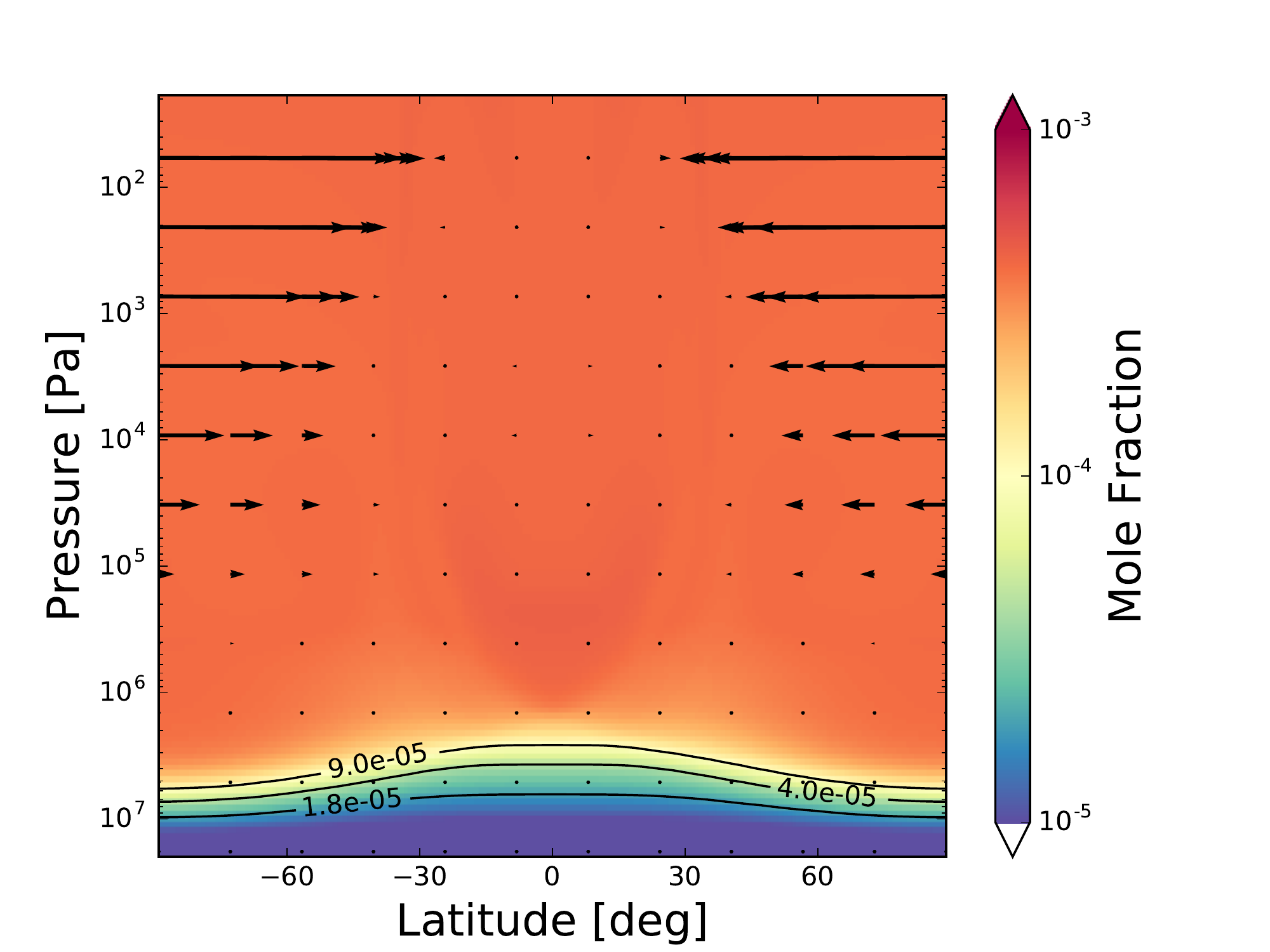} 
    \includegraphics[width=0.32\textwidth]{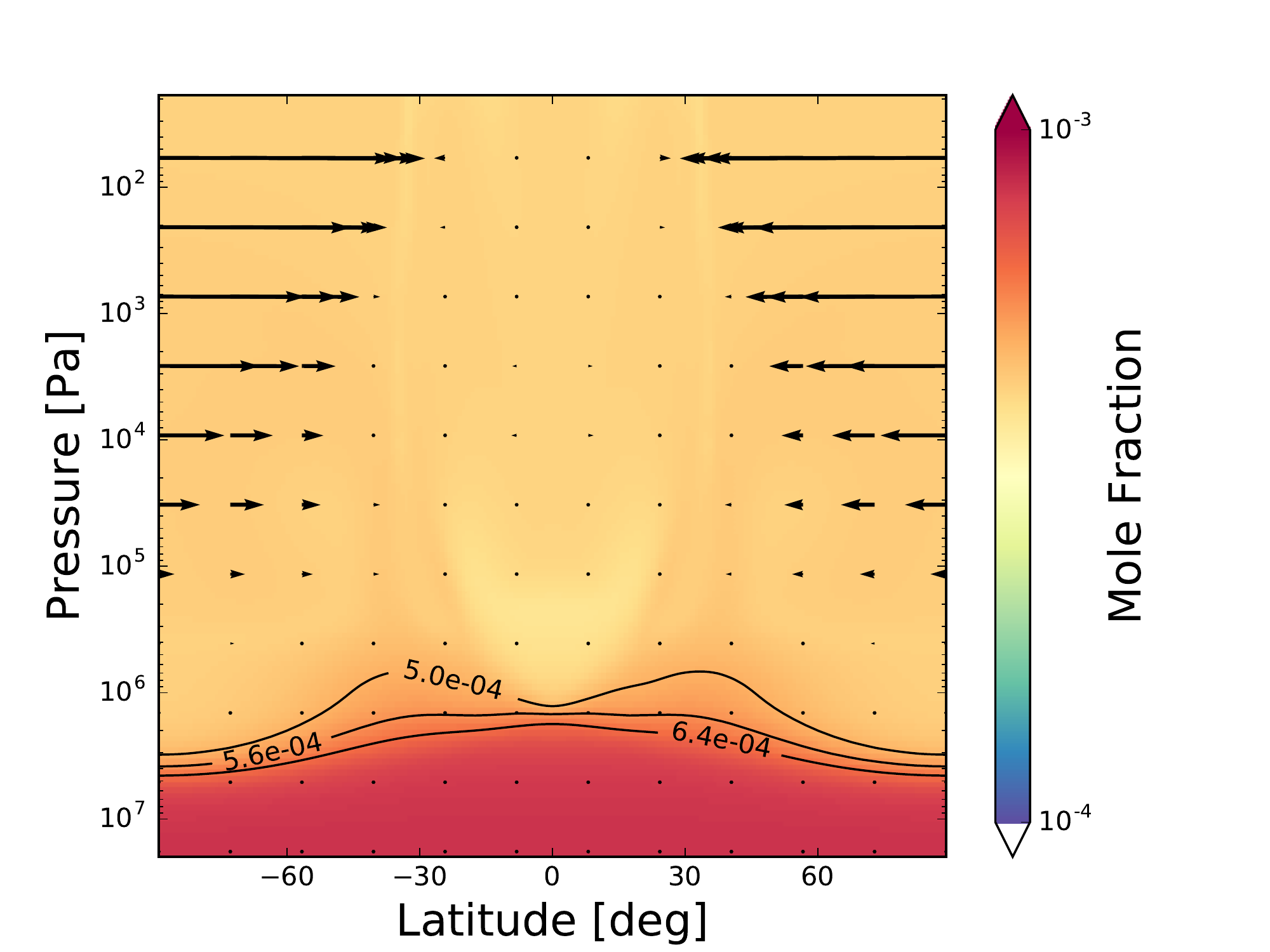} 
    \includegraphics[width=0.32\textwidth]{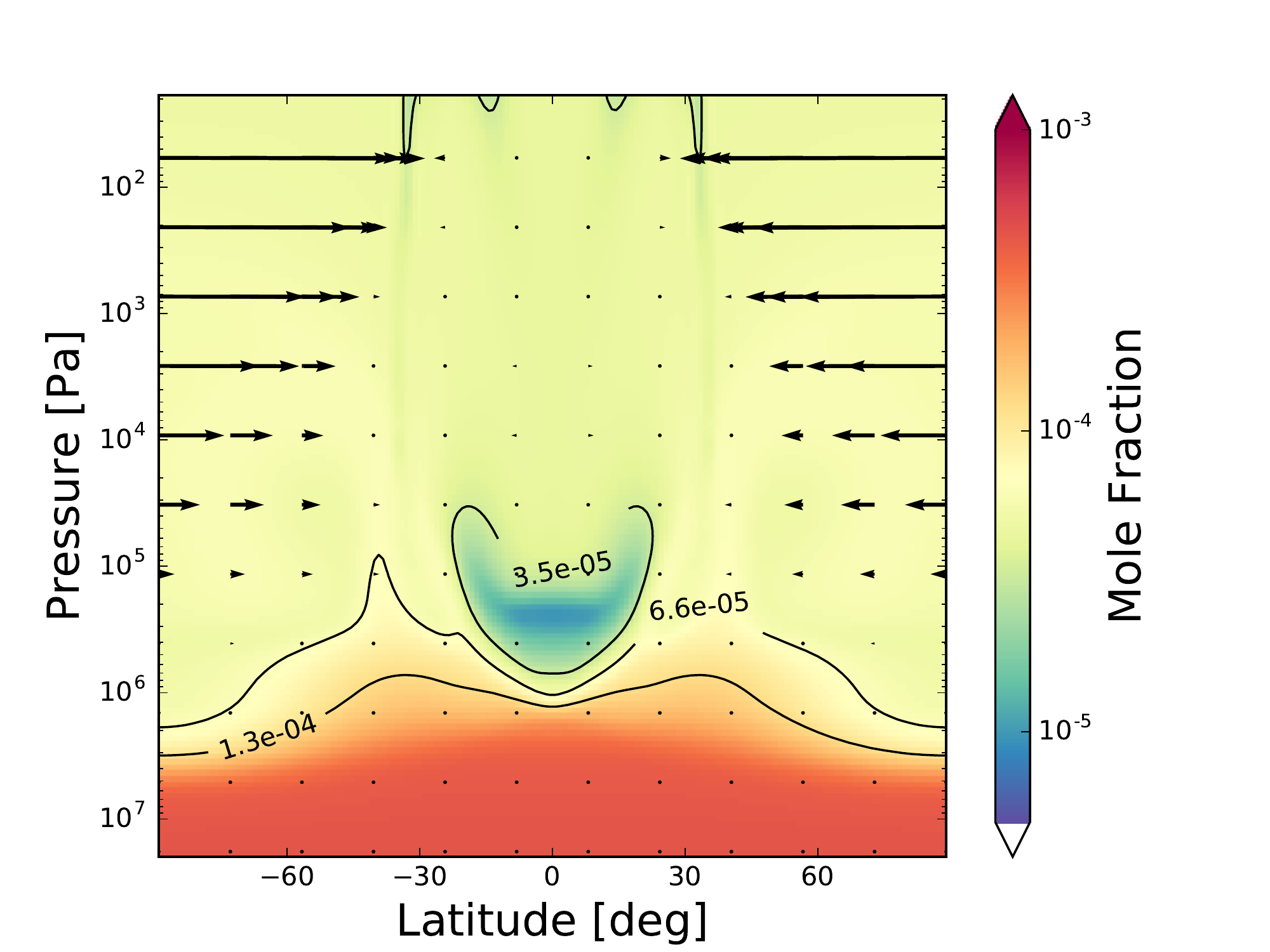} \\ 
    \includegraphics[width=0.32\textwidth]{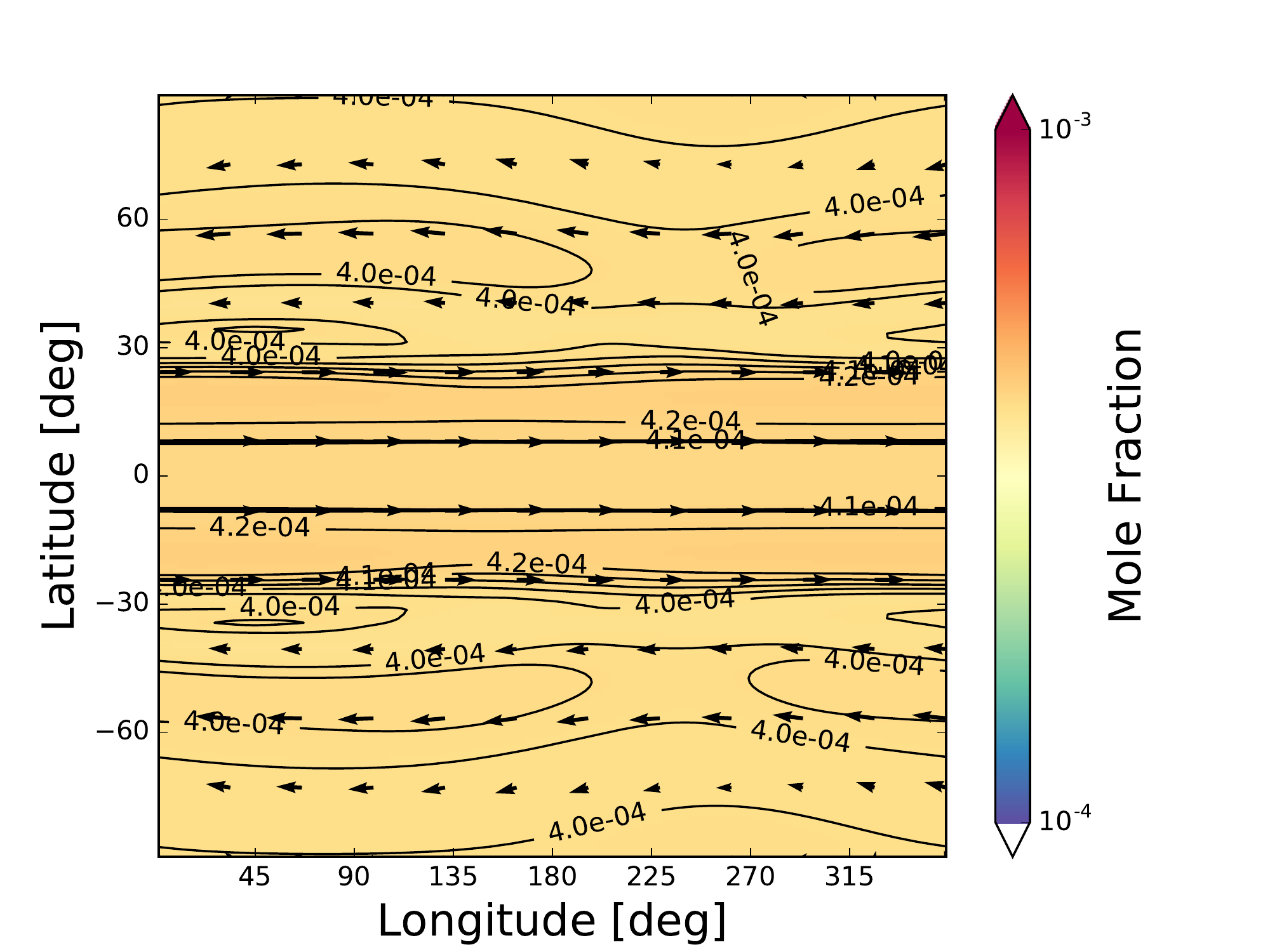} 
    \includegraphics[width=0.32\textwidth]{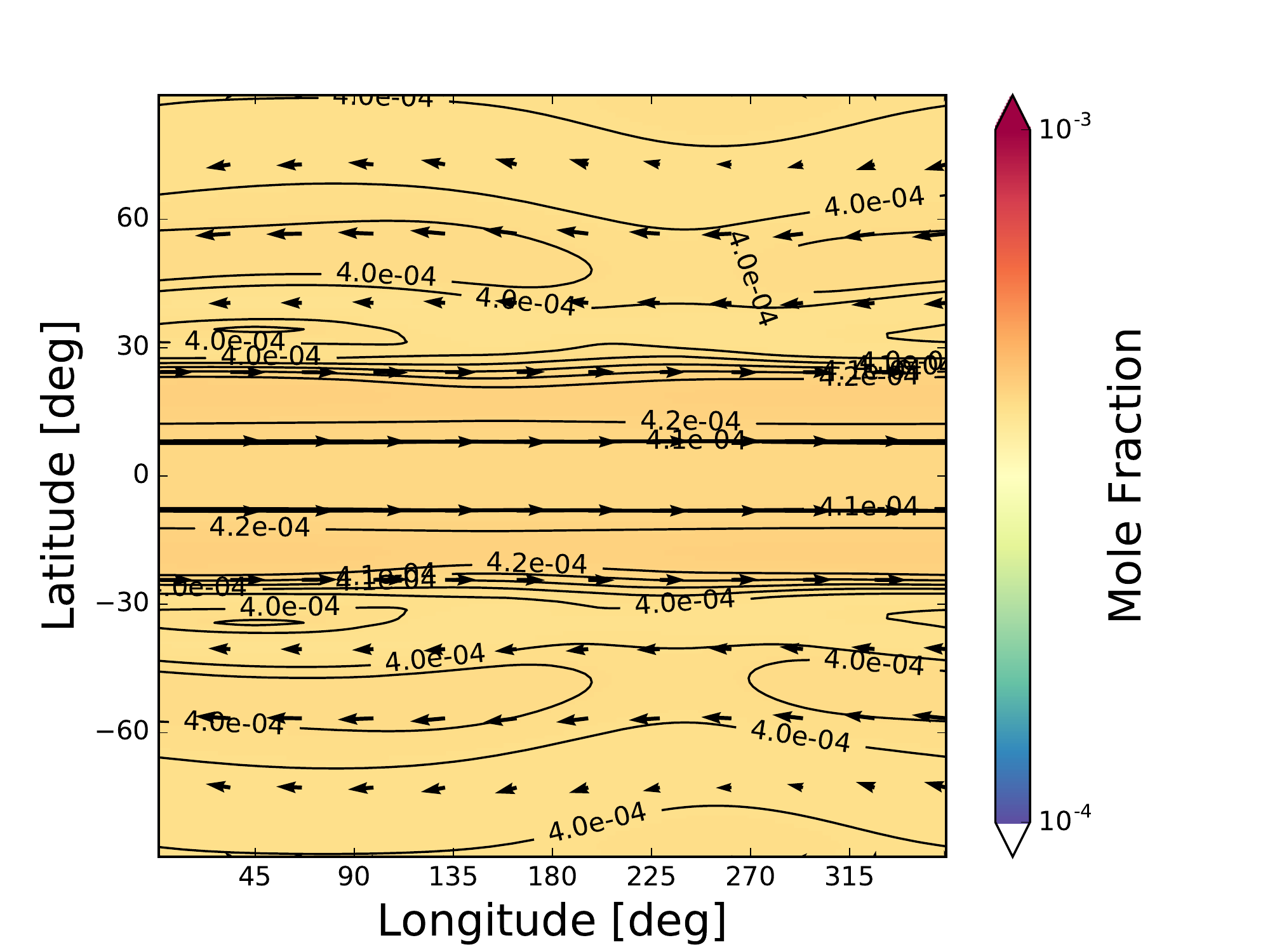} 
    \includegraphics[width=0.32\textwidth]{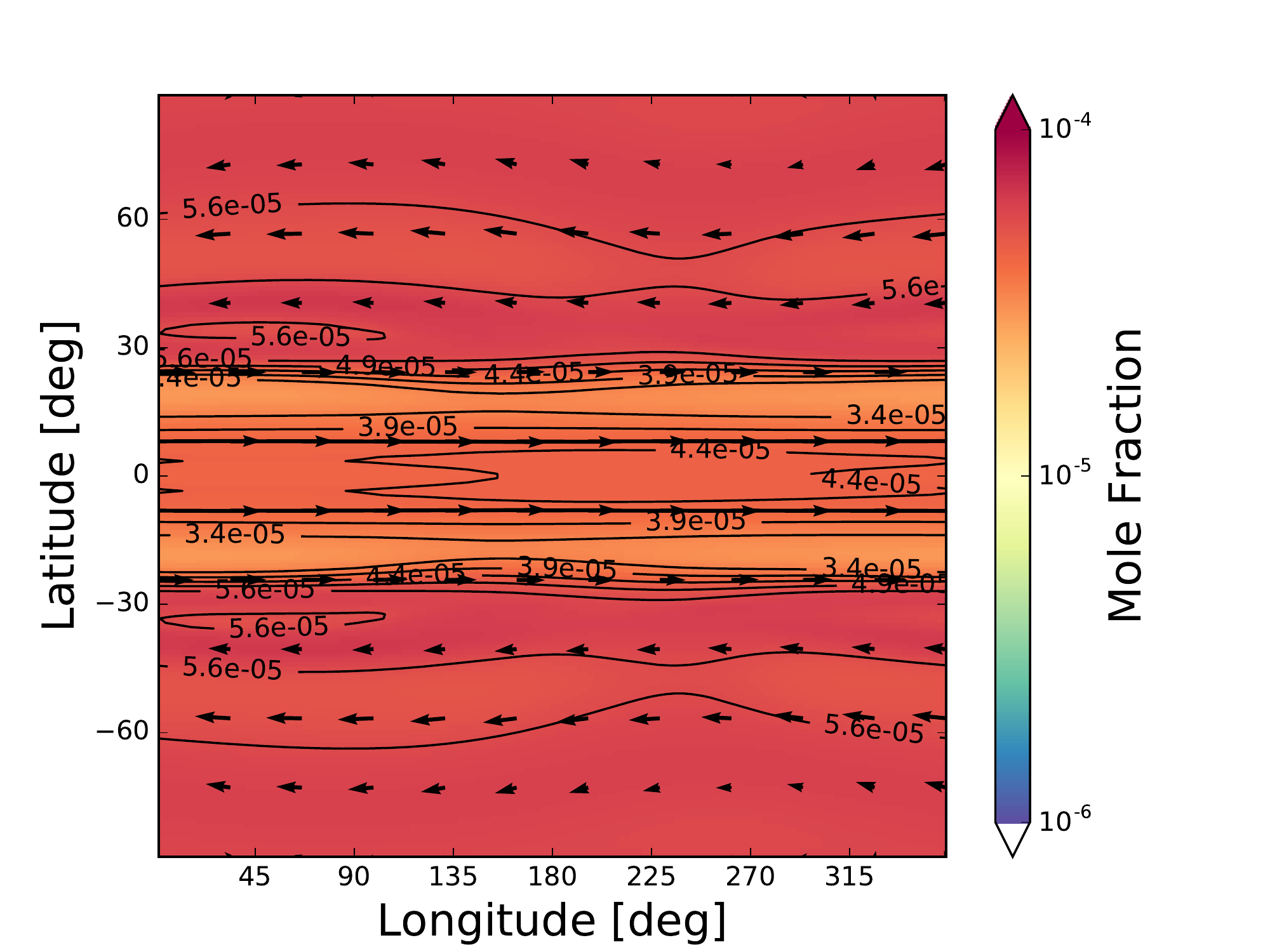} \\
    \includegraphics[width=0.32\textwidth]{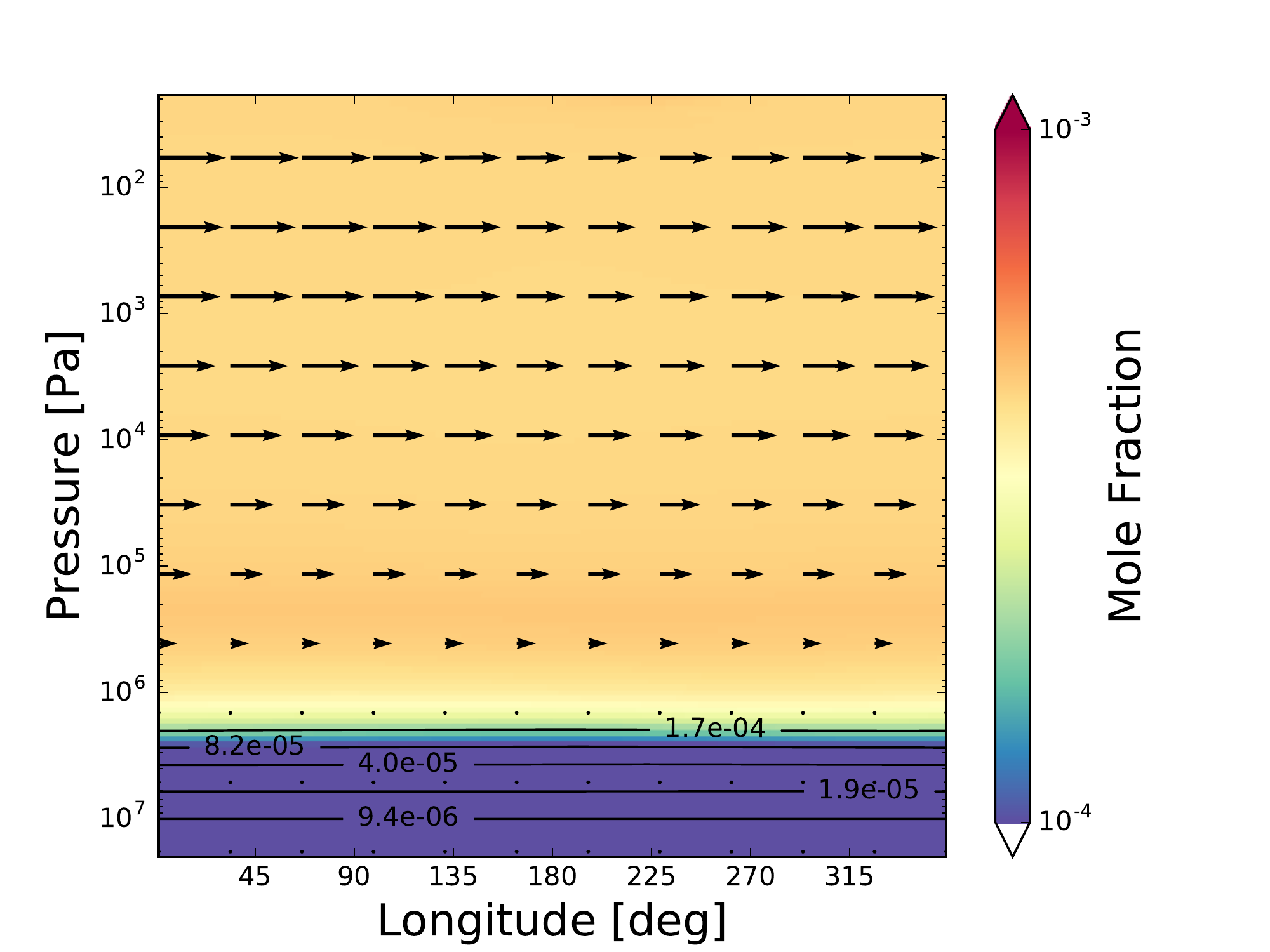} 
    \includegraphics[width=0.32\textwidth]{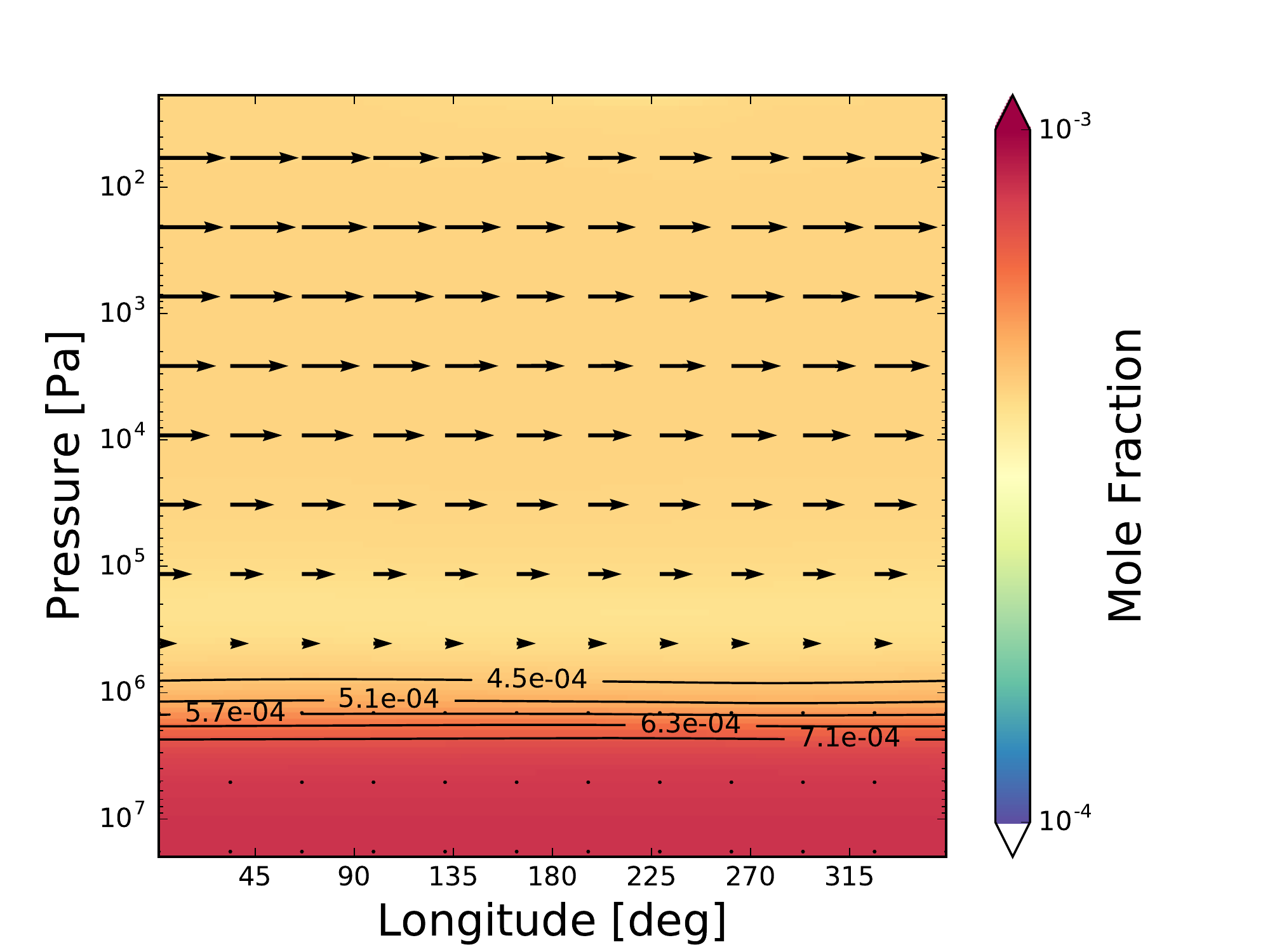} 
    \includegraphics[width=0.32\textwidth]{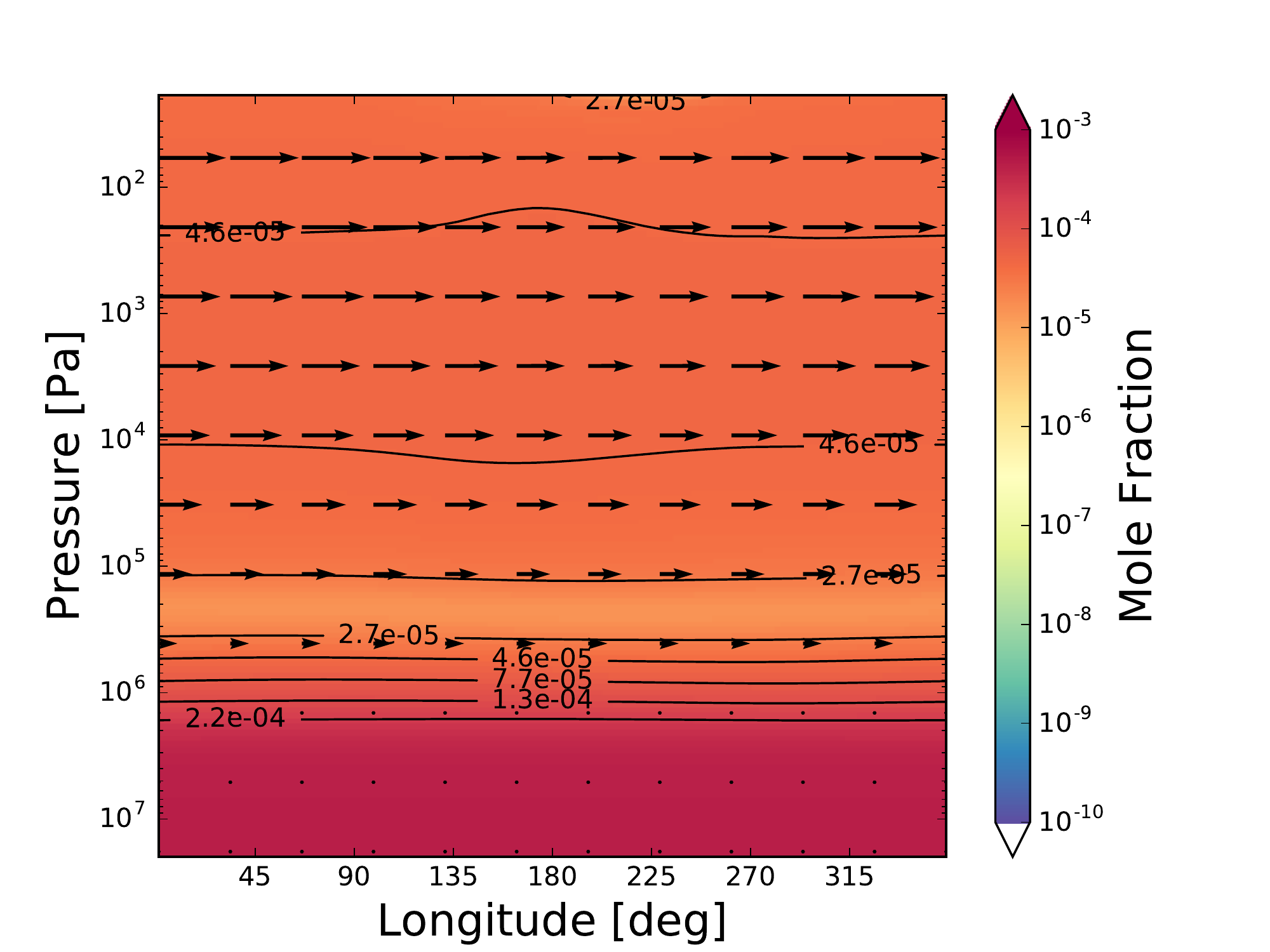} 
  \end{center}
\caption{As \cref{figure:eq} but for the relaxation simulation.}
\label{figure:relax}
\end{figure*}

\cref{figure:eq} shows the mole fractions of carbon monoxide, water and methane for the equilibrium simulation. The distribution of these species exactly traces the temperature structure (compare with \cref{figure:wind_temp}) since in chemical equilibrium the composition is entirely dependent on the local pressure and temperature, for a given mix of elements.

Carbon monoxide is more abundant than methane almost everywhere in the modeled domain, though the methane mole fraction varies significantly between $\sim10^{-4}$ on the nightside and $\sim10^{-10}$ on the warmer dayside. In the mid-latitude region of the nightside (where the atmosphere is the coolest) methane becomes more abundant than carbon monoxide; there is a corresponding increase in the mole fraction of water.

For the relaxation simulation, shown in \cref{figure:relax}, which includes advection due to the resolved wind, the chemistry is homogenised both vertically and horizontally over a large pressure range ($P\lesssim10^5$ Pa), similar to our results for HD~209458b \citep{DruMM18}. Overall this leads to a much larger dayside, equatorial methane abundance compared with the equilibrium simulation. 

In \citet{DruMM18} we identified a mechanism whereby meridional transport leads to an increase in the equatorial methane abundance, compared with chemical equilibrium, for the case of HD~209458b. Using a simple tracer experiment we demonstrated that mass is transported from higher latitudes to the equatorial region. Since the atmosphere is typically cooler at higher latitudes, compared with the equator, the equilibrium abundance of methane is larger. The net result is the transport of mass that has a relatively high methane fraction toward the equator. \cref{figure:wind_temp} (top right panel) clearly shows that significantly large regions of equatorward flow exist. This 3D effect cannot be captured by 1D \citep[e.g.][]{Moses2011,Venot2012,DruTB16} or 2D \citep{AguPV14} models.

\begin{figure}
  \begin{center}
    \includegraphics[width=0.45\textwidth]{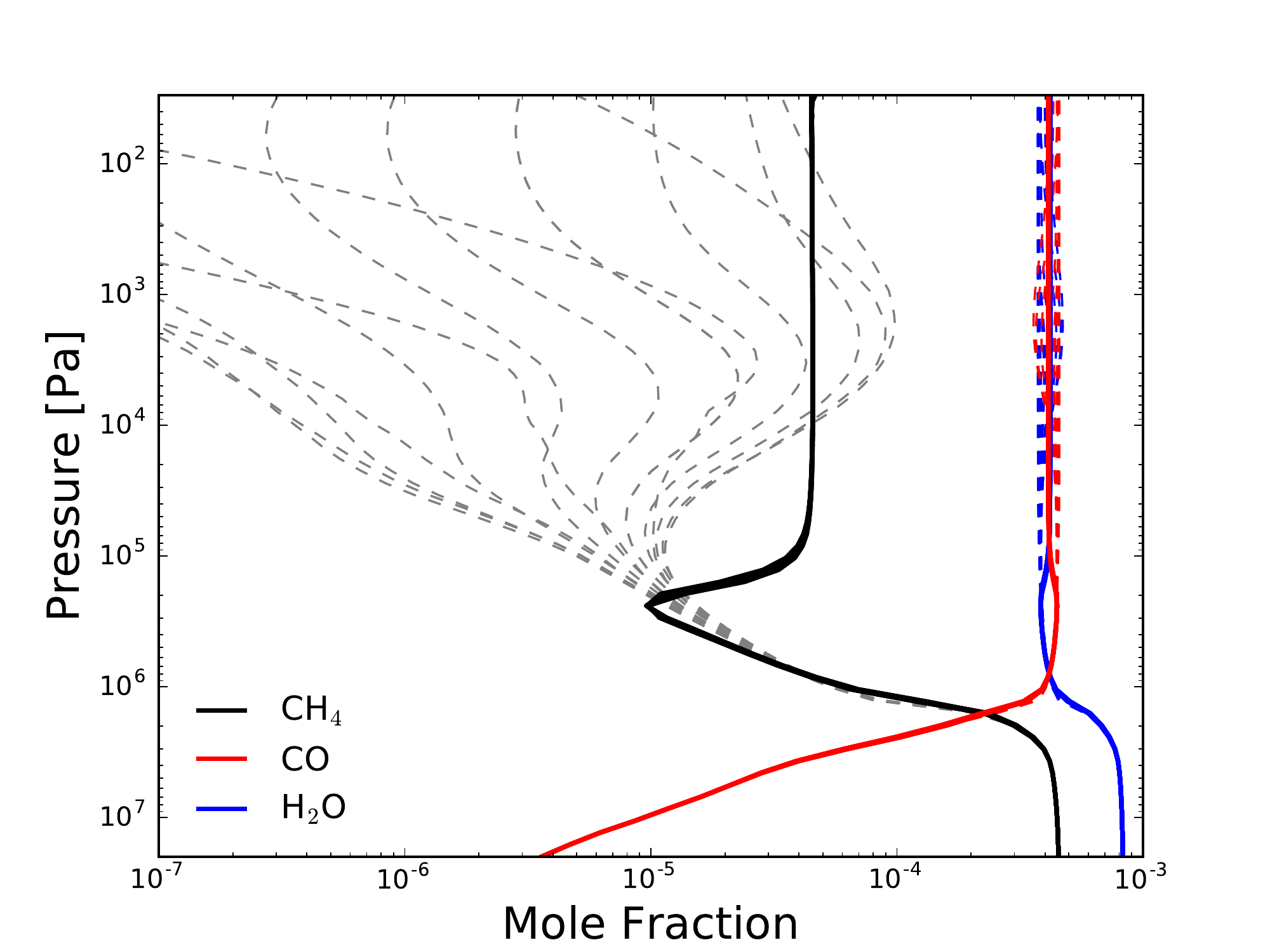}
  \end{center}
  \caption{Vertical profiles of the carbon monoxide (red), water (blue) and methane (black/grey) mole fractions for a number of columns equally spaced in longitude around the equator (at 0$^{\circ}$  latitude) for the chemical equilibrium simulation (dashed) and relaxation simulation (solid).}
  \label{figure:profiles}
\end{figure}

\cref{figure:profiles} shows vertical profiles of the mole fractions of carbon monoxide, water and methane for a number of longitude points around the equator. Methane becomes vertically quenched at $P\sim10^5$ Pa with a mole fraction of $\sim4\times10^{-5}$. Importantly, we find the same meridional transport effect (the sharp increase in methane abundance between $1\times10^{5}$ and $2\times10^5$ Pa) as identified in our simulations of HD~209458b \citep{DruMM18}, which is due to meridional transport from the mid-latitudes toward the equator.

Overall, we find qualitatively the same trends as we previously found for HD~209458b \citep{DruMM18}. The main quantitative differences are the larger mole fractions of methane in the cooler atmosphere of HD~189733b.

%%%%%%%%%%%%%%%%%%%%%%%%%%%%%%%%%%%%
\section{Thermal and dynamical response of the atmosphere}
\label{section:temp}
%%%%%%%%%%%%%%%%%%%%%%%%%%%%%%%%%%%%

In this section we consider the thermal and dynamical response of the atmosphere to the changes in the local chemical composition due to wind--driven chemistry, by comparing the temperature structures of the equilibrium and relaxation simulations. The chemical composition and the thermal structure are linked via the radiative heating rates.

\subsection{Temperature response}
\label{section:temp_response}

\begin{figure}
  \begin{center}
    \includegraphics[width=0.47\textwidth]{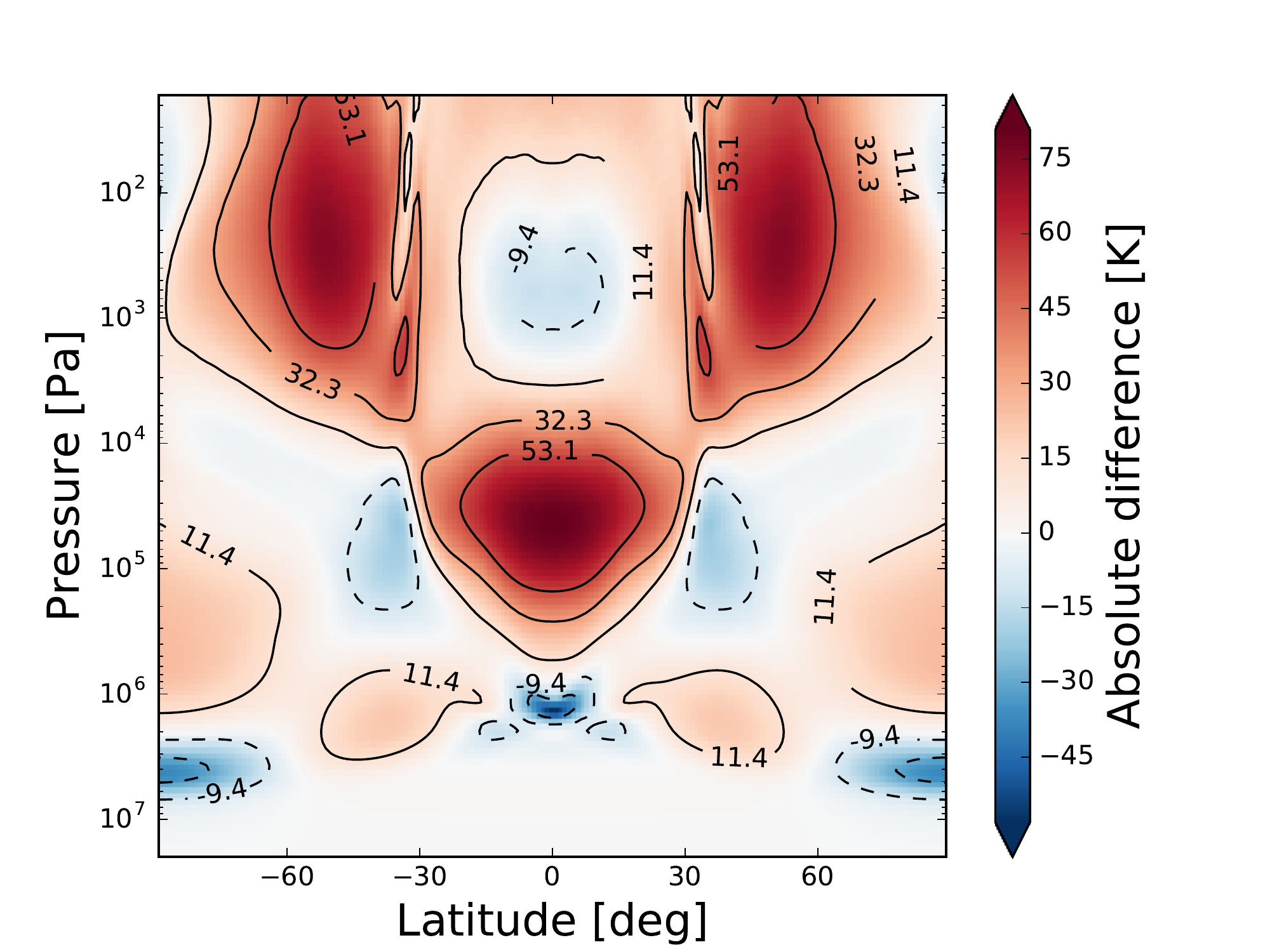} \\
    \includegraphics[width=0.47\textwidth]{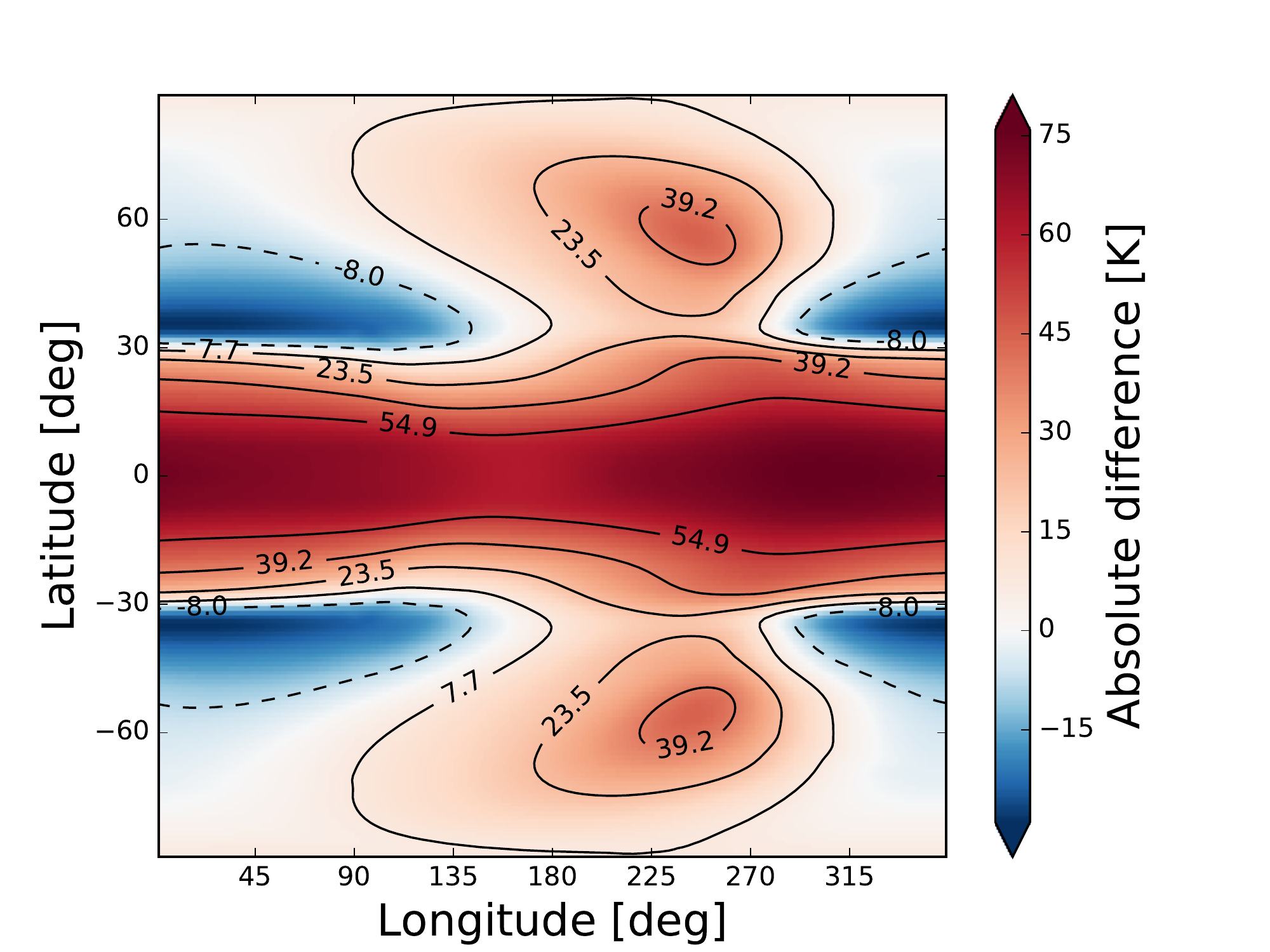} \\
    \includegraphics[width=0.47\textwidth]{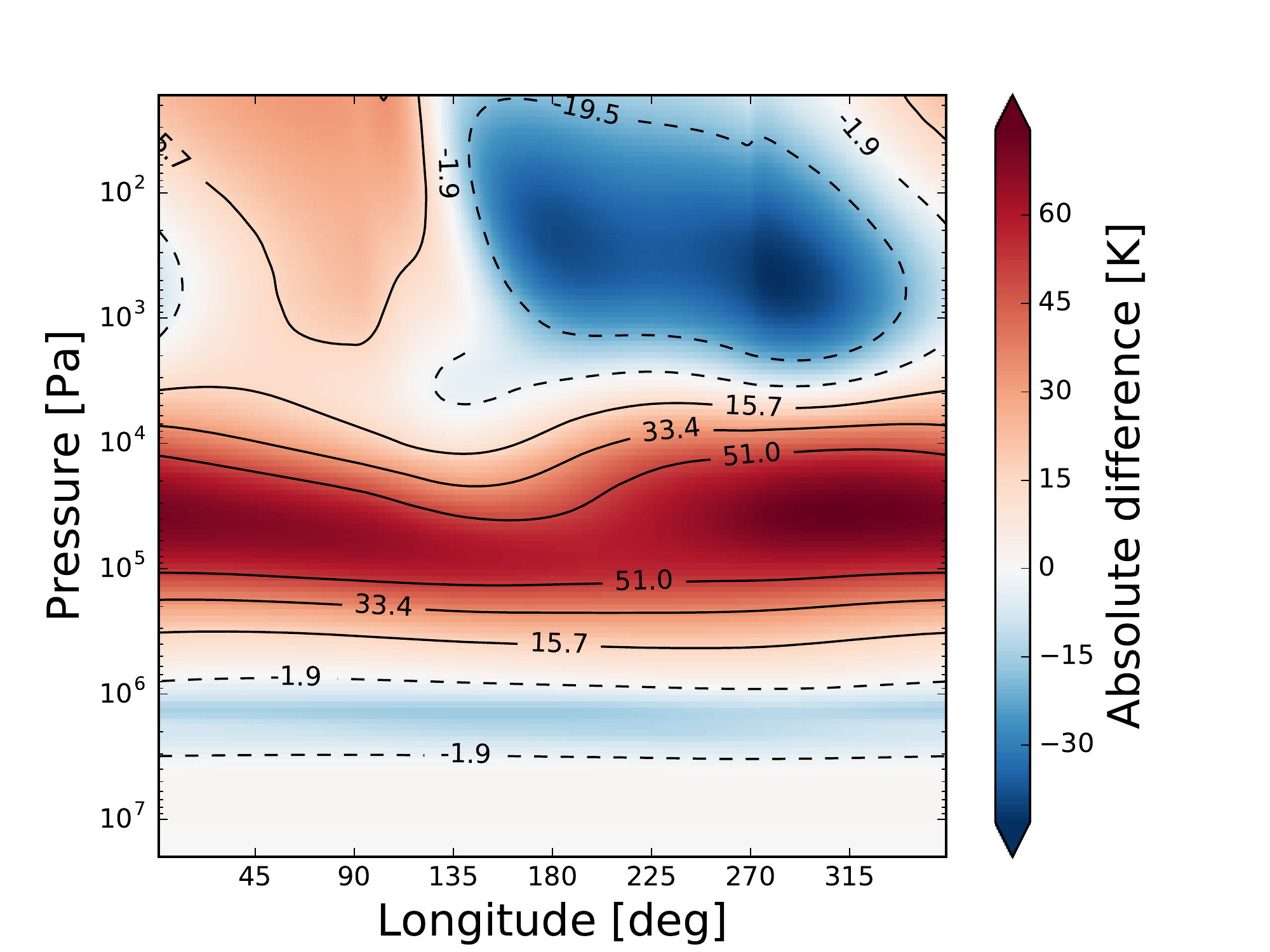}
  \end{center}
\caption{The temperature difference between the relaxation and equilibrium simulations at a longitude of 0$^{\circ}$ (top), on the $P=5\times10^4$ Pa isobaric surface (middle) and an area--weighted meridional--mean between $\pm20^{\circ}$ latitude (bottom). A positive difference indicates a larger temperature in the relaxation simulation.}
\label{figure:diff_temp}
\end{figure}

\cref{figure:diff_temp} shows the absolute temperature difference between the equilibrium and relaxation simulations in various different perspectives. A positive difference indicates a larger temperature in the relaxation simulation. 

For pressures less than $\sim10^4$ Pa the atmosphere is generally cooler on the dayside and warmer on the nightside, for the relaxation simulation. In particular, there is a significant temperature increase of $>80$ K in the mid-latitude regions of the nightside. For larger pressures, particularly between $10^4$ and $10^5$ Pa, the temperature increases significantly in the equatorial region, within the equatorial jet, for all longitudes. 

We compared estimates of both the radiative ($\tau_{\rm rad}$) and dynamical ($\tau_{\rm dyn}$) timescales for the equilibrium simulation. For cases where $\tau_{\rm rad}<\tau_{\rm dyn}$ the atmosphere is expected to be in radiative equilibrium (the atmosphere is radiatively driven) while for $\tau_{\rm rad}>\tau_{\rm dyn}$ advection of heat is expected to be important (the atmosphere is dynamically driven). We used Eq. 10 of \citet{ShoG02} to estimate $\tau_{\rm rad}$ and for $\tau_{\rm dyn}$ assumed (for the upper atmosphere)
\begin{equation}
  \tau_{\rm dyn} \sim \frac{R_{\rm P}}{u}\sim\frac{H}{w} \sim 10^4~{\rm s},
\end{equation}
where $R_{\rm P}\sim10^7$ m is the planet radius, $H\sim10^5$ m is the vertical scale height and $u\sim10^3$ ${\rm m}~{\rm s}^{-1}$ and $w\sim10^1$ m s$^{-1}$ are the horizontal and vertical wind velocities. 

Comparing the timescales we find that the atmosphere is expected to be radiatively driven ($\tau_{\rm rad}<\tau_{\rm dyn}$) for $P<10^4$ Pa and dynamically driven ($\tau_{\rm rad}>\tau_{\rm dyn}$) for $P>10^4$ Pa. This transition gives a hint as to why we find different trends in thermal responses above and below $P\sim10^4$ Pa.

% Temperature and wind response

To unpick the combined effects of the three molecules (methane, carbon monoxide and water) we performed an additional test simulation (not shown) that is identical to the relaxation simulation except the mole fractions of water and carbon monoxide used in the radiative transfer calculations (i.e. to calculate the heating rate) correspond to chemical equilibrium, isolating the effect of methane. The resulting temperature and wind structure is almost identical to the nominal relaxation simulation, indicating that methane is the most important driver of these temperature changes, with water and carbon monoxide making a less important contribution. This is not surprising given the relatively small changes in water and carbon monoxide abundances between the equilibrium and relaxation simulations, despite both species being more abundant than methane.

%%%%%%%%%%%%%%%%%%%%%%%%%%%%%%
\subsection{Top-of-atmosphere (TOA) radiative flux}
%%%%%%%%%%%%%%%%%%%%%%%%%%%%%%

% TOA flux
\begin{figure*}
\begin{center}
  \includegraphics[width=0.45\textwidth]{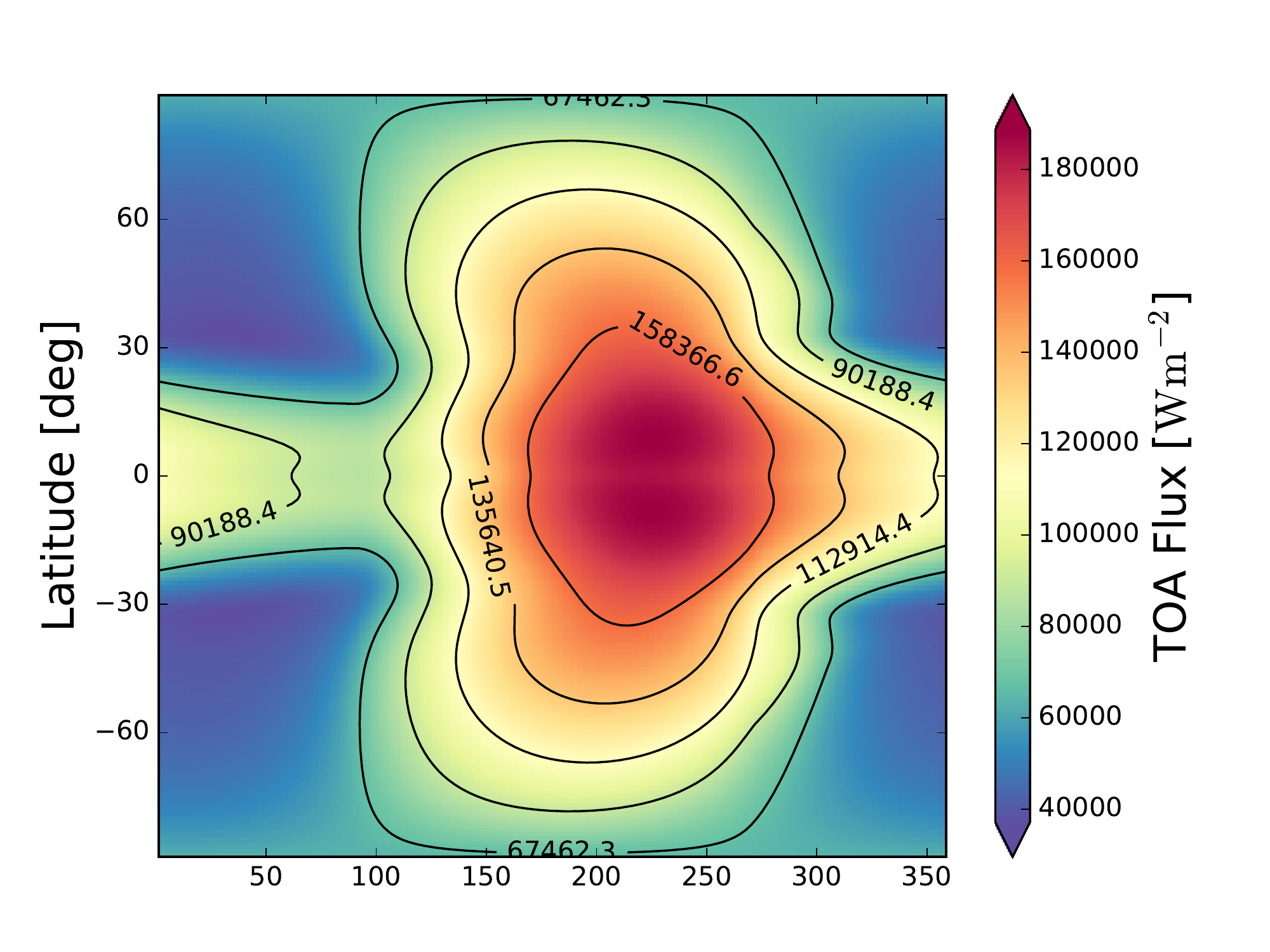} 
    \includegraphics[width=0.45\textwidth]{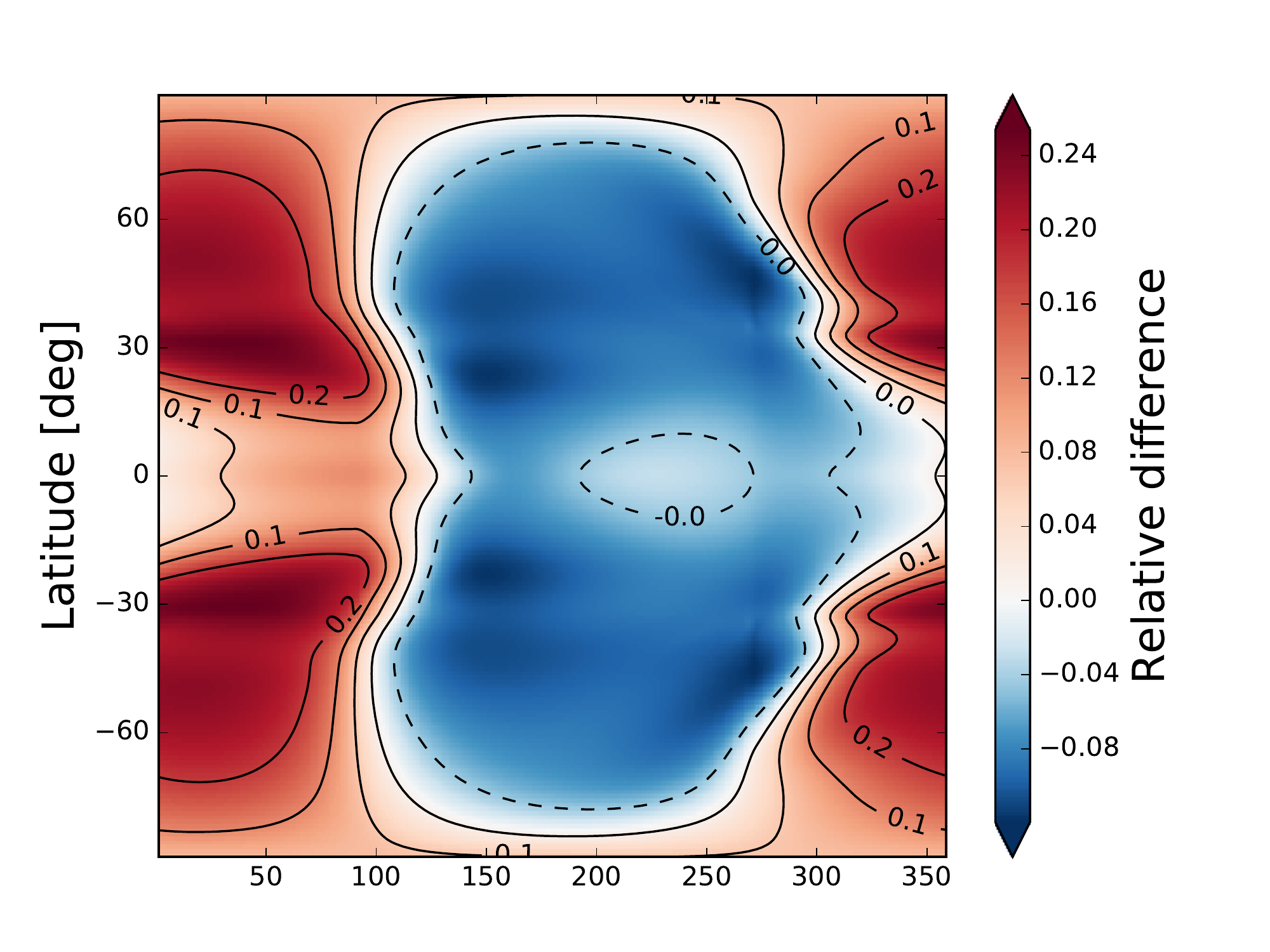} 
\end{center}
\caption{Top-of-atmosphere (TOA) thermal radiative flux for the equilibrium simulation (left) and the relative difference TOA flux between the relaxation and equilibrium simulations (right). A positive difference indicates a larger TOA flux in the relaxation simulation.}
\label{figure:toa}
\end{figure*}

The energy balance of a close--in tidally--locked atmosphere is dominated by stellar (shortwave) heating of the dayside atmosphere due to irradiation by the host star and thermal (longwave) cooling that occurs throughout the atmosphere. Throughout the rest of this paper we will refer to the stellar and thermal components as shortwave and longwave, respectively. The dominant source of heating for the nightside atmosphere is advection of heat from the irradiated dayside. To obtain a global view of the energy balance we consider the outgoing longwave radiative flux at the top-of-atmosphere (TOA), spectrally integrated over all wavelengths, shown in \cref{figure:toa}.

The TOA longwave flux qualitatively traces the temperature structure at $P\sim5\times10^4$ Pa (see \cref{figure:wind_temp}) with the largest emission occuring eastwards of the substellar point, where the atmosphere is the warmest. Comparing the equilibrium and relaxation simulations, it is apparent that the dayside TOA flux is decreased in the relaxation simulation, while on the nightside it is increased. The greatest change occurs in the mid--latitude regions of the nightside where the TOA flux is increased by around $25\%$.

As the primary source of energy for the nightside atmosphere is advection of heat from the irradiated dayside, this indicates an overall increased efficiency of dayside-to-nightside heat transport. This trend agrees with the changes in the temperature explored in the previous section, where the atmosphere was generally found to be warmer on the nightside but cooler on the dayside, for the relaxation simulation, for pressure less than $10^4$ Pa. In the following sections we will show that this is due to changes in the radiative heating rates and an increase in the speed of the equatorial jet, which themselves are not independent of each other.

%%%%%%%%%%%%%%%%%%%%%%%%%%%%%%
\subsection{Heating rates}
%%%%%%%%%%%%%%%%%%%%%%%%%%%%%%

% SW and LW Heating rates
\begin{figure*}
  \begin{center}
    \includegraphics[width=0.45\textwidth]{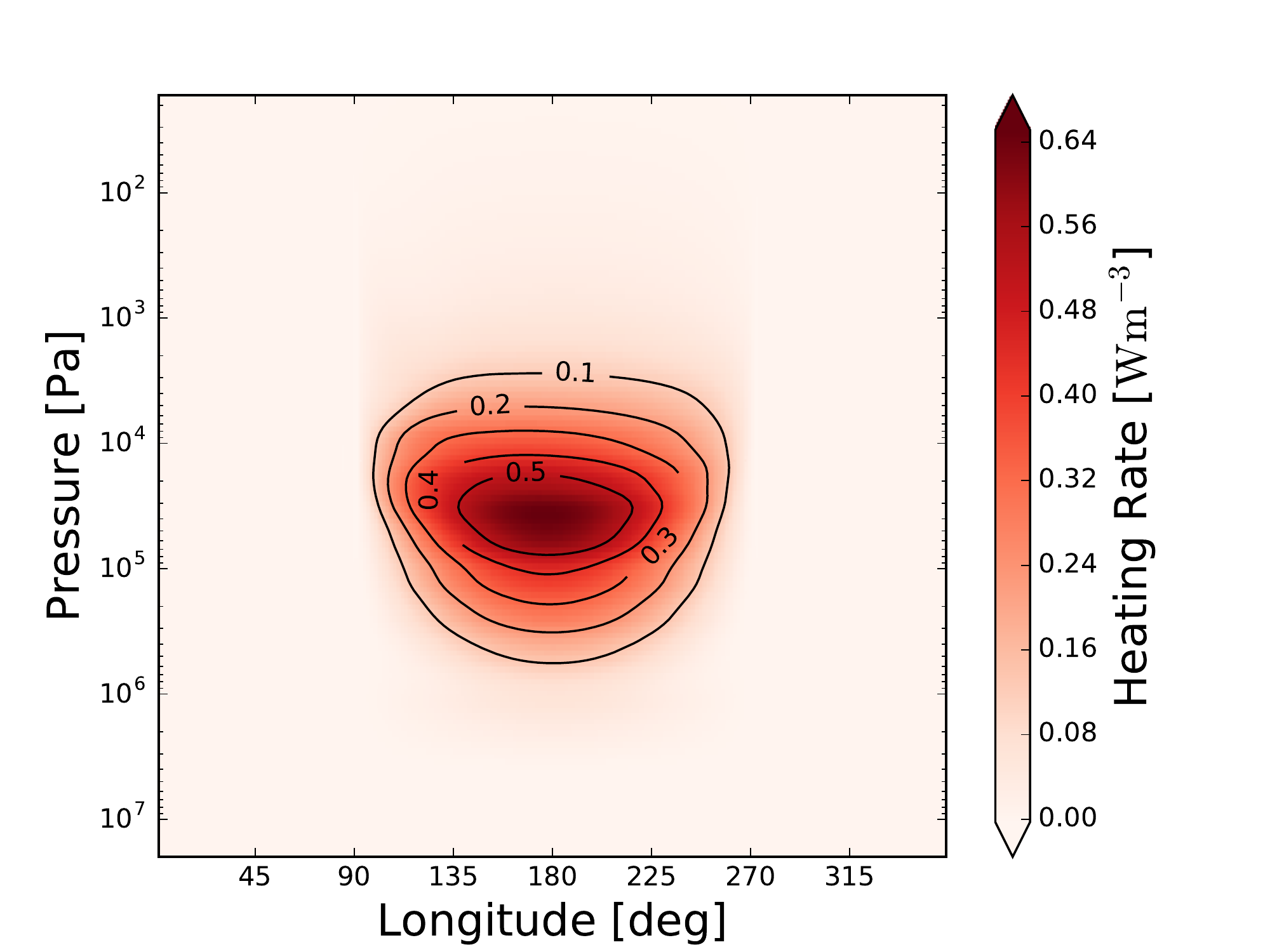}
    \includegraphics[width=0.45\textwidth]{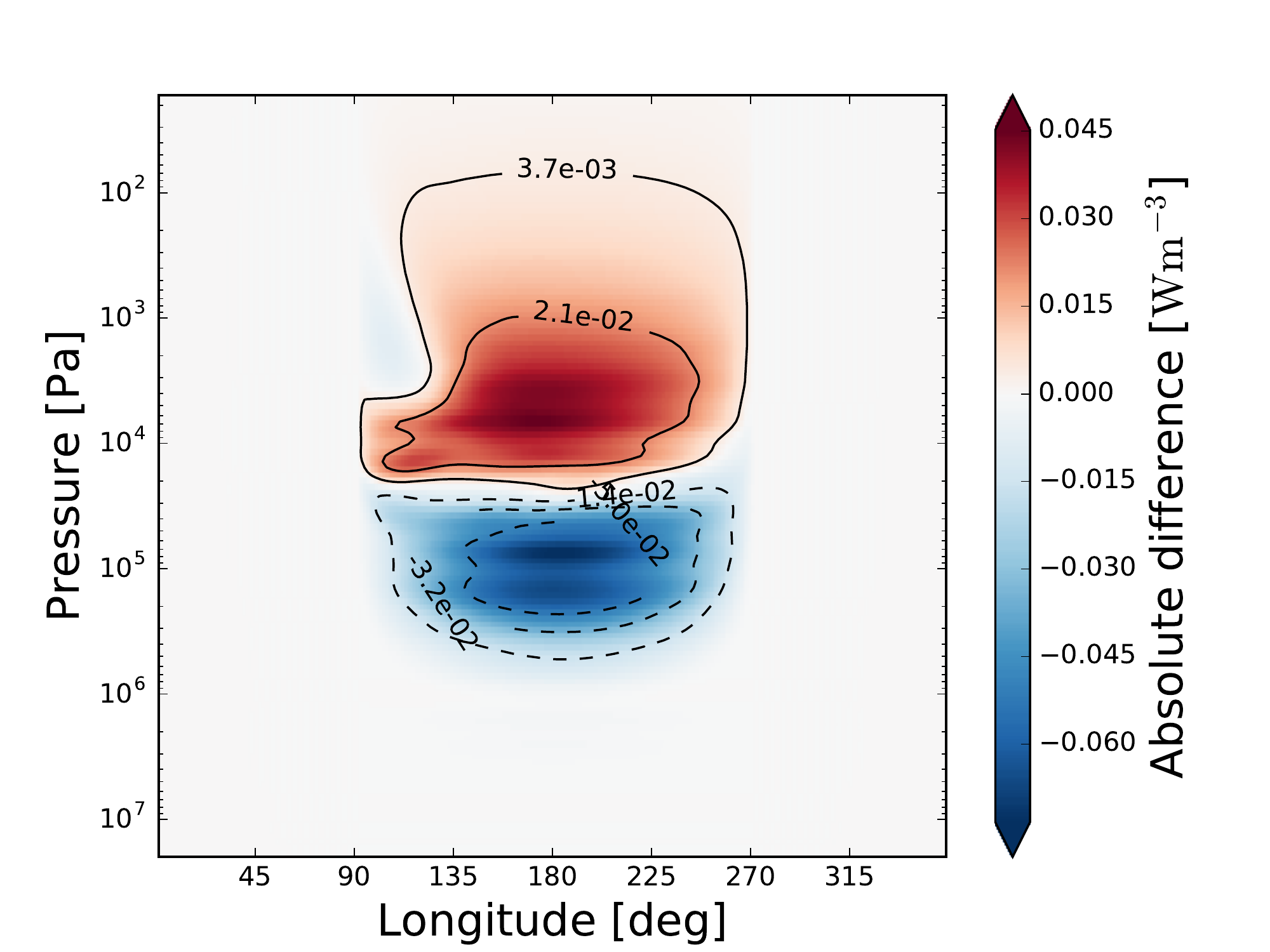} \\
    \includegraphics[width=0.45\textwidth]{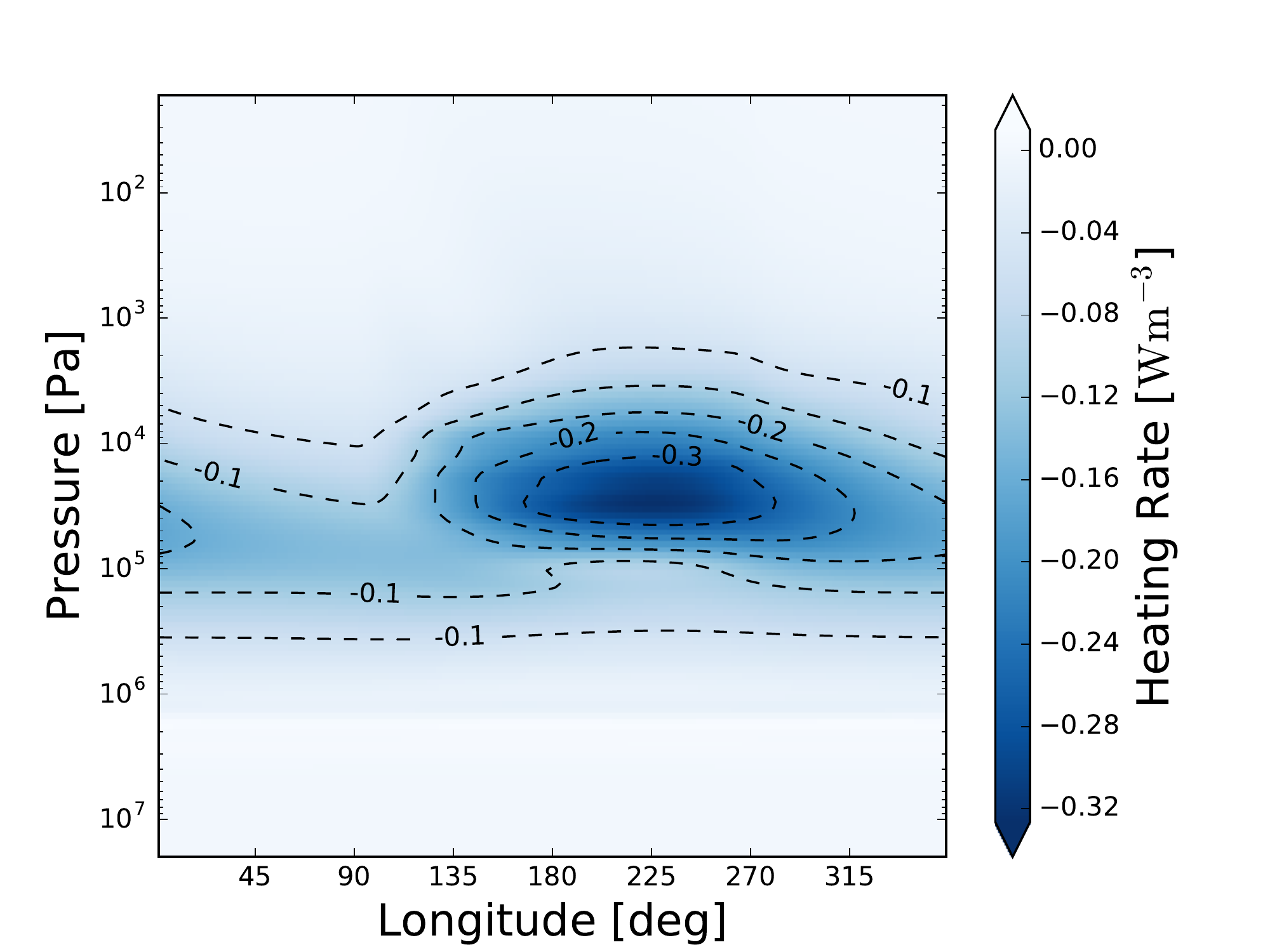} 
    \includegraphics[width=0.45\textwidth]{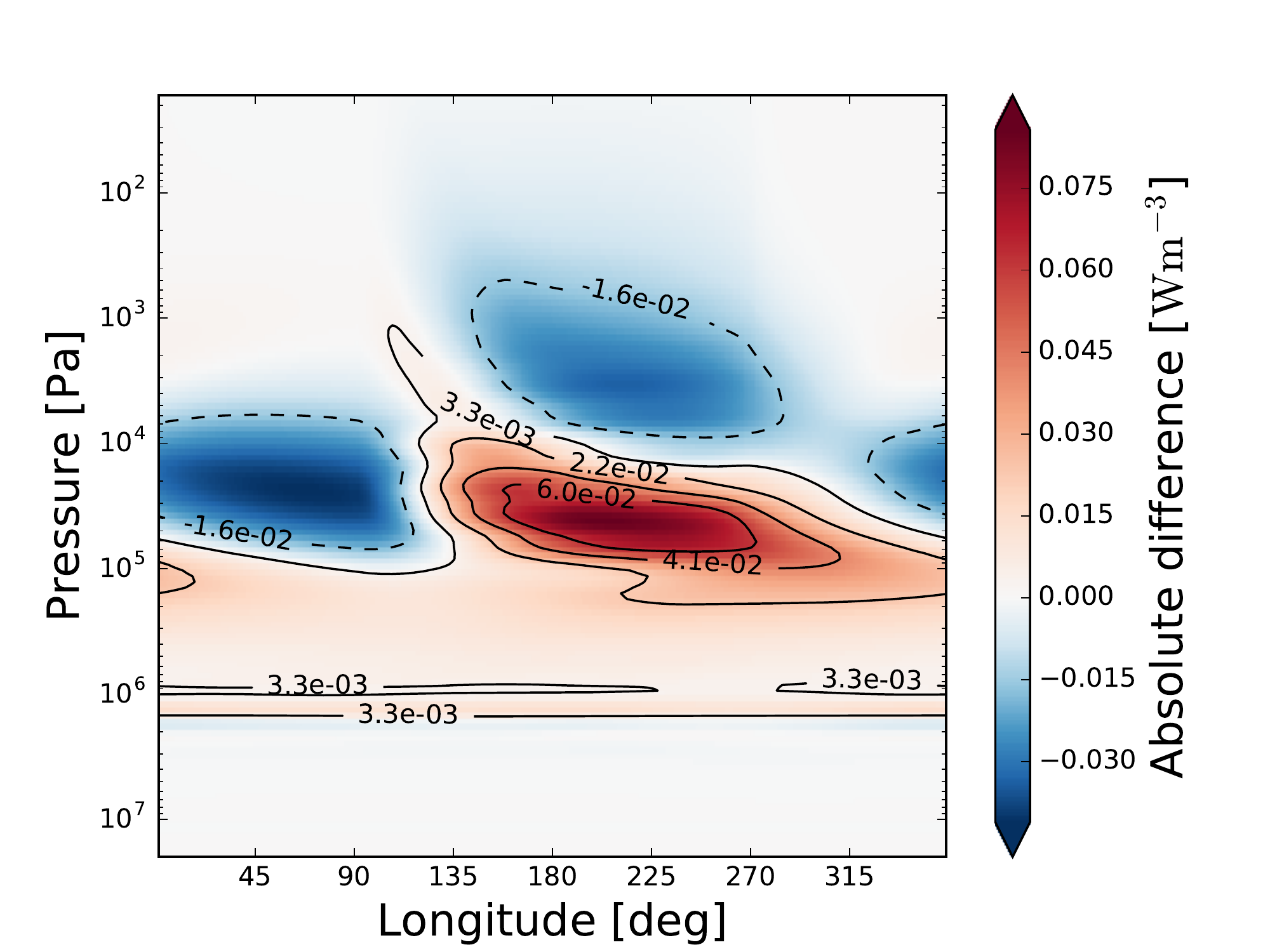} \\
    \includegraphics[width=0.45\textwidth]{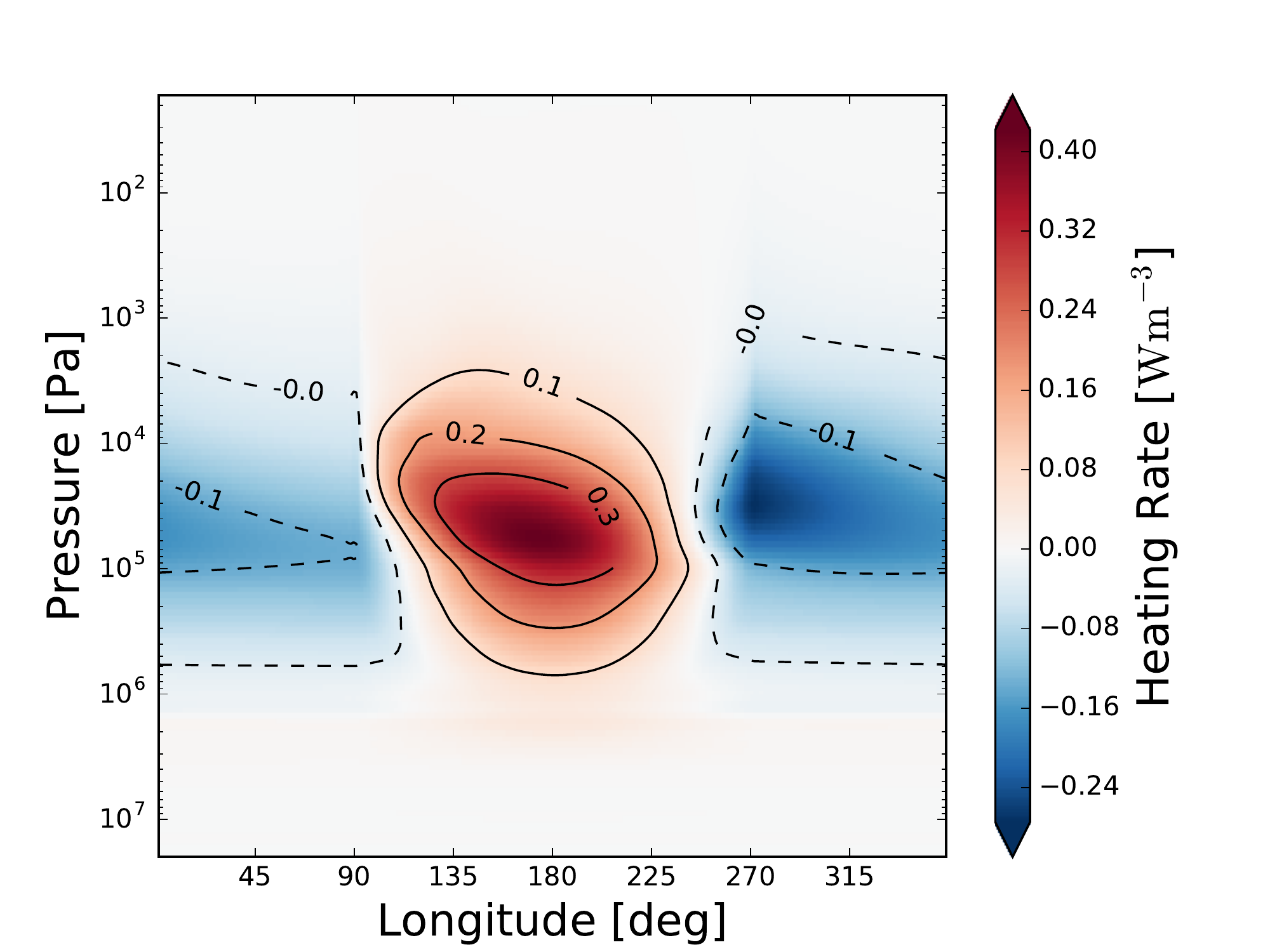}
    \includegraphics[width=0.45\textwidth]{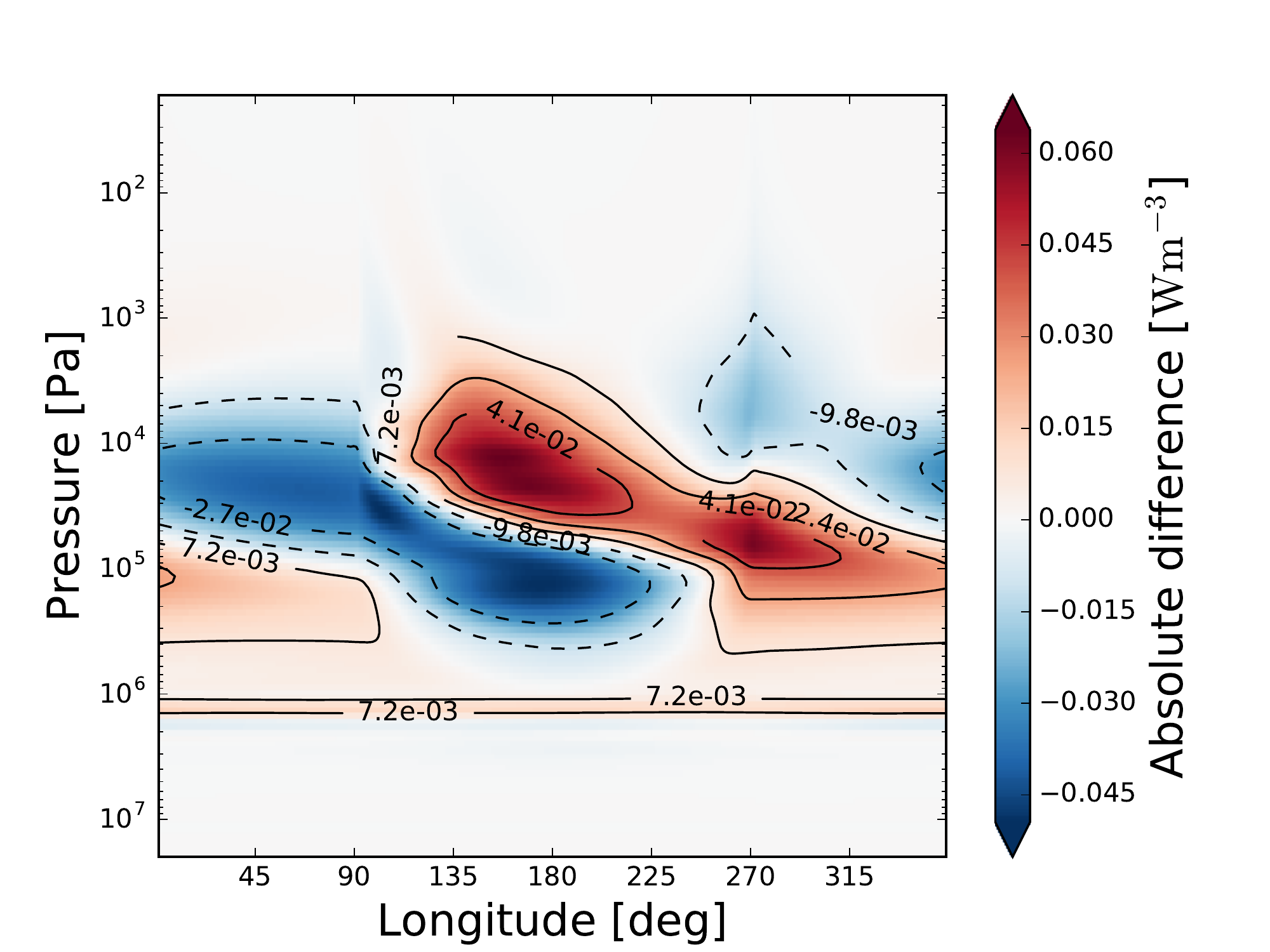} \\
  \end{center}
  \caption{Left column: the shortwave (top), longwave (middle) and net (bottom) heating rates (colour scale and black contours) in ${\rm W}~{\rm m}^{-3}$ for the equilibrium simulation. Right column: absolute differences in the heating rates (colour scale and black contours) between the relaxation and equilibrium simulations for the shortwave (top), longwave (middle) and net (bottom). All panels show area--weighted meridional--means between $\pm20^{\circ}$ latitude. We note that, for the difference plots, a positive difference in the shortwave heating rate (top right panel) means a more positive heating rate (more heating) while a positive change in the longwave (middle right panel) means a less negative heating rate (less cooling). A positive difference in the net heating rate (bottom right panel) means a larger net heating rate (i.e. more net heating or less net cooling).}
  \label{figure:hr}
\end{figure*}

To further understand the temperature response of the atmosphere we consider the radiative heating rates, which are shown in \cref{figure:hr} for the equatorial region. We show separately the shortwave heating (positive heating rate) and the longwave cooling (negative heating rate), as well as the net heating rate which is the sum of the shortwave and longwave components.

We note that in \cref{figure:hr} we show the heating rate $\mathcal{H}$ in units of ${\rm W}~{\rm m}^{-3}$, the rate of change of energy per unit volume \citep[][Eq. 17]{AmuBT14}. This is related to $\mathcal{H}$ in units of ${\rm K}~{\rm s}^{-1}$, the rate of change of temperature, by 
\begin{equation}
\mathcal{H}[{\rm W}~{\rm m}^{-3}] = \rho c_{P}\mathcal{H}[{\rm K}~{\rm s}^{-1}],
\end{equation}
where $\rho$ is the mass density and $c_P$ is the specific heat capacity. In these simulations $c_P$ is a global constant, shown in \cref{table:params}.

\cref{figure:hr} (left column) shows the heating rates for the equilibrium simulation, which are qualitatively very similar to those from the relaxation simulation (not shown). Naturally, the shortwave heating (top left panel) is restricted to the dayside atmosphere and peaks at $P\sim4\times10^4$ Pa at the substellar point. Longwave cooling (middle left panel), which occurs across both the dayside and nightside, peaks at similar pressures although shifted in longitude eastward of the substellar point. The net heating rate (bottom left panel) shows an overall positive heating rate (i.e. net heating) for the dayside and an overall negative heating rate (i.e. net cooling) for the nightside.

\cref{figure:hr} also shows the absolute difference in the heating rates between the relaxation and equilibrium simulations. On the dayside, there is a clear shift to lower pressures for the shortwave heating (top right panel), as stellar flux is absorbed higher in the atmosphere due to the enhanced methane abundance. This results in a relative heating for $P<2\times10^4$ Pa and a relative cooling for larger pressures, in the shortwave.

In the longwave (middle right panel), the peak of the cooling is also shifted to lower pressures, particularly for the region eastward of the substellar point, where the increase in the methane abundance is most significant. In \cref{figure:hr} this is indicated by a negative difference in the longwave heating rate (i.e. more cooling) for $P<10^4$ Pa and a positive difference (i.e. less cooling) for $P>10^4$ Pa. The change in the net heating rate (bottom right panel) shows that the shortwave effects are dominant on the dayside and the longwave effects are dominant on the nightside.

\begin{figure}
 \begin{center}
    \includegraphics[width=0.5\textwidth]{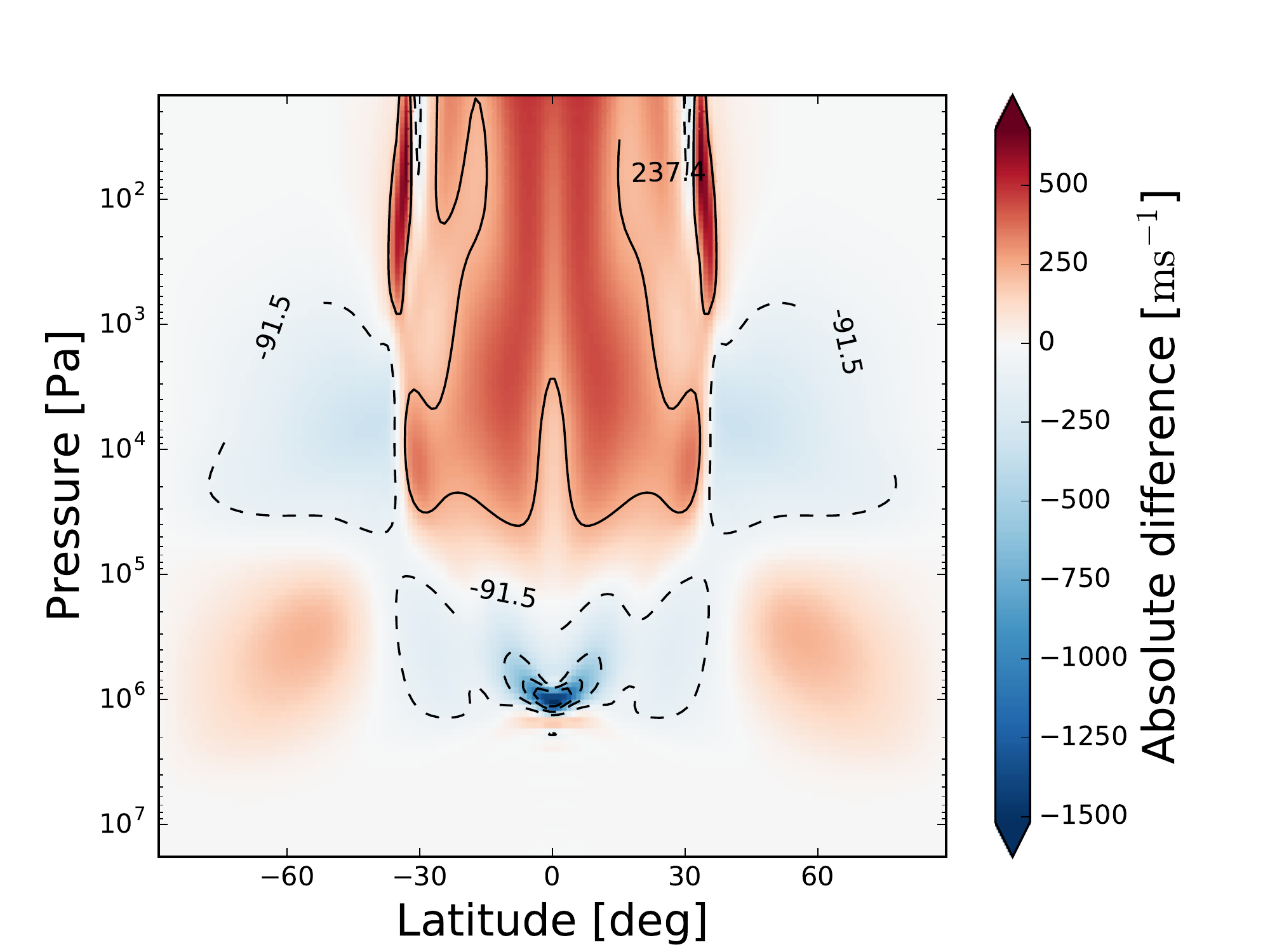}
  \end{center}
  \caption{The absolute difference in the zonal--mean temporal--mean (800--1000 days) zonal wind velocity between the relaxation and equilibrium simulations. A positive difference indicates an increased wind velocity in the relaxation simulation. }
  \label{figure:diff_wind}
\end{figure}

The overall effect of the changes to the shortwave and longwave heating rates is to increase the shortwave heating on the dayside and increase the longwave cooling on the nightside, for the $10^4<P<10^5$ Pa region. This increased differential heating acts to drive a faster equatorial jet. In  \cref{figure:diff_wind} we show the absolute difference in the zonal--mean zonal wind between the relaxation and equilibrium simulations. There is clearly an overall increase in the zonal wind velocity within the equatorial jet of 250--500 ${\rm m}~{\rm s}^{-1}$. Comparing this with \cref{figure:wind_temp} this corresponds to a 5--10\% increase compared with the equilibrium simulation. 

The estimated dynamical and radiative timescales (Section \ref{section:temp_response}) indicate that the atmosphere is radiatively driven for $P<10^4$ Pa and dynamically driven for higher pressures. Advection of heat by the equatorial jet, aided by the increased wind velocities, leads to a heating between $10^4<P<10^5$ Pa all around the planet. For lower pressures, where the atmosphere is radiatively driven, the local temperature change is more spatially dependent.

We note that the longwave cooling is itself dependent on the atmospheric temperature and the temperature changes between the equilibrium and relaxation simulations will result in different steady--state cooling rates. We further note that the net radiative heating rates shown \cref{figure:hr} are approximately balanced by advection of heat in the latter stages of the simulation, where the atmosphere (for less than $P\sim10^6$ Pa) is in an approximate steady-state.

\subsection{Summary}

We find that for pressures less than $10^4$ Pa the atmosphere is generally cooler on the dayside and warmer on the nightside due to wind--driven chemistry, compared with chemical equilibrium. For the pressure region $10^4<P<10^5$ Pa we find a significant warming all around the planet, in the equatorial region. An increased differential heating between the dayside and nightside drives a faster equatorial jet. These temperature changes lead to a decreased TOA radiative flux for the dayside and an increased TOA radiative flux for the nightside. Our testing shows that these changes are primarily due to changes in the abundance of methane.

%%%%%%%%%%%%%%%%%%%%%%%%%%%%%%%%%%%%
\section{Contribution functions}
\label{section:cf}
%%%%%%%%%%%%%%%%%%%%%%%%%%%%%%%%%%%%

To futher understand the effect of wind--driven chemistry on the radiative properties of the atmosphere we consider the contribution function, which quantifies the contribution of a layer to the upwards intensity at the top of the atmosphere. A peak in the contribution function effectively indicates the pressure level of the photosphere. The calculation of the contribution function is described and validated in Appendix \ref{section:app1} and Appendix \ref{section:app2}, respectively. We first discuss the structure of the contribution functions from the equilibrium simulation and compare with previous studies in Section \ref{section:con_fun_eq}. We then show the effect of wind--driven chemistry in Section \ref{section:con_fun_neq}.

\subsection{Chemical equilibrium}
\label{section:con_fun_eq}

\begin{figure*}
  \begin{center}
    \includegraphics[width=0.4\textwidth]{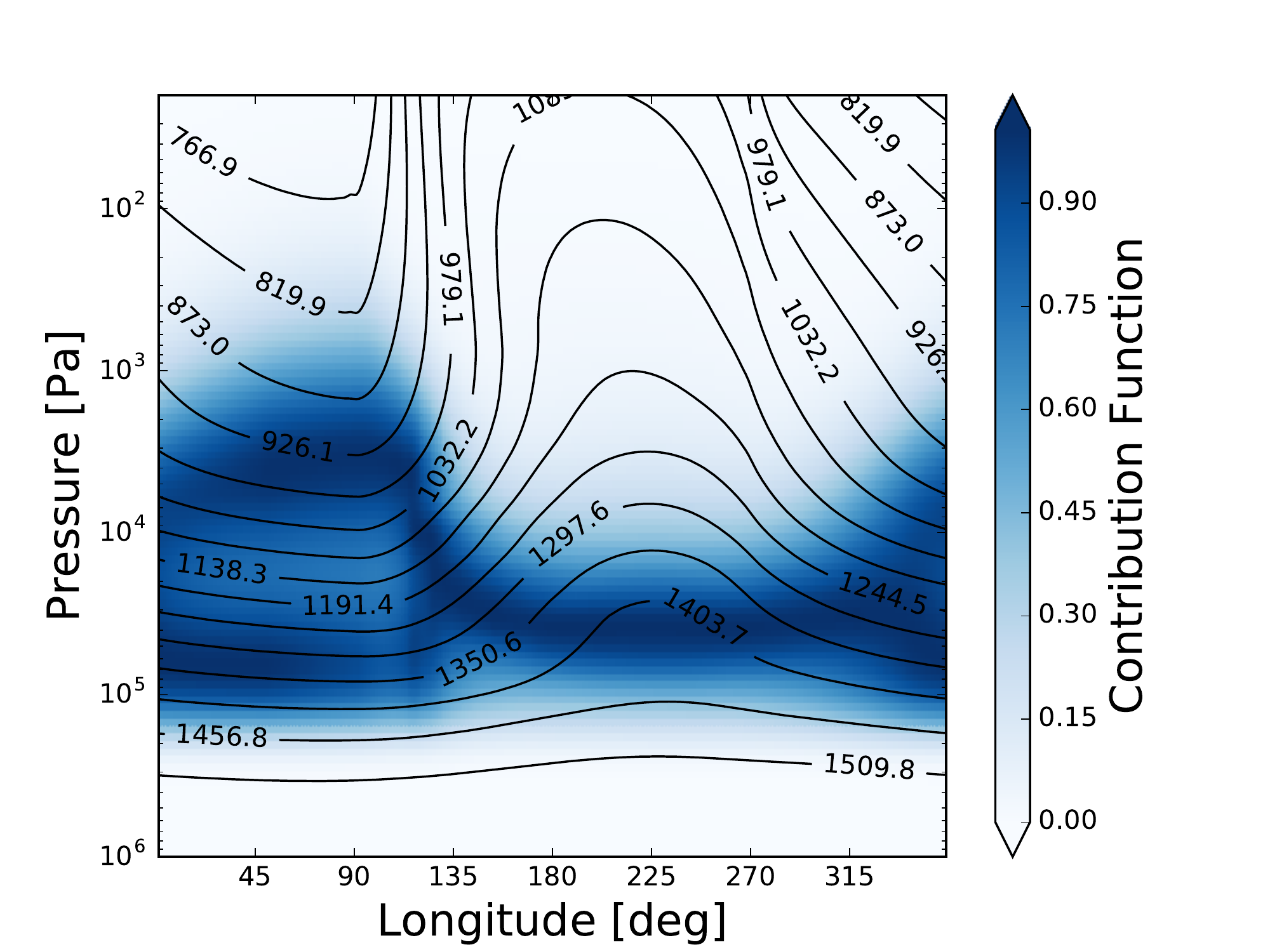}
    \includegraphics[width=0.4\textwidth]{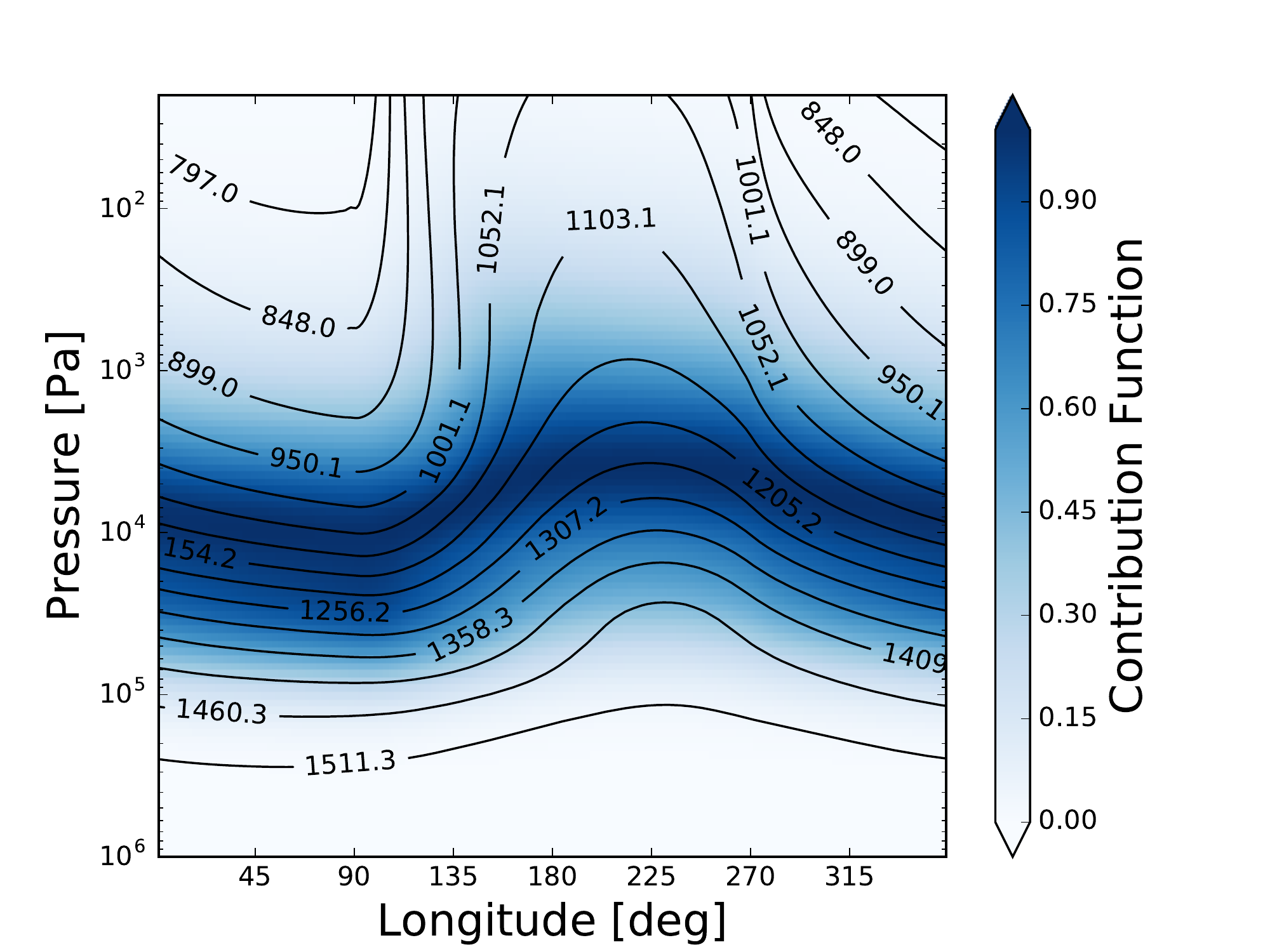} \\
    \includegraphics[width=0.4\textwidth]{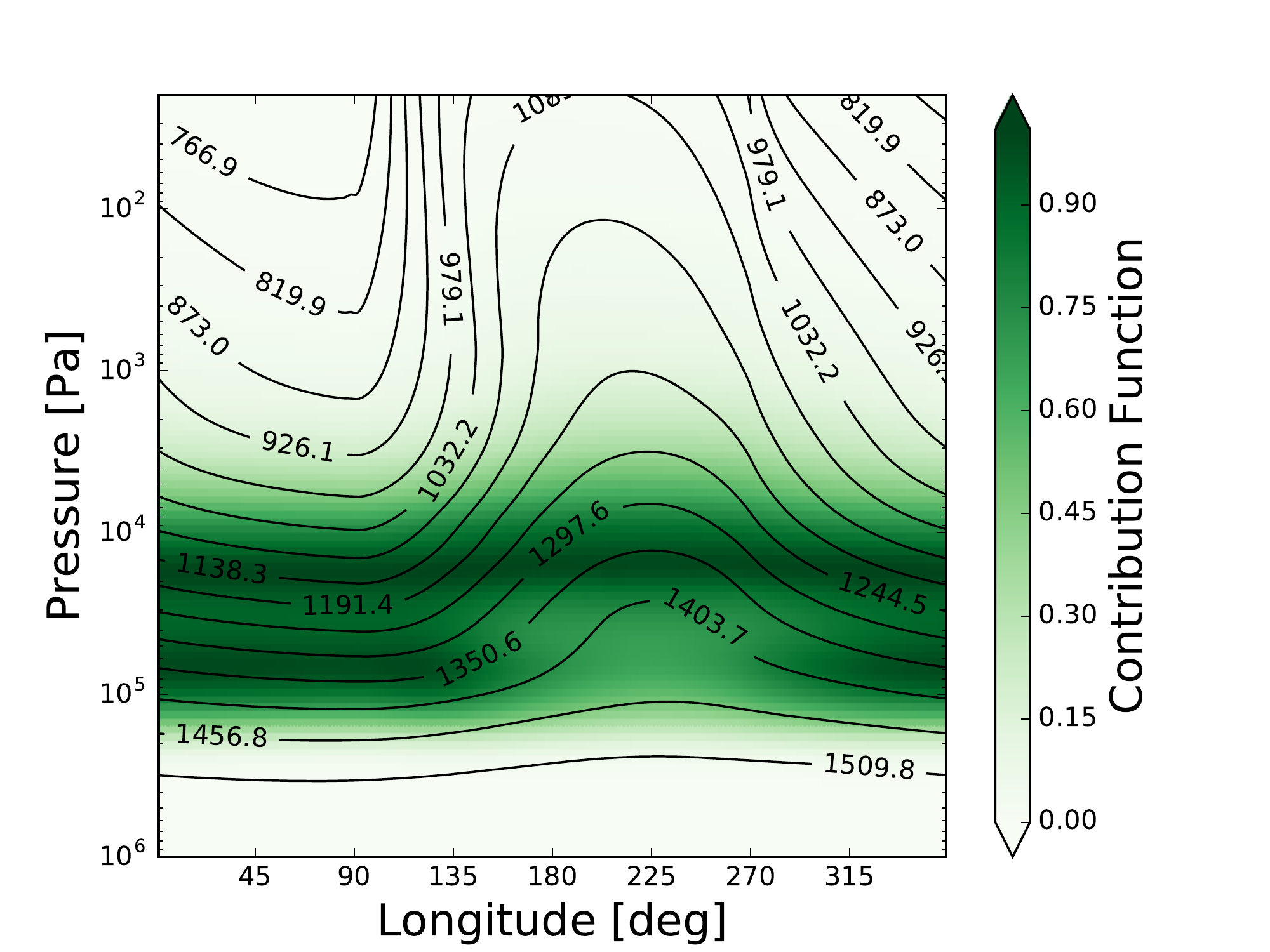}
    \includegraphics[width=0.4\textwidth]{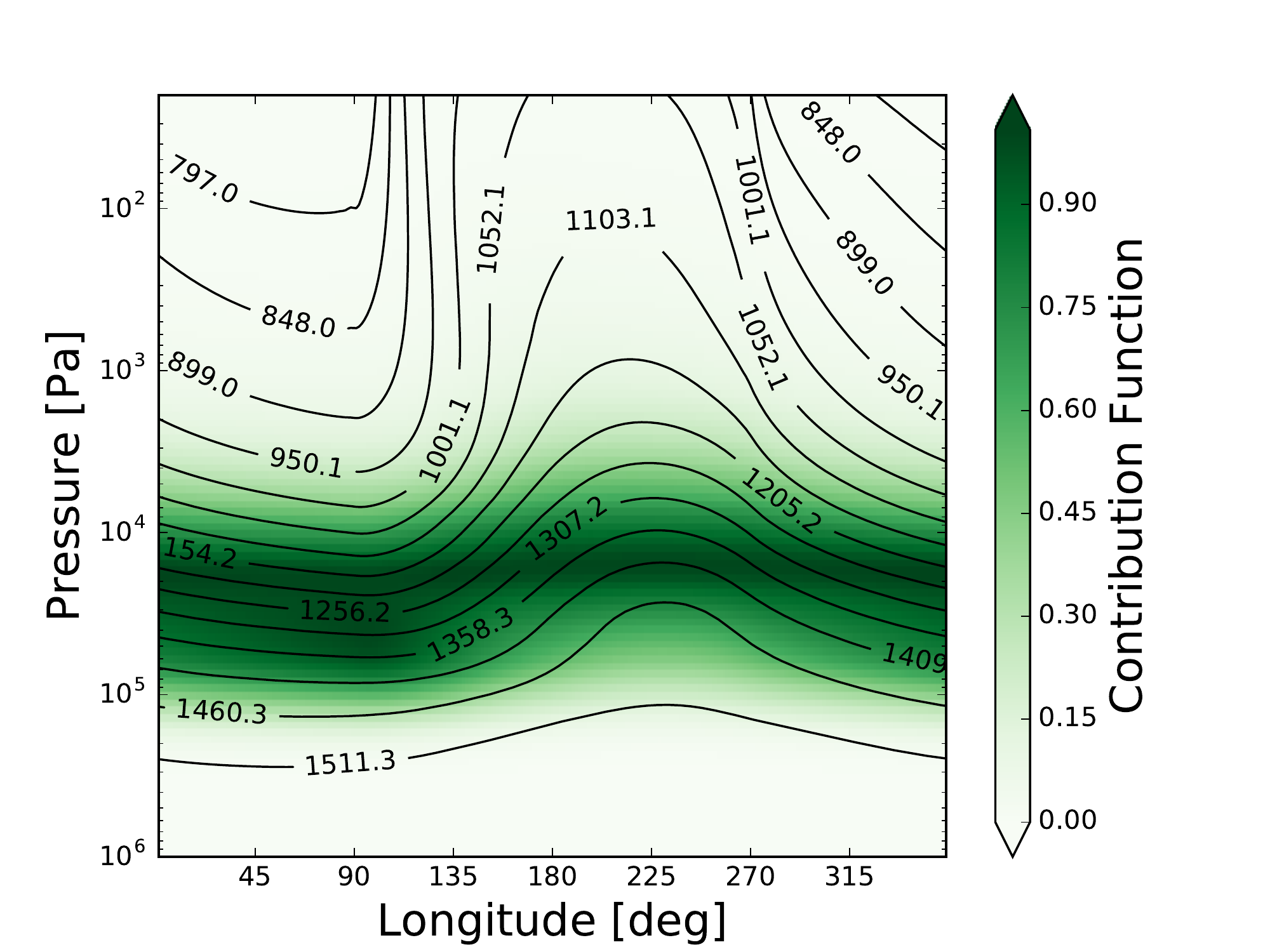} \\
    \includegraphics[width=0.4\textwidth]{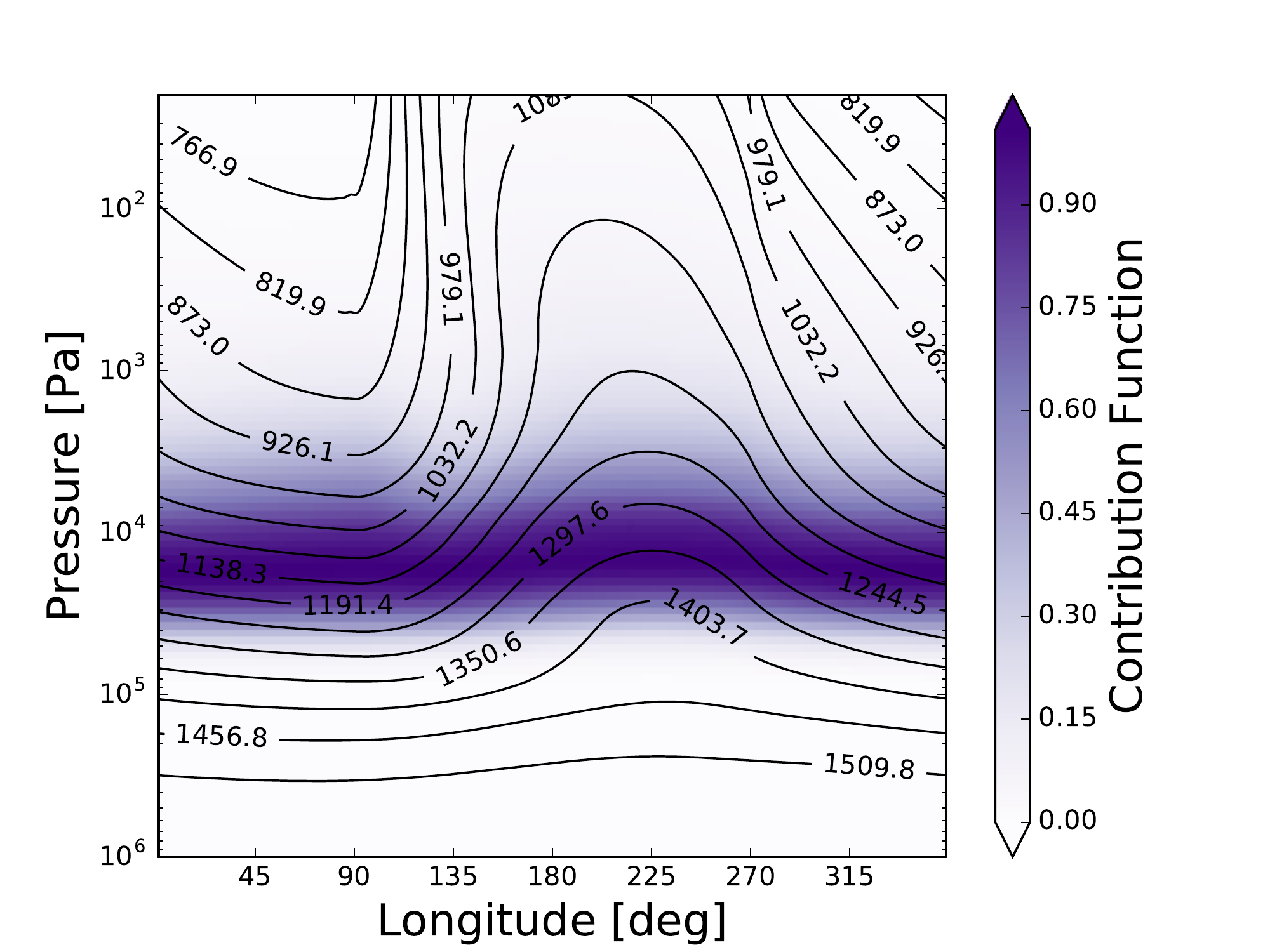}
    \includegraphics[width=0.4\textwidth]{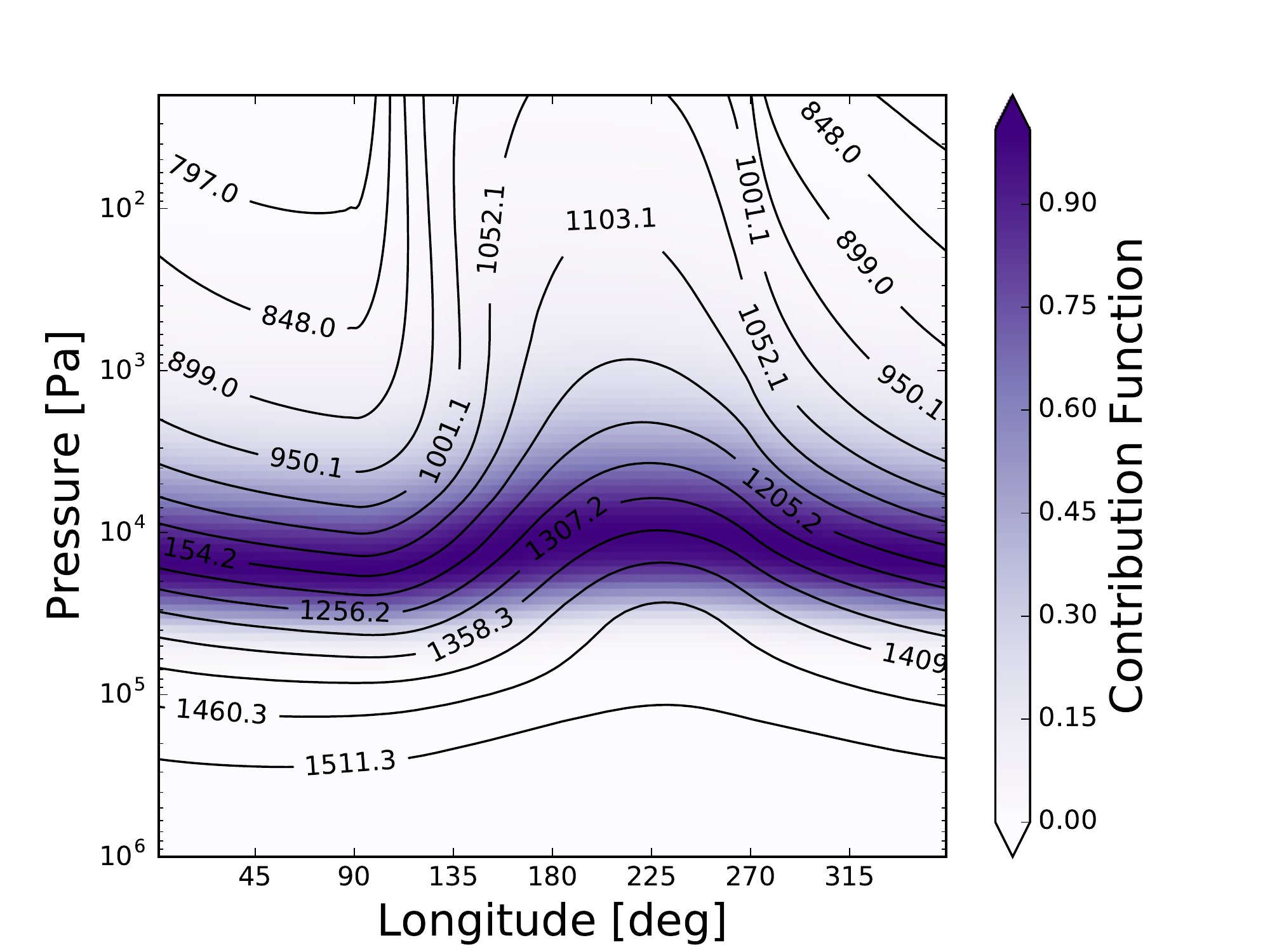} \\
    \includegraphics[width=0.4\textwidth]{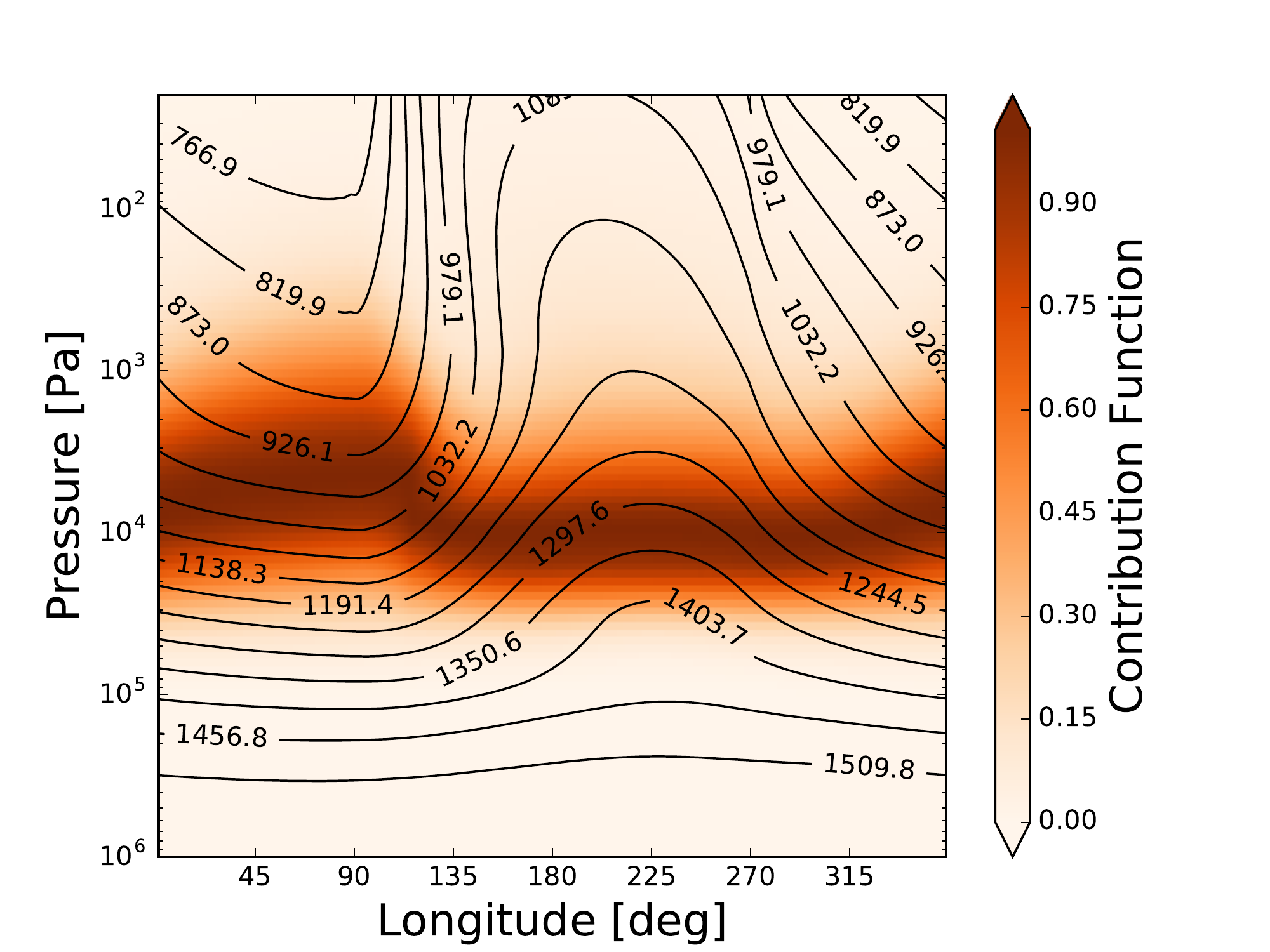}
    \includegraphics[width=0.4\textwidth]{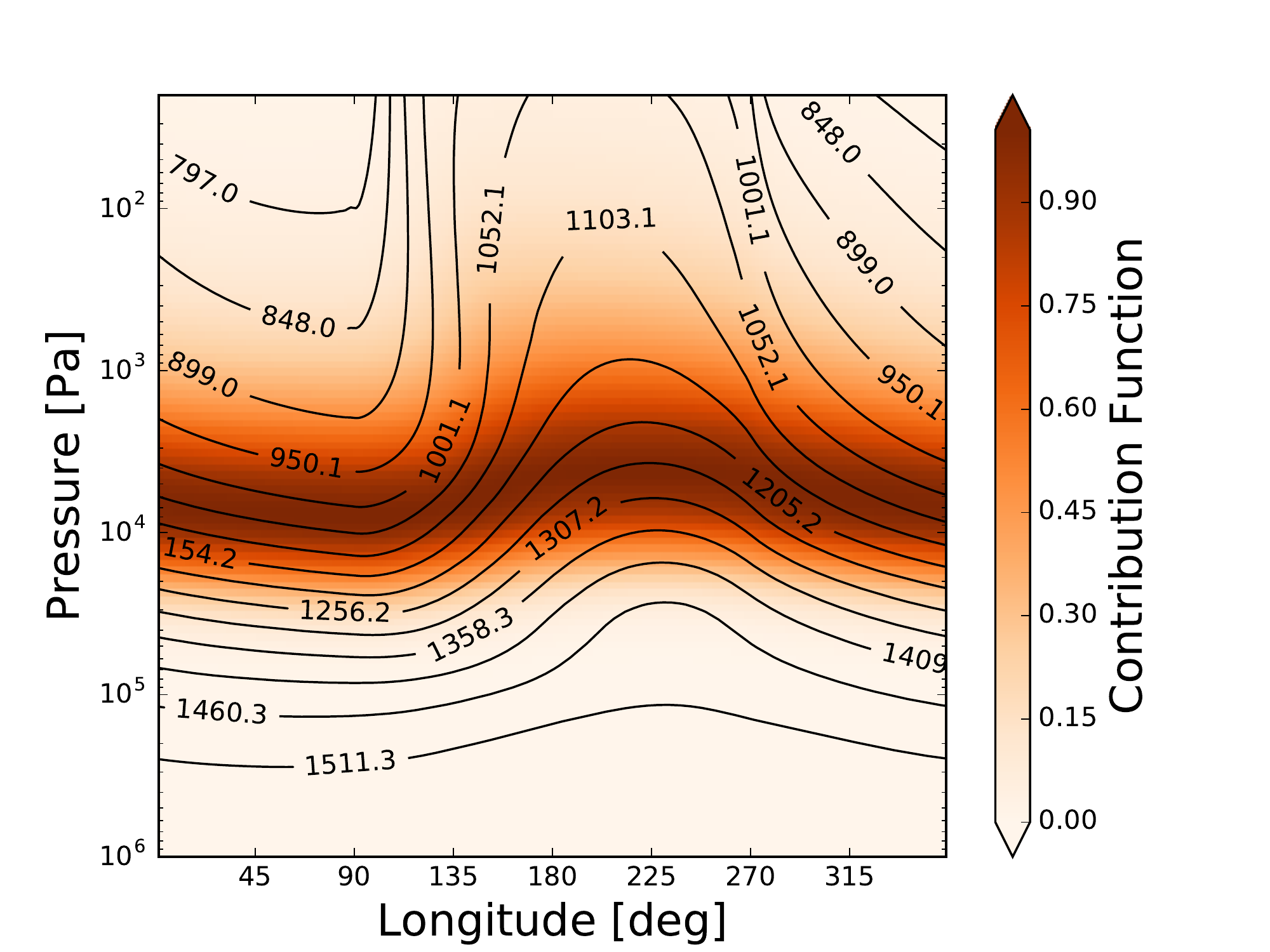} \\
  \end{center}
\caption{Area--weighted meridional-means ($\pm20^{\circ}$ latitude) of the normalised contribution function (colour scale) for different Spitzer/IRAC channels: 3.6 \textmu m (top row), 4.5 \textmu m (second row), 5.8 \textmu m (third row and 8.0 \textmu m (bottom row). The atmospheric temperature is also shown in black contours. Results from the equilibrium simulation are shown in the left column, and from the relaxation simulation in the right column.}
\label{figure:cf}
\end{figure*}

\cref{figure:cf} shows an area--weighted meridional--mean ($\pm20^{\circ}$ latitude) of the contribution function as a function of longitude and pressure for the four Spitzer/IRAC channels centered on 3.6 \textmu m, 4.5 \textmu m, 5.8 \textmu m and 8.0 \textmu m. Methane is a prominant absorber in the 3.6 \textmu m and 8.0 \textmu m channels while water and carbon monoxide are the primary absorbers in the 4.5 \textmu m and 5.8 \textmu m channels, respectively.

% Contribution Functions

In the case of chemical equilibrium (left panels, \cref{figure:cf}), it is immediately apparent that the contribution function varies significantly with longitude for the 3.6 and 8.0 \textmu m channels. The peak of the contribution function in these channels generally occurs at higher pressures on the dayside compared with the nightside.  On the other hand, the contribution function is relatively flat (isobaric) in the 4.5 and 5.8 \textmu m channels. 

The shapes of the 3.6 and 8.0 \textmu m contribution functions are explained by the distribution of methane in chemical equilibrum. The dayside equilibrium abundance of methane is many orders of magnitude less than the nightside equilibrium abundance (Section \ref{section:chemistry}) and therefore the opacity, within these spectral channels, is also smaller for the dayside. The smaller dayside opacity results in a smaller optical depth and hence a ``deeper'' photosphere. On the nightside, where opacity due to methane is significantly larger, the photosphere is shifted to lower pressures. 

The contribution function in the 3.6 \textmu m channel actually contains a ``double peak'' in the nightside region, with a secondary, slightly smaller peak at $P\sim10^5$ Pa. This means that these two pressure regions both make a significant contribution to the intensity emerging from the top of the atmosphere. We find that this double peak is due to the spectral dependence of the opacity within the 3.6 \textmu m Spitzer/IRAC channel, which spans the range 3.08--4.01 \textmu m and corresponds to $\sim75$ spectral bands in the high spectral resolution setup of our model (see Section \ref{sect:mod_param}). For bands where methane has significant absorption the contribution function peaks at lower pressures, but for bands where methane absorption is not so prominent the contribution peaks at higher pressures. When combined in the 3.6 \textmu m channel this leads to two distinct peaks.

The 4.5 and 5.8 \textmu m contribution functions are also consistent with the equilibrium composition. Both water and carbon monoxide show small variations with longitude, compared with methane which varies by orders of magnitude. Therefore, the contribution functions in the spectral regions where these species are dominant absorbers are approximately isobaric around the equatorial region.

We now compare our results from the equilibrium simulation (\cref{figure:cf}) with those of \citet[][see their Fig. 4]{DDC17}, where we find several significant differences. We note that in our figures (\cref{figure:cf}) the substellar point is located at 180$^{\circ}$ longitude, while in Fig. 4 of \citet{DDC17} the substellar point is located at 0$^{\circ}$ longitude. 

Firstly, for the 3.6 \textmu m and 8.0 \textmu m channels, we find that the contribution function peaks at lower pressures on the nightside, compared with the dayside. Intuitively, we expect this to be the case given the larger abundance of methane on the nightside. However, the opposite appears to be true for the results of \citet{DDC17} with the contribution function peaking at lower pressures on the dayside, compared with the nightside. Secondly, for the 4.5 \textmu m and 5.8 \textmu m channels \citet{DDC17} find that the contribution function (generally) shifts to a higher presures on the nightside, in contrast to the approximately isobaric contribution functions that we find in the same channels The temperature structure appears to be very similar between the two models (compare our \cref{figure:wind_temp} with their Fig. 4) and therefore the equilibrium chemical composition should also be similar. The cause of the discrepancy is therefore puzzling. 

Whatever the cause of the difference, the opposing results have important consequences for the interpretation. Taking the 3.6 \textmu m as an example, we find that the peak of the contribution function crosses many isotherms between the dayside and nightside of the planet, since the shape of the contribution function is approximately opposite to that of the isotherms. In contrast, the shape of the contribution functions presented by \citet{DDC17} cause them to approximately follow the isotherms, meaning that the temperature contrast at the photosphere around the planet should be much smaller. 

\subsection{The effect of wind--driven chemistry}
\label{section:con_fun_neq} 

The contribution functions for the relaxation simulations (right panels, \cref{figure:cf}) show significant differences with the equilibrium simulations, particularly for the 3.6 and 8.0 \textmu m channels. Most importantly, we find that the peak of the contribution functions occur at lower pressures on the dayside than on the nightside: opposite to the chemical equilibrium case.

Since the chemical composition is approximately homogenous in the relaxation simulation, the non-isobaric shape of the contribution functions is likely related to the temperature structure. To check, we performed a series of simple tests. Firstly, we fixed the temperature at which the Planck function $B(T)$ (see \cref{equation:final_cfi}) is evaluated and the shape of the contribution functions become flatter, though not isobaric. Secondly, we also fixed the temperature at which the absorption coefficient $\kappa$ is calculated. Fixing the temperature for both $B(T)$ and $\kappa$ yields an isobaric contribution function, indicating that it is the temperature dependence of these two terms that drive the shape of the contribution functions for the relaxation simulations.

Since the temperature at the peak of the contribution function (i.e. the photosphere) effectively determines the TOA flux, changes in the shape of the contribution function will effect the observed emission from the atmosphere. As previously discussed, the shape of the contribution functions in the equilibrium case means the photosphere cuts across many isotherms between the dayside and nightside. In contrast, for the relaxation case the contribution functions approximately follow the shape of the isotherms, leading to a much smaller temperature difference at the pressure level of the photosphere between the dayside and nightside.

With the above argument, this allows us to predict that we should find a larger phase amplitude in the simulated emission phase curve, in the 3.6 \textmu m channel, for the equilibrium simulation compared with the relaxation simulation.

\subsection{Summary}

We have presented contribution functions calculated for both the equilibrium and relaxation simulations of HD~189733b. In the chemical equilibrium case, we find relatively flat (isobaric) contribution functions for the 4.5 \textmu m and 5.8 \textmu m Spitzer/IRAC channels but non-isobaric contribution functions in the 3.6 \textmu m and 8.0 \textmu m channels, where methane is an important absorber.  In the latter channels, the contribution function peaks at lower pressures on the nightside compared with the dayside, opposite to the trends found by \citet{DDC17}. However, when including wind--driven chemistry, this trend reverses with the contribution functions peaking at lower pressures on the dayside in the 3.6 \textmu m channel. This has a significant effect on the temperature at the pressure level of the photosphere.

\section{Emission phase curves}
\label{section:phase}

In this section we present the emission phase curves calculated from both the equilibrium and relaxation simulations of HD~189733b. We compare the results from the equilibrium simulation with results from previous studies and discuss the effect of wind--driven chemistry.

\subsection{Chemical equilibrium}

% Phase curves
\begin{figure}
  \begin{center}
    \includegraphics[width=0.5\textwidth]{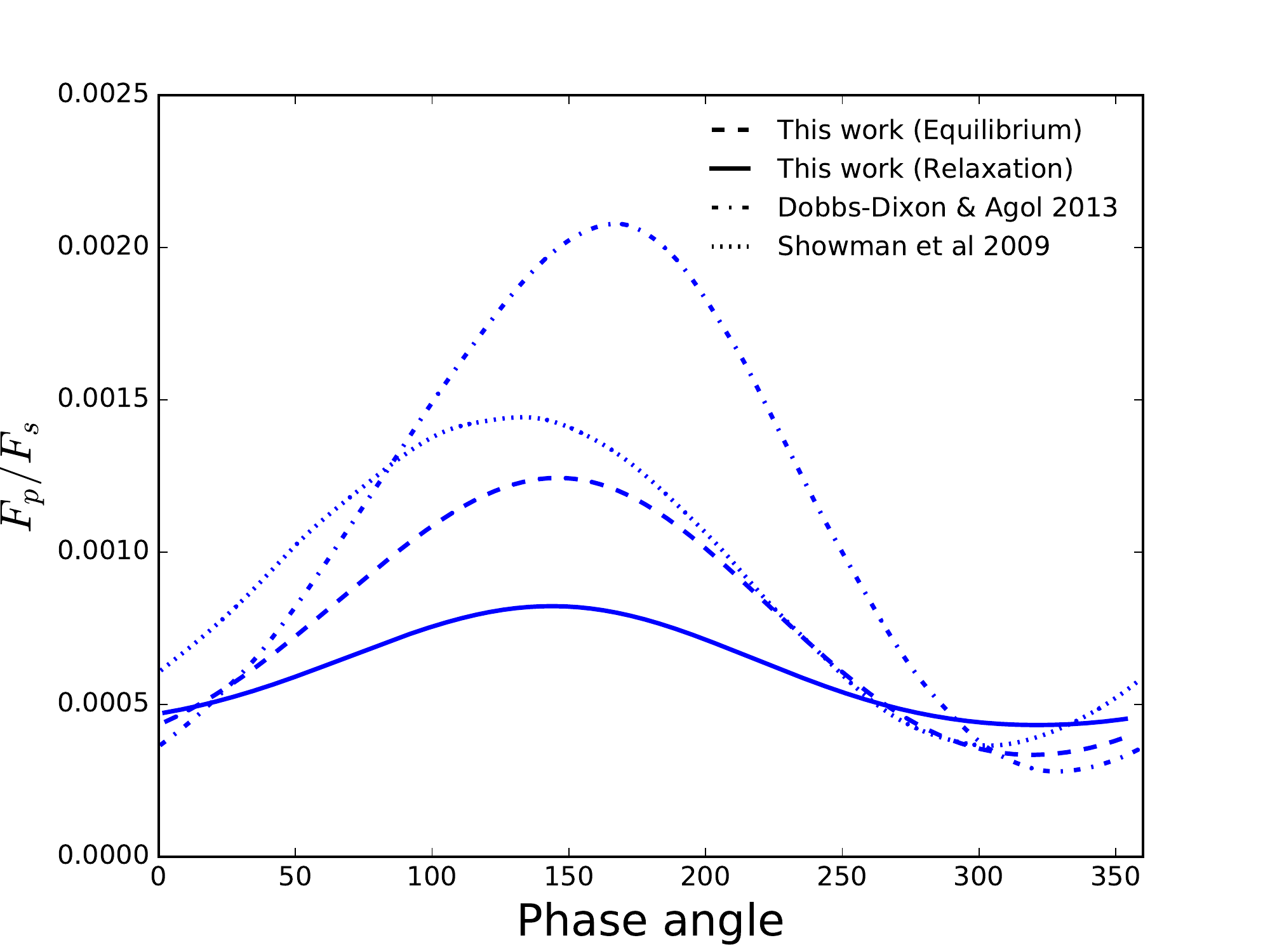} \\
   \includegraphics[width=0.5\textwidth]{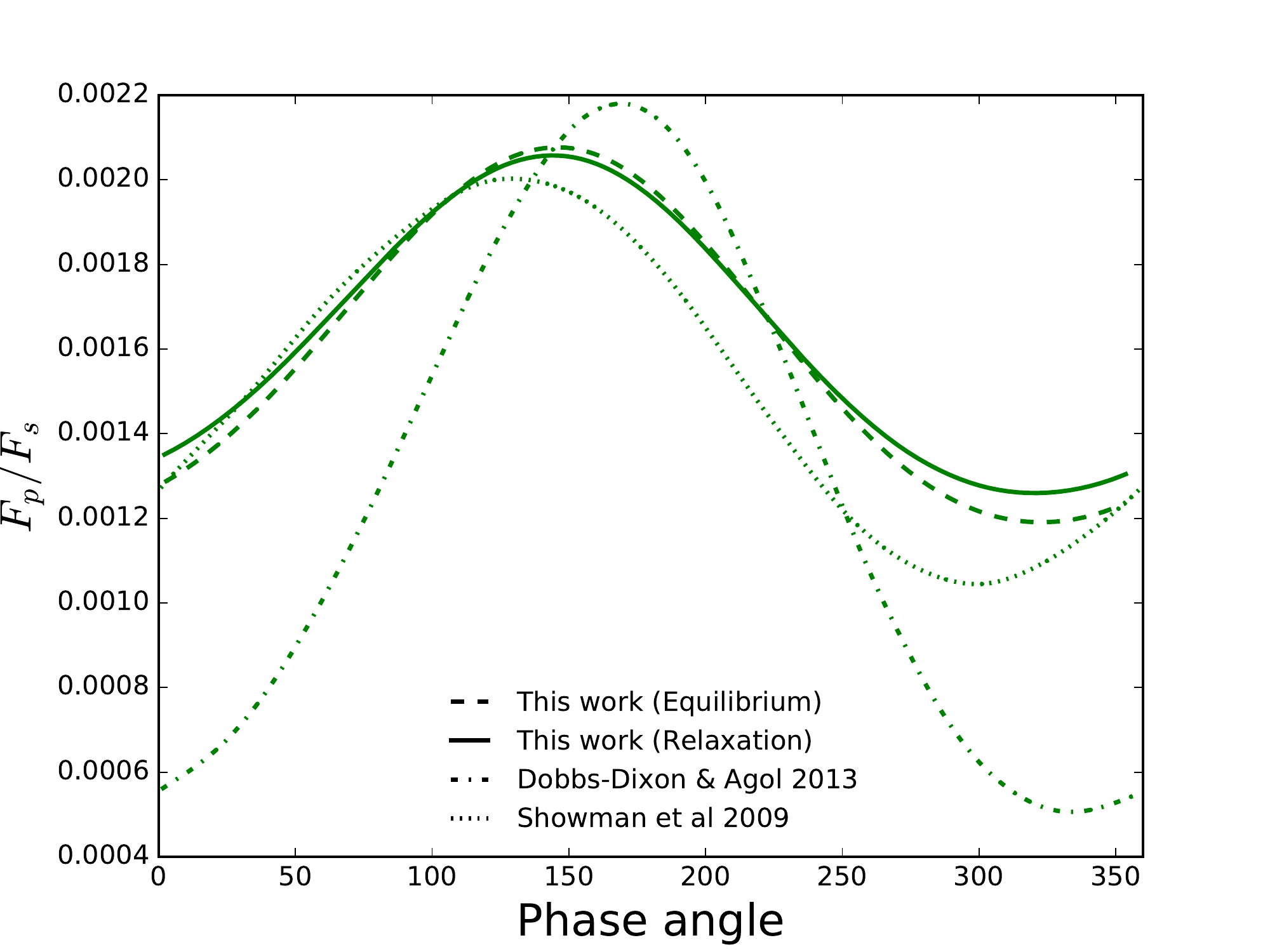}
  \end{center}
\caption{Calculated emission phase curves from our simulations with both chemical equilibrium and chemical relaxation as well as from \citet{DobA13} and \citet{Showman2009} (both assuming chemical equilibrium) for the 3.6 \textmu m (top) and 4.5 \textmu m (bottom) Spitzer/IRAC channels.}
\label{figure:phase_curve}
\end{figure}

\begin{figure}
  \begin{center}
    \includegraphics[width=0.5\textwidth]{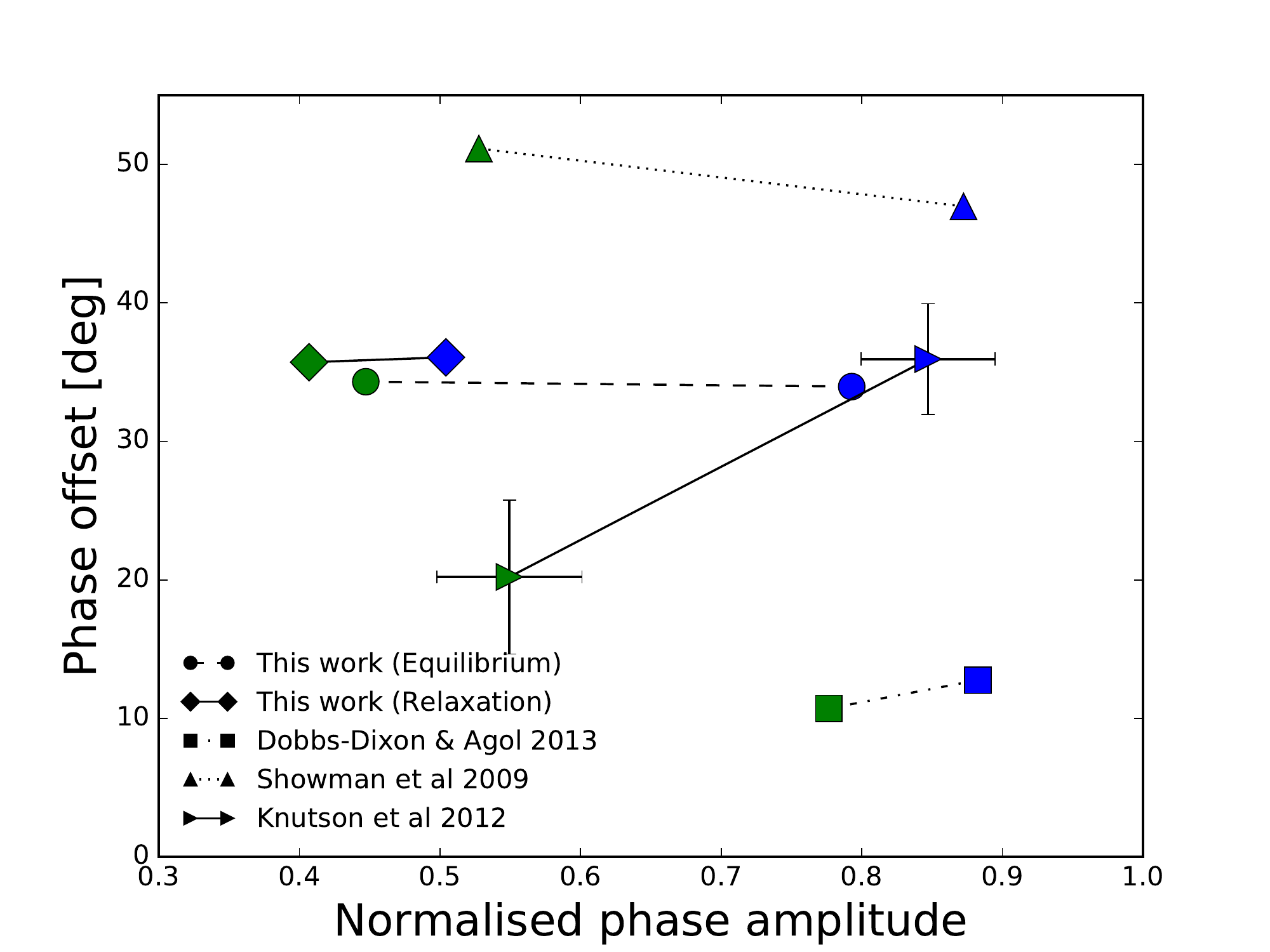}
  \end{center}
\caption{The phase offset versus the normalised phase amplitude for the 3.6 \textmu m (blue) and 4.5 \textmu m (green) Spitzer/IRAC channels from our UM simulations, as well as those of \citet{DobA13} and \citet{Showman2009}. Also shown are the phase offset and normalised phase amplitude derived from the observed phase curves of HD~189733b \citep{KnuLF12}.}
\label{figure:offset_amp}
\end{figure}

\cref{figure:phase_curve} shows the 3.6 \textmu m and 4.5 \textmu m emission phase curves from our simulations, as well as those from previous simulations of \citet{Showman2009} and \citet{DobA13}. We first compare our chemical equilibrium phase curves with those of previous models and focus on the normalised phase amplitude and the phase offset. The normalised phase amplitude quantifies the difference between the maximum and minimum flux ratio normalised by the flux ratio at secondary eclipse \citep[e.g.][]{DDC17} while the phase offset quantifies the angular shift between the maximum flux ratio in the phase curve and the secondary eclipse. The normalised phase amplitudes and phase offsets are shown separately in \cref{figure:offset_amp} which is based on Fig. 1 from \citet{DDC17}.

Firstly, for the 3.6 \textmu m channel, we find that all three models show similar normalised phase amplitudes (see \cref{figure:offset_amp}). However, \cref{figure:phase_curve} shows that the \citet{DDC17} phase curve has a significantly larger actual (i.e. non-normalised) phase amplitude, compared with both our result and that of \citet{Showman2009}. There is a significant phase offset difference between all three models, with an offset of $\sim50^{\circ}$ for \citet{Showman2009}, $\sim35^{\circ}$ for this work and $\sim15^{\circ}$ for \citet{DDC17}.

Secondly, for the 4.5 \textmu m channel, we obtain a similar normalised phase amplitude ($\sim0.5$) to that of \citet{Showman2009}. The normalised phase amplitude of \citet{DDC17} is significantly larger ($\sim0.8$) than both our result and that of \citet{Showman2009}, which is particularly apparent in \cref{figure:phase_curve}. The phase offsets for each model are similar to their values in the 3.6 \textmu m channel.

The trends in the emission phase curves from our simulations, shown in \cref{figure:phase_curve} and \cref{figure:offset_amp}, are consistent with the trends in the contribution functions previously discussed in Section \ref{section:temp}. The equilibrium 3.6 \textmu m contribution function crosses many isotherms as it peaks at lower pressures on the nightside compared with the dayside, leading to a large normalised phase amplitude. In contrast, the 4.5 \textmu m contribution function is approximately isobaric and crosses fewer isotherms, resulting in a smaller emission phase amplitude. However, the contribution functions in both channels peak at similar pressures on the dayside, near the location of the hotspot, giving a similar phase offset.

\cref{figure:offset_amp} also shows the phase offset and normalised phase amplitude derived from observations of HD~189733b by \citet{KnuLF12}. We note that none of the models reproduce the significant phase offset difference between the 3.6 \textmu m and 4.5 \textmu m channels.

% Phase amplitude v phase offset

%%%%%%%%%%%%%%%%%%%%%%%%
\subsection{The effect of wind--driven chemistry}
%%%%%%%%%%%%%%%%%%%%%%%%

We now consider the effect of wind--driven chemistry on the emission phase curve, also shown in \cref{figure:phase_curve}. As expected from previous discussion of the contribution function, we find only a small difference in the 4.5 \textmu m phase curve between the equilibrium and relaxation simulations. The phase offset is slightly increased and the normalised phase amplitude is slightly decreased compared with equilibrium. On the other hand, we find a significant difference in the 3.6 \textmu m channel with a strongly decreased normalised phase amplitude compared with the equilibrium simulation.

This effect on the 3.6 \textmu m channel phase curve is consistent with the previous discussion of the contribution functions (Section \ref{section:cf}). Homogenisation of the methane chemistry effectively inverts the shape of the contribution function (\cref{figure:cf}) with the ultimate effect of significantly reducing the number of isotherms that the photosphere crosses, leading to a smaller emission phase amplitude. We note that the 3.6 \textmu m normalised phase amplitude for the relaxation simulation provides a significantly poorer match to the observed normalised phase amplitude, compared with the equilibrium simulation.

\subsection{Summary}

The emission phase curves calculated from our equilibrium simulation indicate that we find a phase offset somewhere inbetween the previous results of \citet{Showman2009} and \citet{DobA13}. The normalised phase amplitude in the 3.6 \textmu m channel is somewhat similar between all three models, however \citet{DobA13} obtains a significantly larger normalised phase amplitude in the 4.5 \textmu m channel compared with both our results and those of \citet{Showman2009}. The effect of wind--driven chemistry is to significantly decrease the normalised phase amplitude in the 3.6 \textmu m channel, as well as to slightly increase the phase offset in both the 3.6 \textmu m and 4.5 \textmu m channels.

%%%%%%%%%%%%%%%%%%%%%%%%
% CONCLUSIONS
%%%%%%%%%%%%%%%%%%%%%%%%
\section{Discussion and Conclusions}

\subsection{Summary of results}

We have presented results from two simulations of the atmosphere of the hot Jupiter HD~189733b: one with the assumption of local chemical equilibrium and the other with the inclusion of wind--driven advection and chemical relaxation.

The trends in the chemical composition between the equilibrium and relaxation simulations are qualitatively similar to previous results for the warmer atmosphere of HD~209458b \citep{DruMM18}, and further demonstrate the importance of 3D modeling of exoplanet atmospheres. The main difference with HD~209458b is a larger equilibrium abundance of methane due to the lower temperature of the atmosphere, leading to larger methane mole fractions in the relaxation simulation. The net result of wind--driven advection is to increase the methane mole fraction by several orders of magnitude, compared with chemical equilibrium, throughout most of the modeled domain, with the largest effect on the dayside.

In \citet{DruMM18} we found relatively unimportant changes ($\sim$1\%) to the thermal and dynamical structure due to wind--driven chemistry for the atmosphere of HD~209458b. However, for the simulations of HD~189733b we find more significant differences of up to $\sim10\%$ in both temperature and wind velocities. For pressures less than $10^4$ Pa we estimate the radiative timescale to be faster than the dynamical timescale, and in this region we observe a local temperature decrease (increase) where the local methane abundance is increased (decreased), compared with chemical equilibrium. This is dominated by changes in the thermal (longwave) cooling.

For pressures greater than $10^4$ Pa, where advection of heat is expected to be more important, this trend no longer holds. We find a significant temperature increase between $10^4$ and $10^5$ Pa in the equatorial region, due to an increase in the differential heating between the dayside and nightside. This increased differential heating drives a faster equatorial jet. 

Our tests show that these changes to the temperature and circulation are due to the large increase in the methane abundance, compared with chemical equilibrium, with a minor contribution from the relatively small changes in the abundances of carbon monoxide and water.

The temperature changes found in this paper (up to 10\%) using a 3D model are similar in magnitude to the temperature changes that we previously found using a 1D model \citep{DruTB16}, including only vertical transport. One of the main effects of vertical transport in the 1D model is to increase the abundance of methane above chemical equilibrium.

When considering the TOA thermal radiative flux it is clear that there is a total reduction in the amount of energy emitted from the dayside and an increase from the nightside when wind--driven chemistry is taken into account. Overall, this indicates an increased efficiency of heat transport from the dayside to nightside atmosphere. 

We find significant quantitative and qualitative differences in the 3D contribution functions between our chemical equilibrium simulation and the results of \citet{DDC17}. The contribution functions of \citet{DDC17} generally peak at lower pressures on the dayside compared with the nightside for all spectral channels considered. In contrast, our simulations show that the contribution functions in the 4.5 \textmu m and 5.8 \textmu m channels are approximately isobaric, while we find an inverse trend to \citet{DDC17} in the 3.6 \textmu m and 8.0 \textmu m channels with contribution functions that peak at lower pressures on the nightside. Since the temperature structure between our simulation and theirs appears to be similar, the cause of the discrepancy is undetermined. When including the effect of wind--driven chemistry we find that the longitude dependence of the contribution functions in the 3.6 \textmu m and 8.0 \textmu m channels is essentially inverted, peaking at lower pressures on the dayside than the nightside. 

Differences in the shapes of the contribution functions explain the differences that we find in the simulated emission phase curves. The main effect of wind--driven chemistry is to significantly reduce the amplitude of the 3.6 \textmu m phase curve, since the photosphere crosses fewer isotherms in the relaxation simulation compared with the equilibrium simulation. The 4.5 \textmu m phase curve does not change significantly between the equilibrium and relaxation simulations. 

Comparing the phase curves calculated from our equilibrium simulation with previous simulations of HD~189733b we find fairly similar phase amplitudes with \citet{Showman2009} in both the 3.6 \textmu m and 4.5 \textmu m spectral channels, but larger differences in the phase amplitude with \citet{DDC17}. The phase offset in both channels varies significantly between all three models. We find that none of the three models \citep[this work;][]{Showman2009,DDC17} match the observed phase curve characteristics well.

\subsection{The spectrum of HD~189733b and the potential effect of clouds and haze}
\label{section:cloud}

The observed transmission spectrum of HD~189733b shows potential signatures of cloud or haze \citep{PonKF08,PonSG13,SinFN16,BarAI17}. The presence of cloud or haze particles can effect the observed spectrum both directly and indirectly. The direct effect is to change the optical properties of the atmosphere and thereby change the observed transit or eclipse depth. The indirect effect occurs if the presence of cloud or haze particles alters the radiative heating rates, and thereby also the thermal structure and atmospheric circulation.

The structure, composition and effect of clouds on the background atmosphere has been investigated using a number of different 3D models with varying degrees of complexity and levels of approximation \citep{LeeDH16,ParFS16,RomR17,LinMB18}. Recently, \citet{LinMM18} reported a flat transmission spectrum from their 3D simulations of HD~189733b that include a state--of--the--art treatment of cloud formation \citep{HelWT08,LeeDH16}, contrary to the molecular absorption features that have been measured \citep{SinFN16}. This model--observation discrepancy suggests that current theoretical models are missing important physical mechanisms at play in the atmospheres of hot Jupiters \citep{LinMM18}. \citet{LeeWD17} investigated the effect of inhomogenous cloud structures on the reflected light and thermal emission spectra of HD~189733b. Haze particles may also be produced in a HD~189733b--like atmosphere via photochemical processes, with their presence impacting the observed transmission spectrum and, potentially, the thermal profile of the atmosphere \citep{LavK17}.

The simulations presented in this paper assume that clouds and haze are not present in the atmosphere. The inclusion of clouds would likely alter the thermal and dynamical structure of the model atmosphere and changes to the optical properties would also directly impact the emission phase curves that we present. 

In this paper we choose to focus on the gas--phase chemistry and, in particular, how this interacts with the circulation and radiative heating. This is important as absorption from gas--phase chemical species (e.g. water) gives rise to the dominant features in many observed spectra of transiting exoplanets \citep[e.g.][]{SinFN16}. In addition, models that include a sophisticated treatment of cloud formation currently assume a simple treatment of the gas--phase chemistry \citep[i.e. local chemical equilibrium,][]{LeeDH16,LinMB18}. The simulations presented in this paper therefore represent an important step in the hierarchy of model complexity.

\subsection{Future prospects}

In this study we found a larger effect of wind--driven chemistry on the thermal and dynamical structure compared with our previous results for the warmer case of HD~209458b \citep{DruMM18}, with differences in the temperature and wind velocities reaching $\sim10\%$. This is due to the cooler atmosphere of HD~189733b (compared with HD~209458b) having a larger equilibrium methane abundance, and hence larger quenched methane abundance. A natural question to ask is whether this trend will continue with cooler temperature planets. 

It is possible that a ``sweet spot'' exists where the dayside atmosphere is warm enough that carbon monoxide is dominant over methane, in chemical equilibrium, but the nightside is cool enough that methane is dominant over carbon monoxide. In this case, horizontal and/or vertical transport may lead to significant changes in the relative abundances of methane and carbon monoxide, with consequences for the opacity and heating rates. For cooler atmospheres still, it is likely that methane will be the dominant carbon species everywhere, with carbon monoxide present as a trace species. If this is true we might expect a similar process as presented in this paper but with the roles of carbon monoxide and methane reversed.

Our simulations used a chemical relaxation scheme \citep{CooS06} to solve for the chemical evolution. The accuracy of this type of scheme is reliant on an accurate estimation of the chemical timescale \citep{TsaKL17}. Whilst we have validated the \citet{CooS06} chemical relaxation scheme against a full chemical kinetics calculation in a 1D model (Appendix \ref{section:app3}), a full kinetics calculation within a 3D framework has not yet been achieved in the literature. As we have previously discussed in \citet{DruMM18} there remain significant differences in the results between the 3D chemical relaxation approach (complex dynamics with simplified chemistry) and the pseudo--2D chemical kinetics approach (complex chemistry with simplified dynamics) of \citet{AguPV14}. It is also unclear whether it is possible to include a treatment of photochemistry within the chemical relaxation method. For these reasons, a model consistently coupling chemical kinetics calculations within a 3D framework is required.

Using a 2D steady-state circulation model \citet{TreCM17} found that the deep atmosphere ($P\gtrsim1\times10^5$ Pa) of hot Jupiters may be significantly hotter than predicted by 1D radiative-convective equilibrium models, possibly due to the advection of potential temperature. GCMs also show a trend of converging towards a hotter profile in the deep atmosphere \citep{AmuMB16,MayDB17} though, due to computational limitations, the models cannot be integrated for long enough to reach a steady-state at these pressures. 

A hotter deep atmosphere may directly impact atmospheric emission within the 4.5 \textmu m Spitzer/IRAC channel, as the contribution function in this spectral channel peaks near to this region in the simulations presented here. In addition, our results have shown that methane first departs from the equilibrium profile at pressures greater than $P=1\times10^5$ Pa. Therefore, a hotter deep atmosphere may change the equilibrium abundance at the quench point and the pressure level of the quench point itself, with consequences for the chemical abundances at lower pressures.

Our results add to the evidence that solar composition gas--phase models fail to match the observed properties of the emission phase curve (i.e. the phase offset and normalised phase amplitude) within the Spitzer/IRAC 3.6 \textmu m and 4.5 \textmu m channels, both assuming local chemical equilibrium and with the effect of wind--driven chemistry. An important missing ingredient in the simulations presented in this paper is cloud formation (Section \ref{section:cloud}) which can have a significant impact on the thermal structure and observed spectra. Consistently coupling a cloud formation module with gas--phase wind--driven chemistry in a 3D model is a significant challange, but is likely required to understand the atmospheric composition of hot exoplanets. On the other hand, \citet{NikSF18} recently reported the clearest hot gas-giant atmosphere to date, which may prove to be a highly useful observational target for testing the theory of gas--phase chemistry.

\begin{acknowledgements}
The authors thank the annonymous referee for their report that helped to improve the quality of this paper. This work is partly
supported by the European Research Council under the
European Community’s Seventh Framework Programme
(FP7/2007-2013 Grant Agreement No. 336792-CREATES and No. 320478-TOFU). N.J.M. and J.G. are partially funded by a
Leverhulme Trust Research Project Grant. J.M.
acknowledges the support of a Met Office Academic Partnership
secondment. This work was performed using the DiRAC Data Intensive service at Leicester, operated by the University of Leicester IT Services, which forms part of the STFC DiRAC HPC Facility (www.dirac.ac.uk). The equipment was funded by BEIS capital funding via STFC capital grants ST/K000373/1 and ST/R002363/1 and STFC DiRAC Operations grant ST/R001014/1. DiRAC is part of the National e-Infrastructure. This work also used the
University of Exeter Supercomputer ISCA.
\end{acknowledgements}

\appendix

%%%%%%%%%%%%%%%%%%%%%%%%
\section{Deriving the contribution function}
%%%%%%%%%%%%%%%%%%%%%%%%
\label{section:app1}

We start with the half-range (upward) intensity at the top of the atmosphere \citep[see][section 5.4.3]{Thomas1999}
\begin{alignat}{2}
  I_{\nu}^+\left(0,\mu,\phi\right) = I_{\nu}^+\left(\tau^*,\mu,\phi\right){\rm e}^{-\tau^*/\mu} \nonumber \\
  + \int_0^{\tau^*} \frac{d\tau'}{\mu}{\rm e}^{-\tau'/\mu}B_{\nu}\left(\tau'\right),
\label{equation:i+}
\end{alignat}
where $I_{\nu}^+$ is the hemispherical intensity at wavenumber $\nu$, $\mu=\cos\theta$ where $\theta$ is the zenith angle, $\phi$ is the azimuthal angle, $B_{\nu}$ is the Planck function and $\tau$ is the optical depth,
\begin{equation}
  \tau = \int_0^s  ds' \kappa\left(s'\right),
\end{equation}
where $\kappa$ is the absorption coefficient and $s$ is the path.

The top of the atmosphere is denoted $\tau=0$ while the bottom of the atmosphere is denoted $\tau=\tau^*$. The first term on the right in \cref{equation:i+} is the upward intensity from the bottom of the atmosphere that has survived extinction while the second term on the right is the thermal contribution from the atmosphere. For a gas--giant atmosphere with no solid surface we can assume $\tau^*=\infty$ which reduces \cref{equation:i+} to
\begin{equation}
 I_{\nu}^+\left(0,\mu,\phi\right) = \int_0^{\infty} \frac{d\tau'}{\mu}{\rm e}^{-\tau'/\mu}B_{\nu}\left(\tau'\right).
\end{equation}

Changing the limits of the integral, we can write the intensity contribution of a particular layer (dropping the $\nu$ notation) as
\begin{equation}
   I^+\left(0,\mu,\phi\right) \equiv \mathcal{CF}_I\left(\mu,\phi\right) = \int_{\tau_1}^{\tau_2} \frac{d\tau'}{\mu}{\rm e}^{-\tau'/\mu}B\left(\tau'\right)
\end{equation}
where we have defined the intensity contribution function $\mathcal{CF}_I$. Evaluating the integral we obtain
\begin{alignat}{2}
  \mathcal{CF}_I\left(\mu,\phi\right) &= B\left(\tau_{1,2}\right)\left[{\rm e}^{-\tau_1/\mu}-{\rm e}^{-\tau_2/\mu}\right] \nonumber \\
\mathcal{CF}_I\left(\mu,\phi\right) &= B\left(\tau_{1,2}\right)d\left[{\rm e}^{-\tau/\mu}\right],
\label{equation:cfi}
\end{alignat}
where $B\left(\tau_{1,2}\right)$ is the Planck function of the layer $\tau_1\rightarrow\tau_2$, which is assumed to be constant across the layer.

Discretising \cref{equation:cfi} onto a model grid would give the contribution of a model layer to the emergent intensity. To generalise this, we can introduce a factor $1/d\log P$ \citep[e.g.][]{ChaH87} to give the contribution per decade in pressure. Finally then, we have
\begin{equation}
\mathcal{CF}_I\left(\mu,\phi\right) = B\left(\tau_{1,2}\right)\frac{d\left[{\rm e}^{-\tau/\mu}\right]}{d\left[\log P\right]}.
\label{equation:final_cfi}
\end{equation}

It is clear that \cref{equation:final_cfi} is equivalent to \citet[][their Eq. 2]{Knutson2009} if $\mu=1$. For the calculations presented in this paper $\mu=1$. However, our equation is different to that presented in both \citet{Griffith1998} and \citet{DDC17}, where the negative sign is missing in the exponential factor; we assume that this negative sign is taken care of elsewhere in their implementation.

Importantly we note that $\mathcal{CF}_I$ is a physical quantity with units ${\rm W}~{\rm m}^{-2}~{\rm ster}^{-1}$. However, it is more common to present the {\it normalised} contribution function ($\bar{\mathcal{CF}}_I$) such that the layer that contributes the most to the top-of-atmosphere intensity has a value $\bar{\mathcal{CF}}_I=1$. In the main body of this paper we present the quantity $\bar{\mathcal{CF}}_I$.

%%%%%%%%%%%%%%%%%%%%%%%%
\section{Validating the contribution function calculation}
%%%%%%%%%%%%%%%%%%%%%%%%

\label{section:app2}

To validate the implementation of the contribution function calculation we first derive the {\it flux} contribution function ($\mathcal{CF}_F$). Integrating \cref{equation:cfi} over $\mu$ and $\phi$ \citep[see][Eq. 5.48]{Thomas1999}
\begin{alignat}{2}
  \mathcal{CF}_F &= \int_0^{2\pi} d\phi \int_0^1 d\mu \mu \mathcal{CF}_I\left(\mu,\phi\right) \nonumber \\
  \mathcal{CF}_F &= 2\pi B\left(\tau_{1,2}\right) \int_0^1d\mu\left[\mu{\rm e}^{-\tau_1/\mu} - \mu{\rm e}^{-\tau_2/\mu}\right]. \label{eq:cff_int}
\end{alignat}

To evalute the integral in \cref{eq:cff_int} we use the diffusivity approximation \citep[see][section 11.2.5]{Thomas1999}
\begin{equation}
  \int_0^1d\mu \mu{\rm e}^{-\tau/\mu} = \frac{{\rm e}^{-D\tau}}{D} \nonumber
\end{equation}
where $D$ is the diffusivity factor. Finally, we now have
\begin{alignat}{2}
  \mathcal{CF}_F &= \frac{2\pi B{\left(\tau_{1,2}\right)}}{D}\left[{\rm e}^{-D\tau_1}-{\rm e}^{-D\tau_2}\right] \nonumber \\
\mathcal{CF}_F &= \frac{2\pi B{\left(\tau_{1,2}\right)}}{D}d\left[{\rm e}^{-D\tau}\right]. \label{equation:cff}
\end{alignat}

$\mathcal{CF}_F$ is the amount of flux that escapes to space from a particular layer of the atmosphere. For the special case of an isothermal atmosphere this will be equal to the divergence of the radiative flux ($\nabla F$) in the layer, as exchange with other layers of the atmosphere will be zero \citep[see][section 11.2.7]{Thomas1999}. We note that $\nabla F$ is the main output of the radiative scheme and is extensively tested. In the isothermal case $D\sim2$ \citep{Edw96}.

\begin{figure}
  \begin{center}
     \includegraphics[width=0.5\textwidth]{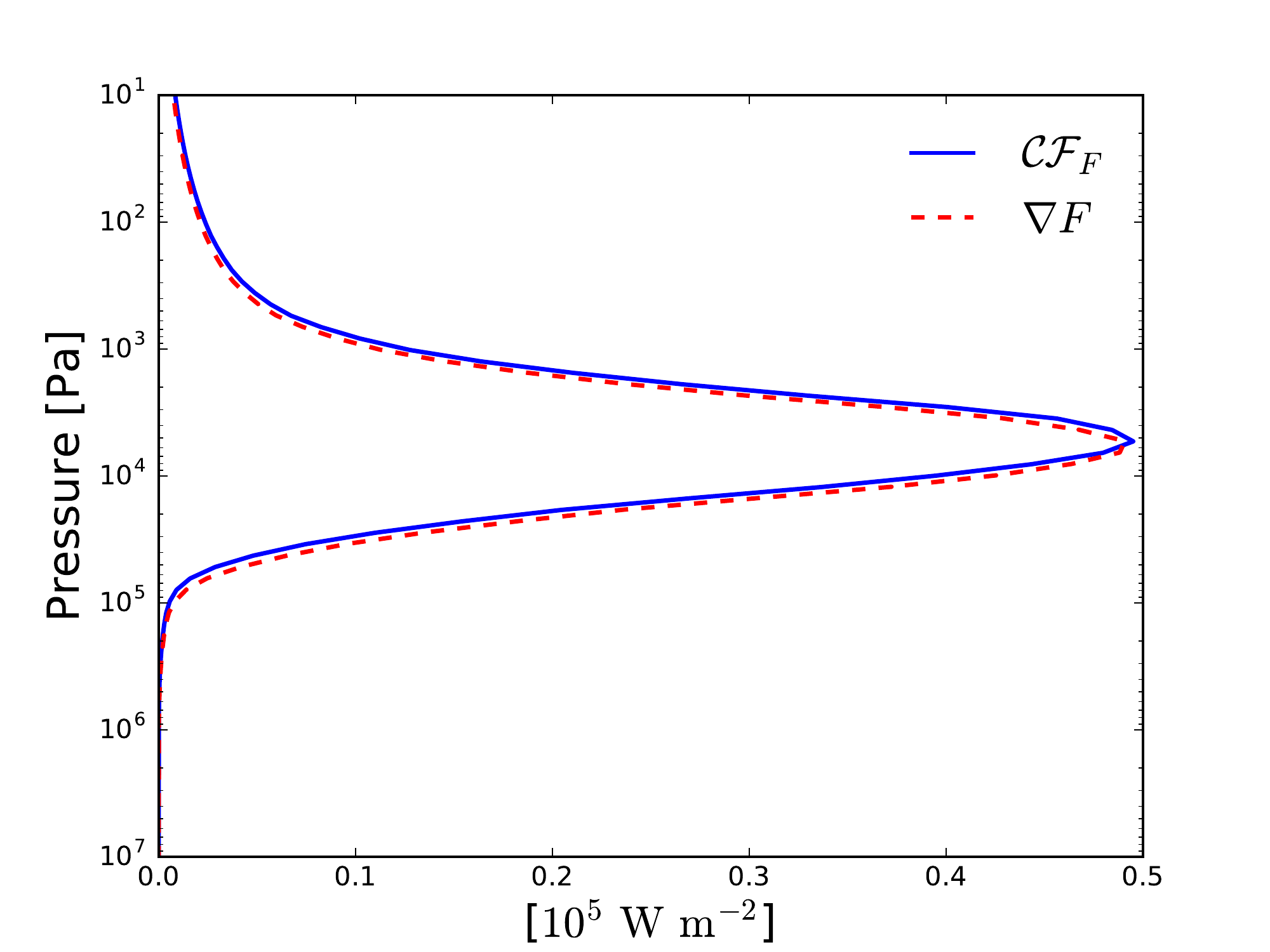}
  \end{center}
  \caption{The flux contribution function $\mathcal{CF}_F$ and the net flux divergence $\nabla F$ across the layer for a 1000 K isothermal atmosphere.}
  \label{figure:test}
\end{figure}

We compare $\mathcal{CF}_F$ and $\nabla F$ for a 1000 K isothermal atmosphere (where we have used $D=2.0$) in \cref{figure:test} and we find excellent agreement between the two quantities, validating our implementation of the contribution function calculation.

\section{Validating the chemical relaxation scheme}\label{section:app3}

Chemical relaxation methods rely on the accurate estimation and paramaterisation of chemical timescale that is based on results from a full chemical kinetics calculation \citep{TsaKL17}. To assess the accuracy of the chemical relaxation scheme that we use in this study \citep{CooS06} we compare it against a full chemical kinetics calculation using two different chemical networks: \citet{Venot2012} and \citet{TsaLG2017}. We used the chemical equilibrium pressure--temperature profile and model parameters for HD~189733b from \citet{DruTB16} and tested a range of eddy diffusion coefficients ($10^8<K_{zz}<10^{11}$ cm$^2$~s$^{-1}$) that are constant with pressure.

\begin{figure}
  \begin{center}
    \includegraphics[width=0.5\textwidth]{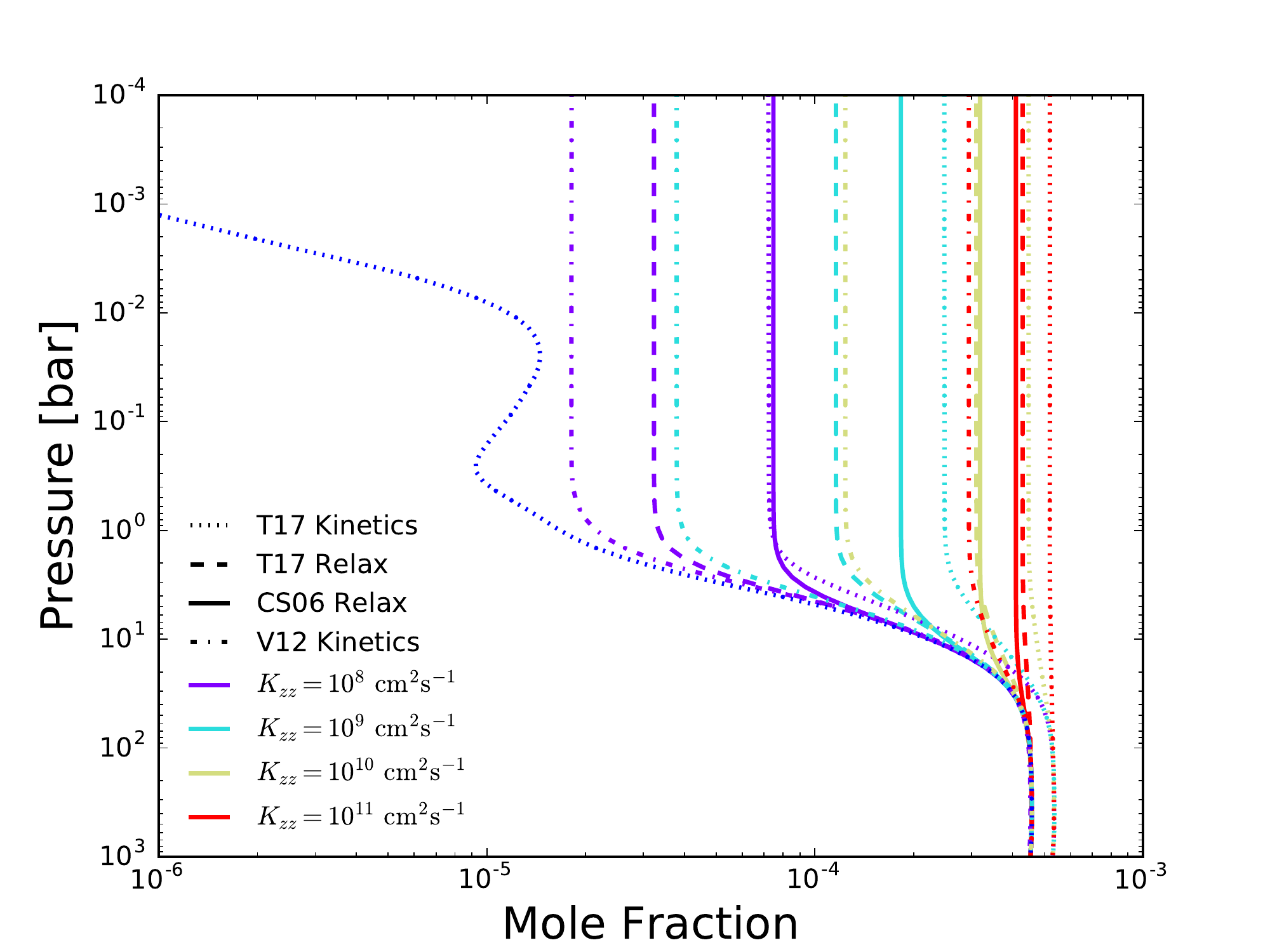}
  \end{center}
  \caption{The mole fraction of methane along a 1D profile with different values of the $K_{zz}$. We compare the steady-state abundances obtained from the \citet{CooS06} chemical relaxation scheme (CS06 Relax) with the \citet{TsaKL17} chemical relaxation scheme (T17 Relax) as well as chemical kinetics calculations using both the \citet{Venot2012} and \citet{TsaLG2017} chemical networks (V12 Kinetics and T17 Kinetics, respectively). The chemical equilibrium mole fraction is shown in dotted blue. The calculations are performed with the pressure-temperature profile and planetary parameters (e.g. surface gravity) of HD~189733b.}
 \label{figure:1d_test}
\end{figure}

\cref{figure:1d_test} shows the comparison between the \citet{CooS06} chemical relaxation with the \citet{Venot2012} and \citet{TsaLG2017} chemical kinetics schemes, as well as the \citet{TsaKL17} relaxation scheme. For the smallest $K_{zz}$ value, the \citet{CooS06} relaxation result matches very well with the \citet{TsaLG2017} kinetics result, while for larger values the  \citet{CooS06} relaxation result is somewhere in between the \citet{TsaLG2017} and \citet{Venot2012} kinetics results.

We conclude that the \citet{CooS06} chemical relaxation method has an acceptable level of accuracy, compared with a full chemical kinetics calculation, for the current application. Importantly, we note that the result obtained with the \citet{CooS06} relaxation scheme typically lies somewhere in between the result obtained with the \citet{TsaLG2017} and \citet{Venot2012} kinetics schemes, depending on the value of $K_{zz}$. Therefore, potential inaccuracies of the present chemical relaxation method are within the expected uncertainties of currently available chemical networks, for HD~189733b conditions.

%Relax v kinetics

%%%%%%%%%%%%%%%%%%%%%%%%%%%%%%%%%%%%
\section{Sensitivity to the chemical timescale}
\label{section:test}
%%%%%%%%%%%%%%%%%%%%%%%%%%%%%%%%%%%%

\begin{figure}
  \begin{center}
    \includegraphics[width=0.48\textwidth]{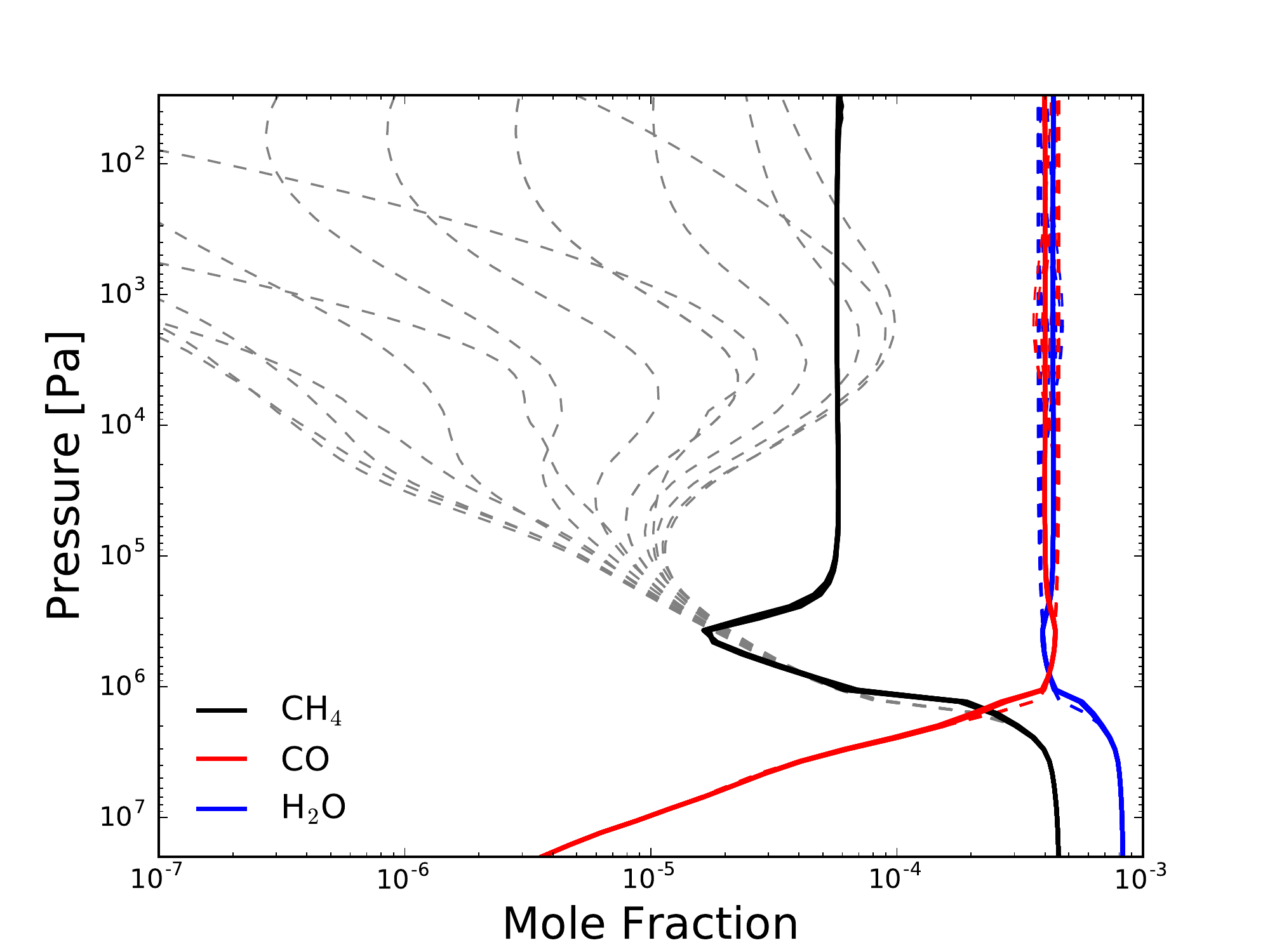} 
    \includegraphics[width=0.48\textwidth]{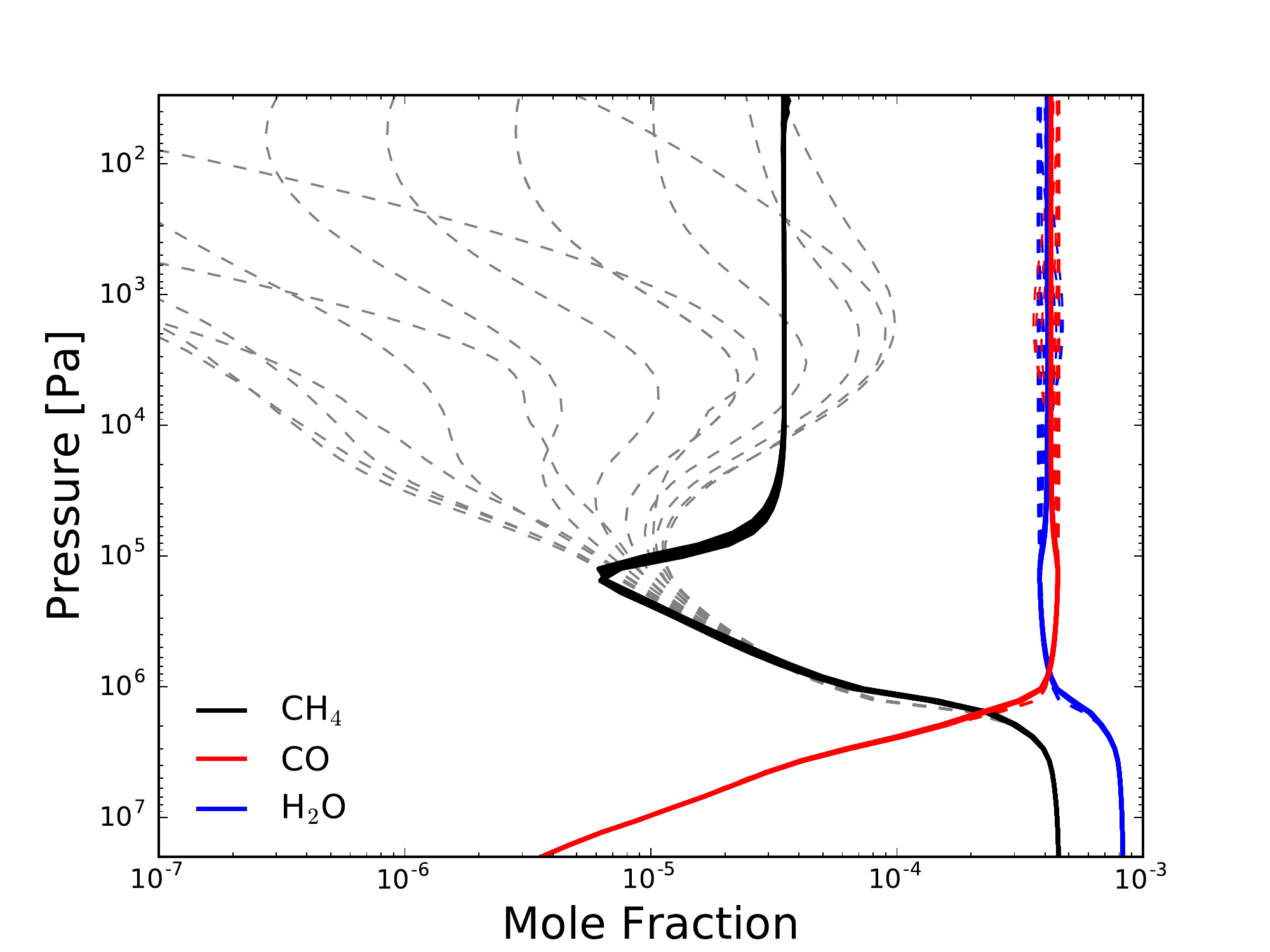} \\
  \end{center}
\caption{Vertical profiles of the chemical equilibrium (dashed) and chemical relaxation (solid) mole fractions of methane, carbon monoxide and water for several longitude points around the equator, for simulations with the \citet{CooS06} chemical timescale multiplied by a factor 10 (left panel) and the \citet{CooS06} chemical timescale divided by a factor 10 (right panel). This figure should be compared with \cref{figure:profiles} which shows the simulation with the nominal \citet{CooS06} chemical timescale.}
\label{figure:vert_profiles_test}
\end{figure}

To test the sensitivity of the model to the value of the chemical timescale we performed additional simulations that are identical to the relaxation simulation but with an artificially increased/decreased chemical timescale by a factor 10. The vertical mole fraction profiles around the equator for these tests are shown in \cref{figure:vert_profiles_test} and should be compared with the results using the nominal chemical timescale in \cref{figure:profiles}.

% Vertical profiles of mole fractions

We find that an increase (decrease) in the chemical timescale leads to an increase (decrease) in the pressure level of the ``turnoff'' point, where the methane abundance begins the increase with decreasing pressure. The location of the quench point depends on the ratio of the transport to chemical timescales. The chemical timescale typically increases with decreasing pressure due to the pressure and temperature dependence of the reaction rate, while the transport timescale depends on the local wind velocities and the relevant length scale \citep[see][Fig. 3]{DruMM18}. An increase in the chemical timescale will therefore shift the quench point to higher pressures, while decreasing the chemical timescale will shift the quench point to lower pressures. 

While changing the chemical timescale clearly alters the pressure level at which meridional transport becomes important, the final vertically quenched methane abundance shows only a small variation between the three cases. The quenched mole fraction of methane using the nominal \citet{CooS06} timescale is $\sim4.5\times10^{-5}$, while in the $\times10$ and $\div10$ cases this changes to $\sim6\times10^{-5}$ and $\sim3.5\times10^{-5}$, respectively. This test suggests that uncertainties or errors in the value of the chosen timescale do not lead to large differences in our results.

%%%%%%%%%%%%%%%%%%%%%%%%%%%%%%%%%%%%
\section{Conservation of global mass of the elements}
\label{section:cons}
%%%%%%%%%%%%%%%%%%%%%%%%%%%%%%%%%%%%

As a basic test of the advection scheme we check the conservation of the global mass of carbon and oxygen. We note that this test does not necessarily validate the accuracy of the advection scheme but does provide a basic check that the advection scheme is not artificially gaining or losing mass. 

The total mass of the atmosphere $M$ can written,
\begin{equation}
  M = \sum_{k} \rho_{k} V_{\rm k},
\end{equation}
where $\rho_k$ and $V_k$ are the mass density and volume of the cell $k$, respectively, and the sum is over the total number of cells in the model grid. Similarly, the total mass of an element $i$ can be written as,
\begin{equation}
  M_{i} = \sum_k \sum_j \alpha_{i,j}w_{j,k}\rho_kV_k =\frac{1}{\mu} \sum_k \sum_j \alpha_{i,j}\mu_jf_{j,k}\rho_kV_k
\end{equation}
where $w_{j,k}$ and $f_{j,k}$ are the mass fraction and mole fraction of the species $j$, respectively, $\alpha_{i,j}$ is the fractional mass of element $i$ in species $j$, $\mu_{j}$ is the molar mass of the species $j$ and $\mu$ is the mean molar mass. Specifically for the simulations presented in this paper, that include methane, water and carbon monoxide, we can write the global mass of carbon as,
\begin{alignat}{2}
  M_{\rm C} = & \frac{1}{\mu}\sum_k\rho_kV_k\left[\alpha_{\rm C, CO}\mu_{\rm CO}f_{{\rm CO},k} + \alpha_{\rm C, CH_4}\mu_{\rm CH_4}f_{{\rm CH_4},k} \right], \nonumber \\
  M_{\rm C} = & \frac{1}{\mu}\sum_k\rho_kV_k\left[\alpha_{\rm C, CO}\mu_{\rm CO}f_{{\rm CO},k} + \alpha_{\rm C, CH_4}\mu_{\rm CH_4}\left(A_{\rm C} - f_{{\rm CO},k}\right) \right],
\end{alignat}
where we have replaced $f_{{\rm CH_4},k}$ with $f_{{\rm CO},k}$ using the mass balance approach and $A_{\rm C}$ is a constant \citep{CooS06,DruMM18}.
Similarly for oxygen we have, 
\begin{alignat}{2}
  M_{\rm O} = & \frac{1}{\mu}\sum_k\rho_kV_k\left[\alpha_{\rm O, CO}\mu_{\rm CO}f_{{\rm CO},k} + \alpha_{\rm O, H_2O}\mu_{\rm H_2O}f_{{\rm H_2O},k} \right], \nonumber \\
  M_{\rm O} = & \frac{1}{\mu}\sum_k\rho_kV_k\left[\alpha_{\rm O, CO}\mu_{\rm CO}f_{{\rm CO},k} + \alpha_{\rm O, H_2O}\mu_{\rm H_2O}\left(A_{\rm O} - f_{{\rm CO},k}\right) \right].
\end{alignat}
As before, we have replaced $f_{{\rm H_2O},k}$ with $f_{{\rm CO},k}$ by assuming mass balance and $A_{\rm O}$ is a constant. We note that it is the mole fraction of carbon monoxide ($f_{\rm CO}$) that is advected as a tracer is these simulations. 

\begin{figure}
  \begin{center}
    \includegraphics[width=0.48\textwidth]{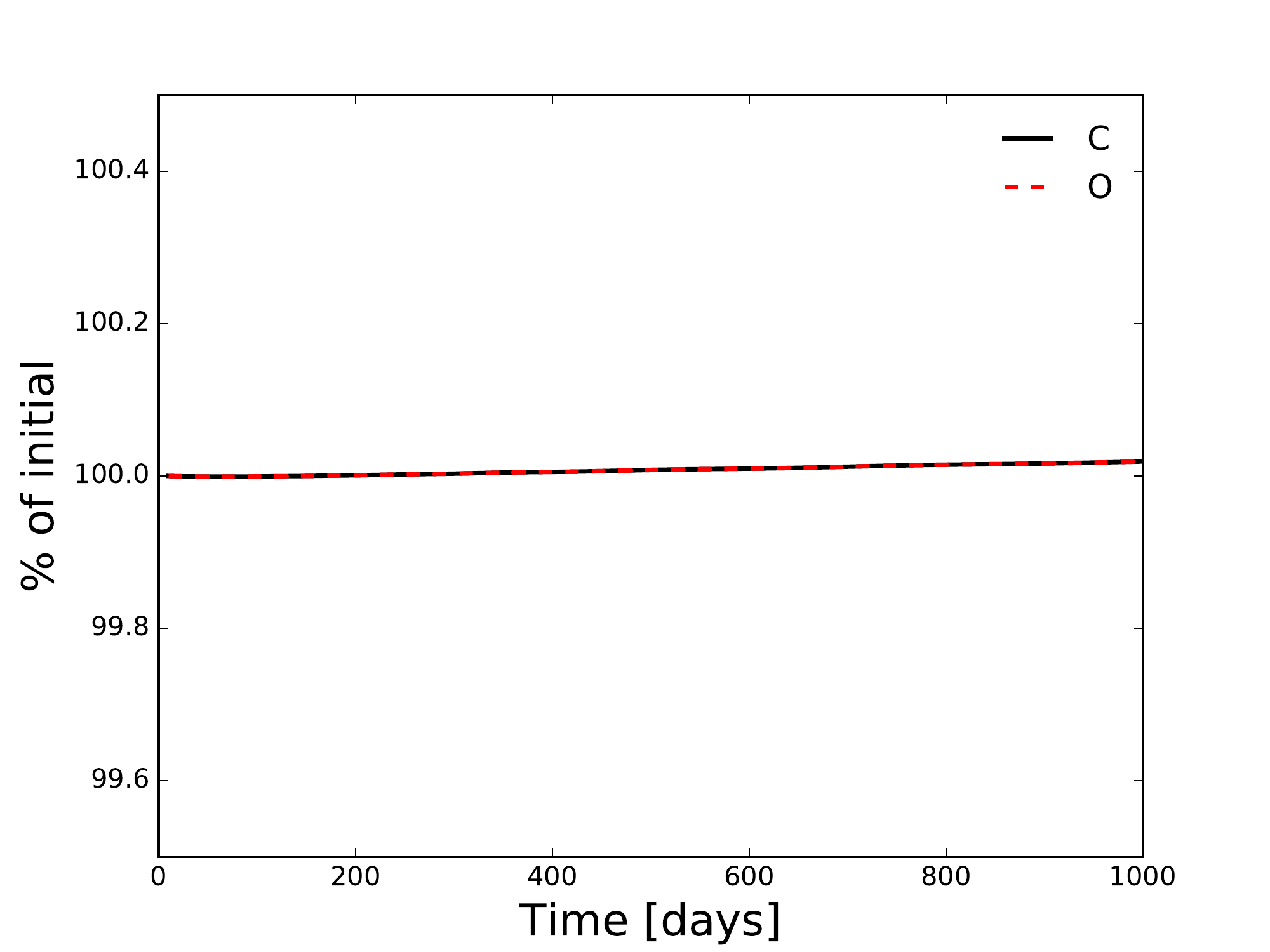}
    \includegraphics[width=0.48\textwidth]{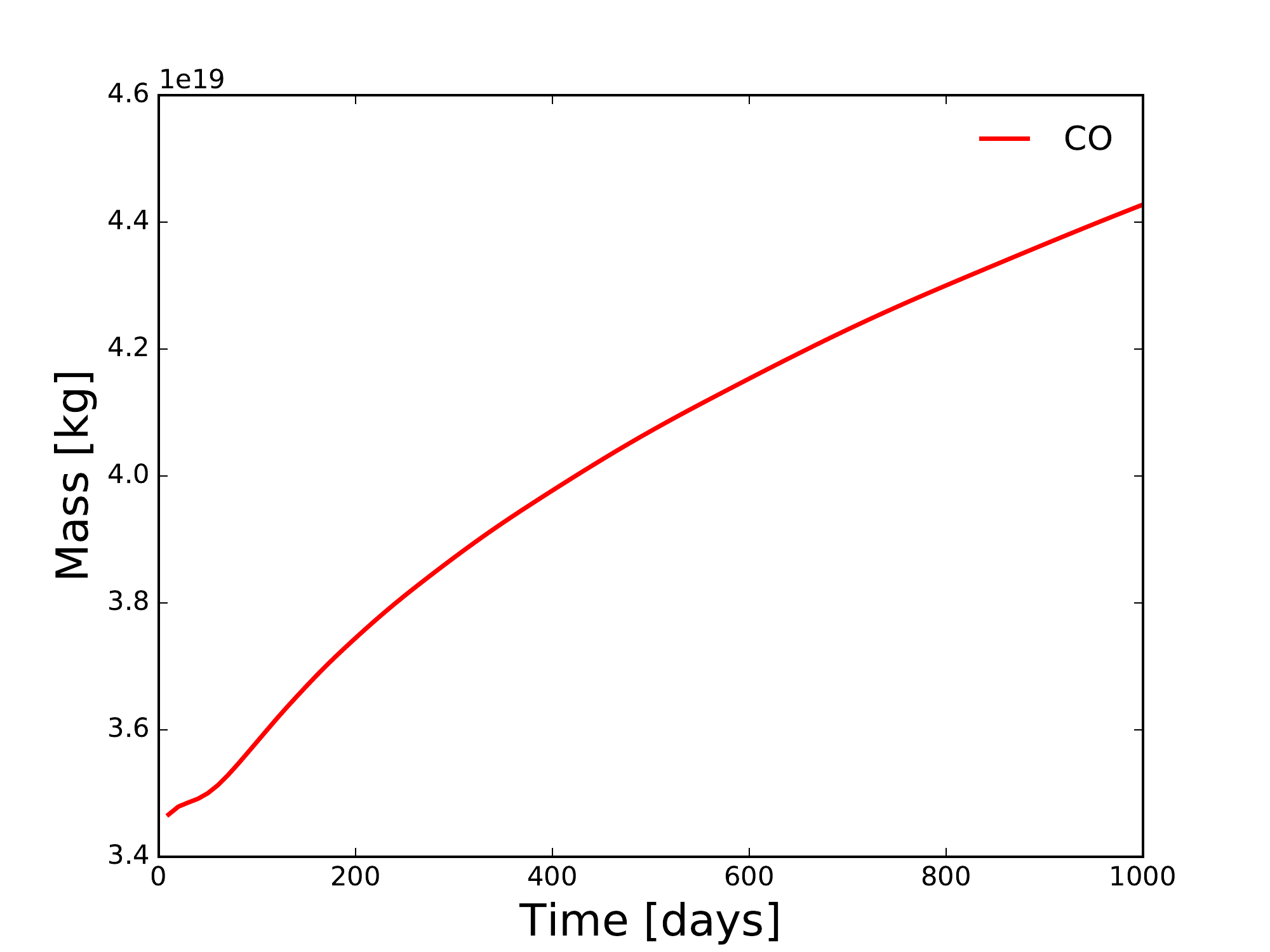}
    \caption{Left panel: the percentage of the initial global mass of carbon and oxygen throughout the simulation. Right panel: the global mass of carbon monoxide throughout the simulation.}
    \label{figure:conservation}
  \end{center}
\end{figure}

\cref{figure:conservation} shows the conservation of the global mass of carbon and oxygen as a function of simulation time. The global mass of carbon and oxygen is conserved to within significantly better than $99.9\%$ over 1000 days. As a point of comparison, \cref{figure:conservation} also shows the change in the global mass of carbon monoxide which increases by approximately 25\%. 

The fractional change in the mass of carbon monoxide is much larger than the errors in the conservation of the global mass of elemental carbon and oxygen. This suggests that the change in the mass is due to the physics schemes included in the model and not due to numerical gain or loss of mass from the advection scheme.

\bibliographystyle{apj}

\begin{thebibliography}{}
\expandafter\ifx\csname natexlab\endcsname\relax\def\natexlab#1{#1}\fi

\bibitem[{{Ag{\'u}ndez} {et~al.}(2014){Ag{\'u}ndez}, {Parmentier}, {Venot},
  {Hersant}, \& {Selsis}}]{AguPV14}
{Ag{\'u}ndez}, M., {Parmentier}, V., {Venot}, O., {Hersant}, F., \& {Selsis},
  F. 2014, \aap, 564, A73

\bibitem[{{Amundsen} {et~al.}(2014){Amundsen}, {Baraffe}, {Tremblin},
  {Manners}, {Hayek}, {Mayne}, \& {Acreman}}]{AmuBT14}
{Amundsen}, D.~S., {Baraffe}, I., {Tremblin}, P., {et~al.} 2014, \aap, 564, A59

\bibitem[{{Amundsen} {et~al.}(2017){Amundsen}, {Tremblin}, {Manners},
  {Baraffe}, \& {Mayne}}]{AmuTM17}
{Amundsen}, D.~S., {Tremblin}, P., {Manners}, J., {Baraffe}, I., \& {Mayne},
  N.~J. 2017, \aap, 598, A97

\bibitem[{{Amundsen} {et~al.}(2016){Amundsen}, {Mayne}, {Baraffe}, {Manners},
  {Tremblin}, {Drummond}, {Smith}, {Acreman}, \& {Homeier}}]{AmuMB16}
{Amundsen}, D.~S., {Mayne}, N.~J., {Baraffe}, I., {et~al.} 2016, \aap, 595, A36

\bibitem[{{Asplund} {et~al.}(2009){Asplund}, {Grevesse}, {Sauval}, \&
  {Scott}}]{Asplund2009}
{Asplund}, M., {Grevesse}, N., {Sauval}, A.~J., \& {Scott}, P. 2009, \araa, 47,
  481

\bibitem[{{Barman} {et~al.}(2005){Barman}, {Hauschildt}, \&
  {Allard}}]{Barman2005}
{Barman}, T.~S., {Hauschildt}, P.~H., \& {Allard}, F. 2005, \apj, 632, 1132

\bibitem[{{Barstow} {et~al.}(2017){Barstow}, {Aigrain}, {Irwin}, \&
  {Sing}}]{BarAI17}
{Barstow}, J.~K., {Aigrain}, S., {Irwin}, P.~G.~J., \& {Sing}, D.~K. 2017,
  \apj, 834, 50

\bibitem[{{Boutle} {et~al.}(2017){Boutle}, {Mayne}, {Drummond}, {Manners},
  {Goyal}, {Hugo Lambert}, {Acreman}, \& {Earnshaw}}]{BouMD17}
{Boutle}, I.~A., {Mayne}, N.~J., {Drummond}, B., {et~al.} 2017, \aap, 601, A120

\bibitem[{{Chamberlain} \& {Hunten}(1987)}]{ChaH87}
{Chamberlain}, J.~W., \& {Hunten}, D.~M. 1987, {Theory of planetary
  atmospheres. An introduction to their physics andchemistry.}

\bibitem[{{Cooper} \& {Showman}(2006)}]{CooS06}
{Cooper}, C.~S., \& {Showman}, A.~P. 2006, \apj, 649, 1048

\bibitem[{{Dang} {et~al.}(2018){Dang}, {Cowan}, {Schwartz}, {Rauscher},
  {Zhang}, {Knutson}, {Line}, {Dobbs-Dixon}, {Deming}, {Sundararajan},
  {Fortney}, \& {Zhao}}]{DanCS18}
{Dang}, L., {Cowan}, N.~B., {Schwartz}, J.~C., {et~al.} 2018, Nature Astronomy,
  2, 220

\bibitem[{{Dobbs-Dixon} \& {Agol}(2013)}]{DobA13}
{Dobbs-Dixon}, I., \& {Agol}, E. 2013, \mnras, 435, 3159

\bibitem[{{Dobbs-Dixon} \& {Cowan}(2017)}]{DDC17}
{Dobbs-Dixon}, I., \& {Cowan}, N.~B. 2017, \apjl, 851, L26

\bibitem[{{Dobbs-Dixon} \& {Lin}(2008)}]{DobL08}
{Dobbs-Dixon}, I., \& {Lin}, D.~N.~C. 2008, \apj, 673, 513

\bibitem[{{Drummond} {et~al.}(2018{\natexlab{a}}){Drummond}, {Mayne},
  {Baraffe}, {Tremblin}, {Manners}, {Amundsen}, {Goyal}, \&
  {Acreman}}]{DruMB18}
{Drummond}, B., {Mayne}, N.~J., {Baraffe}, I., {et~al.} 2018{\natexlab{a}},
  \aap, 612, A105

\bibitem[{{Drummond} {et~al.}(2016){Drummond}, {Tremblin}, {Baraffe},
  {Amundsen}, {Mayne}, {Venot}, \& {Goyal}}]{DruTB16}
{Drummond}, B., {Tremblin}, P., {Baraffe}, I., {et~al.} 2016, \aap, 594, A69

\bibitem[{{Drummond} {et~al.}(2018{\natexlab{b}}){Drummond}, {Mayne},
  {Manners}, {Carter}, {Boutle}, {Baraffe}, {H{\'e}brard}, {Tremblin}, {Sing},
  {Amundsen}, \& {Acreman}}]{DruMM18}
{Drummond}, B., {Mayne}, N.~J., {Manners}, J., {et~al.} 2018{\natexlab{b}},
  \apjl, 855, L31

\bibitem[{Edwards(1996)}]{Edw96}
Edwards, J.~M. 1996, Journal of the Atmospheric Sciences, 53, 1921

\bibitem[{Edwards \& Slingo(1996)}]{EdwS96}
Edwards, J.~M., \& Slingo, A. 1996, Quarterly Journal of the Royal
  Meteorological Society, 122, 689

\bibitem[{{Fortney} {et~al.}(2005){Fortney}, {Marley}, {Lodders}, {Saumon}, \&
  {Freedman}}]{ForML05}
{Fortney}, J.~J., {Marley}, M.~S., {Lodders}, K., {Saumon}, D., \& {Freedman},
  R. 2005, \apjl, 627, L69

\bibitem[{{Goyal} {et~al.}(2018){Goyal}, {Mayne}, {Sing}, {Drummond},
  {Tremblin}, {Amundsen}, {Evans}, {Carter}, {Spake}, {Baraffe}, {Nikolov},
  {Manners}, {Chabrier}, \& {Hebrard}}]{GoyMS18}
{Goyal}, J.~M., {Mayne}, N., {Sing}, D.~K., {et~al.} 2018, \mnras, 474, 5158

\bibitem[{{Griffith} {et~al.}(1998){Griffith}, {Yelle}, \&
  {Marley}}]{Griffith1998}
{Griffith}, C.~A., {Yelle}, R.~V., \& {Marley}, M.~S. 1998, Science, 282, 2063

\bibitem[{{Helling} {et~al.}(2008){Helling}, {Woitke}, \& {Thi}}]{HelWT08}
{Helling}, C., {Woitke}, P., \& {Thi}, W.-F. 2008, \aap, 485, 547

\bibitem[{{Heng} {et~al.}(2011){Heng}, {Menou}, \& {Phillipps}}]{HenMP11}
{Heng}, K., {Menou}, K., \& {Phillipps}, P.~J. 2011, \mnras, 413, 2380

\bibitem[{{Iro} {et~al.}(2005){Iro}, {B{\'e}zard}, \& {Guillot}}]{Iro2005}
{Iro}, N., {B{\'e}zard}, B., \& {Guillot}, T. 2005, \aap, 436, 719

\bibitem[{{Irwin} {et~al.}(2008){Irwin}, {Teanby}, {de Kok}, {Fletcher},
  {Howett}, {Tsang}, {Wilson}, {Calcutt}, {Nixon}, \& {Parrish}}]{IrwTd08}
{Irwin}, P.~G.~J., {Teanby}, N.~A., {de Kok}, R., {et~al.} 2008, \jqsrt, 109,
  1136

\bibitem[{{Kataria} {et~al.}(2016){Kataria}, {Sing}, {Lewis}, {Visscher},
  {Showman}, {Fortney}, \& {Marley}}]{KatSL16}
{Kataria}, T., {Sing}, D.~K., {Lewis}, N.~K., {et~al.} 2016, \apj, 821, 9

\bibitem[{{Knutson} {et~al.}(2009){Knutson}, {Charbonneau}, {Cowan}, {Fortney},
  {Showman}, {Agol}, {Henry}, {Everett}, \& {Allen}}]{Knutson2009}
{Knutson}, H.~A., {Charbonneau}, D., {Cowan}, N.~B., {et~al.} 2009, \apj, 690,
  822

\bibitem[{{Knutson} {et~al.}(2012){Knutson}, {Lewis}, {Fortney}, {Burrows},
  {Showman}, {Cowan}, {Agol}, {Aigrain}, {Charbonneau}, {Deming}, {D{\'e}sert},
  {Henry}, {Langton}, \& {Laughlin}}]{KnuLF12}
{Knutson}, H.~A., {Lewis}, N., {Fortney}, J.~J., {et~al.} 2012, \apj, 754, 22

\bibitem[{{Lavvas} \& {Koskinen}(2017)}]{LavK17}
{Lavvas}, P., \& {Koskinen}, T. 2017, \apj, 847, 32

\bibitem[{{Lee} {et~al.}(2016){Lee}, {Dobbs-Dixon}, {Helling}, {Bognar}, \&
  {Woitke}}]{LeeDH16}
{Lee}, G., {Dobbs-Dixon}, I., {Helling}, C., {Bognar}, K., \& {Woitke}, P.
  2016, \aap, 594, A48

\bibitem[{{Lee} {et~al.}(2017){Lee}, {Wood}, {Dobbs-Dixon}, {Rice}, \&
  {Helling}}]{LeeWD17}
{Lee}, G.~K.~H., {Wood}, K., {Dobbs-Dixon}, I., {Rice}, A., \& {Helling}, C.
  2017, \aap, 601, A22

\bibitem[{{Lewis} {et~al.}(2018){Lewis}, {Lambert}, {Boutle}, {Mayne},
  {Manners}, \& {Acreman}}]{LewLB18}
{Lewis}, N.~T., {Lambert}, F.~H., {Boutle}, I.~A., {et~al.} 2018, \apj, 854,
  171

\bibitem[{{Line} {et~al.}(2010){Line}, {Liang}, \& {Yung}}]{LinLY2010}
{Line}, M.~R., {Liang}, M.~C., \& {Yung}, Y.~L. 2010, \apj, 717, 496

\bibitem[{{Lines} {et~al.}(2018{\natexlab{a}}){Lines}, {Manners}, {Mayne},
  {Goyal}, {Carter}, {Boutle}, {Lee}, {Helling}, {Drummond}, {Acreman}, \&
  {Sing}}]{LinMM18}
{Lines}, S., {Manners}, J., {Mayne}, N.~J., {et~al.} 2018{\natexlab{a}},
  \mnras, 481, 194

\bibitem[{{Lines} {et~al.}(2018{\natexlab{b}}){Lines}, {Mayne}, {Boutle},
  {Manners}, {Lee}, {Helling}, {Drummond}, {Amundsen}, {Goyal}, {Acreman},
  {Tremblin}, \& {Kerslake}}]{LinMB18}
{Lines}, S., {Mayne}, N.~J., {Boutle}, I.~A., {et~al.} 2018{\natexlab{b}},
  \aap, 615, A97

\bibitem[{{Louden} \& {Wheatley}(2015)}]{Louden2015}
{Louden}, T., \& {Wheatley}, P.~J. 2015, \apjl, 814, L24

\bibitem[{{Madhusudhan} {et~al.}(2011){Madhusudhan}, {Burrows}, \&
  {Currie}}]{MadBC11}
{Madhusudhan}, N., {Burrows}, A., \& {Currie}, T. 2011, \apj, 737, 34

\bibitem[{{Madhusudhan} \& {Seager}(2009)}]{MadS09}
{Madhusudhan}, N., \& {Seager}, S. 2009, \apj, 707, 24

\bibitem[{{Mayne} {et~al.}(2014{\natexlab{a}}){Mayne}, {Baraffe}, {Acreman},
  {Smith}, {Wood}, {Amundsen}, {Thuburn}, \& {Jackson}}]{MayBA14b}
{Mayne}, N.~J., {Baraffe}, I., {Acreman}, D.~M., {et~al.} 2014{\natexlab{a}},
  Geoscientific Model Development, 7, 3059

\bibitem[{{Mayne} {et~al.}(2014{\natexlab{b}}){Mayne}, {Baraffe}, {Acreman},
  {Smith}, {Browning}, {Sk{\aa}lid Amundsen}, {Wood}, {Thuburn}, \&
  {Jackson}}]{MayBA14}
{Mayne}, N.~J., {Baraffe}, I., {Acreman}, D.~M., {et~al.} 2014{\natexlab{b}}, \aap, 561, A1

\bibitem[{{Mayne} {et~al.}(2017){Mayne}, {Debras}, {Baraffe}, {Thuburn},
  {Amundsen}, {Acreman}, {Smith}, {Browning}, {Manners}, \& {Wood}}]{MayDB17}
{Mayne}, N.~J., {Debras}, F., {Baraffe}, I., {et~al.} 2017, \aap, 604, A79

\bibitem[{{Menou} \& {Rauscher}(2009)}]{MenR09}
{Menou}, K., \& {Rauscher}, E. 2009, \apj, 700, 887

\bibitem[{{Moses} {et~al.}(2011){Moses}, {Visscher}, {Fortney}, {Showman},
  {Lewis}, {Griffith}, {Klippenstein}, {Shabram}, {Friedson}, {Marley}, \&
  {Freedman}}]{Moses2011}
{Moses}, J.~I., {Visscher}, C., {Fortney}, J.~J., {et~al.} 2011, \apj, 737, 15

\bibitem[{{Nikolov} {et~al.}(2018){Nikolov}, {Sing}, {Fortney}, {Goyal},
  {Drummond}, {Evans}, {Gibson}, {De Mooij}, {Rustamkulov}, {Wakeford},
  {Smalley}, {Burgasser}, {Hellier}, {Helling}, {Mayne}, {Madhusudhan},
  {Kataria}, {Baines}, {Carter}, {Ballester}, {Barstow}, {McCleery}, \&
  {Spake}}]{NikSF18}
{Nikolov}, N., {Sing}, D.~K., {Fortney}, J.~J., {et~al.} 2018, \nat, 557, 526

\bibitem[{{Parmentier} {et~al.}(2016){Parmentier}, {Fortney}, {Showman},
  {Morley}, \& {Marley}}]{ParFS16}
{Parmentier}, V., {Fortney}, J.~J., {Showman}, A.~P., {Morley}, C., \&
  {Marley}, M.~S. 2016, \apj, 828, 22

\bibitem[{{Pont} {et~al.}(2008){Pont}, {Knutson}, {Gilliland}, {Moutou}, \&
  {Charbonneau}}]{PonKF08}
{Pont}, F., {Knutson}, H., {Gilliland}, R.~L., {Moutou}, C., \& {Charbonneau},
  D. 2008, \mnras, 385, 109

\bibitem[{{Pont} {et~al.}(2013){Pont}, {Sing}, {Gibson}, {Aigrain}, {Henry}, \&
  {Husnoo}}]{PonSG13}
{Pont}, F., {Sing}, D.~K., {Gibson}, N.~P., {et~al.} 2013, \mnras, 432, 2917

\bibitem[{{Rauscher} \& {Menou}(2013)}]{RauM13}
{Rauscher}, E., \& {Menou}, K. 2013, \apj, 764, 103

\bibitem[{{Roman} \& {Rauscher}(2017)}]{RomR17}
{Roman}, M., \& {Rauscher}, E. 2017, \apj, 850, 17

\bibitem[{{Showman} {et~al.}(2009){Showman}, {Fortney}, {Lian}, {Marley},
  {Freedman}, {Knutson}, \& {Charbonneau}}]{Showman2009}
{Showman}, A.~P., {Fortney}, J.~J., {Lian}, Y., {et~al.} 2009, \apj, 699, 564

\bibitem[{{Showman} \& {Guillot}(2002)}]{ShoG02}
{Showman}, A.~P., \& {Guillot}, T. 2002, \aap, 385, 166

\bibitem[{{Sing} {et~al.}(2016){Sing}, {Fortney}, {Nikolov}, {Wakeford},
  {Kataria}, {Evans}, {Aigrain}, {Ballester}, {Burrows}, {Deming},
  {D{\'e}sert}, {Gibson}, {Henry}, {Huitson}, {Knutson}, {Etangs}, {Pont},
  {Showman}, {Vidal-Madjar}, {Williamson}, \& {Wilson}}]{SinFN16}
{Sing}, D.~K., {Fortney}, J.~J., {Nikolov}, N., {et~al.} 2016, \nat, 529, 59

\bibitem[{{Snellen} {et~al.}(2010){Snellen}, {de Kok}, {de Mooij}, \&
  {Albrecht}}]{Snedd10}
{Snellen}, I.~A.~G., {de Kok}, R.~J., {de Mooij}, E.~J.~W., \& {Albrecht}, S.
  2010, \nat, 465, 1049

\bibitem[{{Southworth}(2010)}]{Southworth2010}
{Southworth}, J. 2010, \mnras, 408, 1689

\bibitem[{Thomas \& Stammes(1999)}]{Thomas1999}
Thomas, G.~E., \& Stammes, K. 1999, {Radiative Transfer in the Atmosphere and
  Ocean} (Cambridge University Press)

\bibitem[{{Tremblin} {et~al.}(2017){Tremblin}, {Chabrier}, {Mayne}, {Amundsen},
  {Baraffe}, {Debras}, {Drummond}, {Manners}, \& {Fromang}}]{TreCM17}
{Tremblin}, P., {Chabrier}, G., {Mayne}, N.~J., {et~al.} 2017, \apj, 841, 30

\bibitem[{{Tsai} {et~al.}(2017{\natexlab{a}}){Tsai}, {Kitzmann}, {Lyons},
  {Mendon{\c c}a}, {Grimm}, \& {Heng}}]{TsaKL17}
{Tsai}, S.-M., {Kitzmann}, D., {Lyons}, J.~R., {et~al.} 2017{\natexlab{a}},
  ArXiv e-prints, arXiv:1711.08492

\bibitem[{{Tsai} {et~al.}(2017{\natexlab{b}}){Tsai}, {Lyons}, {Grosheintz},
  {Rimmer}, {Kitzmann}, \& {Heng}}]{TsaLG2017}
{Tsai}, S.-M., {Lyons}, J.~R., {Grosheintz}, L., {et~al.} 2017{\natexlab{b}},
  \apjs, 228, 20

\bibitem[{{Venot} {et~al.}(2012){Venot}, {H{\'e}brard}, {Ag{\'u}ndez},
  {Dobrijevic}, {Selsis}, {Hersant}, {Iro}, \& {Bounaceur}}]{Venot2012}
{Venot}, O., {H{\'e}brard}, E., {Ag{\'u}ndez}, M., {et~al.} 2012, \aap, 546,
  A43

\bibitem[{{Waldmann} {et~al.}(2015){Waldmann}, {Tinetti}, {Rocchetto},
  {Barton}, {Yurchenko}, \& {Tennyson}}]{WalTR15}
{Waldmann}, I.~P., {Tinetti}, G., {Rocchetto}, M., {et~al.} 2015, \apj, 802,
  107

\bibitem[{Wood {et~al.}(2014)Wood, Staniforth, White, Allen, Diamantakis,
  Gross, Melvin, Smith, Vosper, Zerroukat, \& Thuburn}]{WooSW14}
Wood, N., Staniforth, A., White, A., {et~al.} 2014, Quarterly Journal of the
  Royal Meteorological Society, 140, 1505

\bibitem[{{Zahnle} \& {Marley}(2014)}]{Zahnle2014}
{Zahnle}, K.~J., \& {Marley}, M.~S. 2014, \apj, 797, 41

\bibitem[{{Zellem} {et~al.}(2014){Zellem}, {Lewis}, {Knutson}, {Griffith},
  {Showman}, {Fortney}, {Cowan}, {Agol}, {Burrows}, {Charbonneau}, {Deming},
  {Laughlin}, \& {Langton}}]{ZelLK14}
{Zellem}, R.~T., {Lewis}, N.~K., {Knutson}, H.~A., {et~al.} 2014, \apj, 790, 53

\end{thebibliography}

\end{document}